\documentclass{aa}  
\usepackage{graphicx}
\usepackage[varg]{txfonts}
\begin{document}
   \title{SiO and CH$_3$CCH abundances and dust emission in high-mass star-forming cores
 \thanks{Based on observations collected at the European Southern Observatory, La Silla, Chile.}}

   \author{O. Miettinen \inst{1}, J. Harju \inst{1},
          L.K. Haikala \inst{1}, \and
          C. Pomr\'en \inst{1}}

   \offprints{O. Miettinen (osmietti@astro.helsinki.fi)}

   \institute{\inst{1} Observatory, P.O. Box 14,
               FI-00014 University of Helsinki, Finland}

   \date{Received ...; accepted ...}

% \abstract{}{}{}{}{} 
% 5 {} token are mandatory
\authorrunning{Miettinen et al.}
\titlerunning{SiO and CH$_3$CCH in GMC cores}

 \abstract
  % context heading (optional)
  % {} leave it empty if necessary  
   {}
  % aims heading (mandatory)
   {The main goal of the present study is to determine the fractional SiO
abundance in high-mass star-forming cores, and to investigate
its dependence on the physical conditions. In this way we wish to provide
constraints on the chemistry models concerning the formation of
SiO in the gas phase or via grain mantle evaporation. The work addresses
also CH$_3$CCH chemistry as the kinetic temperature is determined using this
molecule.}
  % methods heading (mandatory)
   {We estimate the physical conditions of 15 high-mass
star-forming cores and derive the fractional SiO and CH$_3$CCH
abundances in them by using spectral line and dust continuum
observations with the SEST.}
  % results heading (mandatory)
   {The kinetic temperatures as derived from
CH$_3$CCH range from 25 to 39 K, the average being 33 K. The average
gas density in the cores is $4.5 \cdot 10^6$ cm$^{-3}$. The SiO emission
regions are extended and typically half of the integrated line
emission comes from the velocity range traced out by CH$_3$CCH
emission. The upper limit of SiO abundance in this 'quiescent' gas
component is $\sim 10^{-10}$. The average CH$_3$CCH abundance is about 
$7 \cdot 10^{-9}$. It shows a shallow, positive correlation with
the temperature, whereas SiO shows the opposite tendency.}
  % conclusions heading (optional), leave it empty if necessary 
   {We suggest that the high CH$_3$CCH abundance and its
possible increase when the clouds get warmer is related to
the intensified desorption of the chemical precursors
of the molecule from grain surfaces.
In contrast, the observed tendency of SiO does not support the idea
that the evaporation of Si-containing species from the grain mantles
would be important, and it contradicts with the models
where neutral reactions with activation
barriers dominate the SiO production. A possible explanation for the decrease
is that warmer cores represent more evolved stages of core evolution
with fewer high-velocity shocks and thus less efficient SiO replenishment.}


   \maketitle
%
%________________________________________________________________

\section{Introduction}

High-mass star-forming regions with large molecular column densities and
bright molecular lines offer excellent opportunities to study interstellar
chemisty. Results from chemistry models can be combined with
observational data on the physical conditions and molecular
abundances. This information can help us to
link the properties of massive dense cores,
their chemical characteristics, and phenomena like massive
outflows, ultracompact (UC) \ion{H}{ii} regions and
molecular masers into a coherent evolutionary sequence.
This is useful since the process of high-mass star formation
is not yet well understood
(see, e.g., \cite{garay1999}; \cite{fontani2002}; \cite{thompson2003}).

In this paper we derive the abundances of silicon monoxide, SiO, and
methyl acetylene, CH$_3$CCH, in a sample of massive molecular cloud
cores, and discuss the relation of these abundances to the physical
conditions and the probable evolutionary stages of the cores. The
purpose is to contribute towards a better understanding of the
interplay between the processes associated with massive star 
formation and the physical and chemical properties of the surrounding 
dense core.
 
SiO is believed to trace exclusively shocked gas as its spectral lines  
have wings and are often shifted relative to the emission from the 
ambient gas (see, e.g., \cite{martin1992}; \cite{schilke1997}).
Current understanding is that SiO is evaporated from the dust grains when
the shock velocity is greater than about 20 km s$^{-1}$.
This, however, depends on the grain-mantle composition
in the preshock gas (e.g., \cite{schilke1997}).
In dense and warm giant molecular cloud (GMC) cores, SiO seems to
present also in the quiescent gas component. In few interferometric
maps available extended SiO emission at ambient cloud velocity is
either seen in the vicinity of high-velocity outflows
(\cite{hatchell2001}) or totally separate from them
(\cite{lefloch1998}; \cite{codella1999}; \cite{shepherd2004}; 
\cite{fuente2005}). The fractional abundance of 'quiescent SiO' is
estimated to be about $10^{-10}$ in GMC cores. SiO absorption
measurements towards dormant GMCs, so called 'spiral arm clouds', give
similar abundances (\cite{greaves1996}).
The present observational evidence is insufficient to determine
whether the presence of SiO is caused by thermal evaporation,
enhanced neutral-neutral production pathways,
photon-induced reactions, or by shock removal. 
To resolve this indistinctness it seems beneficial to determine
fractional SiO abundances and kinetic
temperatures in a representative sample of GMC cores,
and to correlate these also with the kinematic information
contained in the spectral lines.

CH$_3$CCH is an organic molecule observed rather frequently towards 
dense cores. Its abundance is on the order of $10^{-9}$.
The rotational temperature, $T_{\rm rot}$, derived from
a series of $J_K-(J-1)_K$ rotational lines
of CH$_3$CCH is considered a good estimate of
the gas kinetic temperature, $T_{\rm kin}$, in molecular clouds
(e.g., \cite{askne1984}; \cite{bergin1994}).
Since CH$_3$CCH has a relatively low dipole
moment ($\mu=0.78$ D, \cite{bauer1979}; \cite{burrell1980}),
its rotational levels are thermalized at densities
$n({\rm H_2}) \sim 10^4 \, {\rm cm}^{-3}$ or higher
(e.g., \cite{kuiper1984}). This property together with
the assumption that CH$_3$CCH emission is optically thin ($\tau \ll 1$),
allow us to assume local thermodynamic equilibrium (LTE),
so that $T_{\rm kin}$ may be derived.
Besides for $T_{\rm kin}$, we use $T_{\rm rot}$ for
estimating the dust temperature, $T_{\rm d}$, which is needed to
derive the total gas column densities ($\sim N({\rm H_2}$)) from dust
continuum observations.

The source list consists of 15 high-mass star-forming cores associated
with OH, H$_2$O, and CH$_3$OH masers, UC \ion{H}{ii} regions and
bright, thermal SiO rotational line emission.
The sources were selected from the SiO survey of
\cite{harju1998} (hereafter HLBZ98). The positions selected for the
SiO and CH$_3$CCH line observations are listed in Table
\ref{table:sources}. These positions served as map centres in the
continuum observations.

The SiO and CH$_3$CCH column densities and the kinetic temperatures
in the present paper are derived from spectral line observations with the
SEST. The estimates for the molecular hydrogen column densities,
$N({\rm H_2})$, and the core masses are derived from dust continuum maps
obtained with the SIMBA bolometer at SEST.  The observations and
the data reduction procedures are described in Sect. 2.
The direct observational results are presented in Sect. 3.
In Sect. 4 we describe the methods used to derive
the physical/chemical properties of the GMC cores. In Sect. 5
we discuss the results, and finally, in Sect. 6 we summarize our 
major conclusions.

\section{Observations and data reduction}

\subsection{Molecular lines}

The spectral line observations were made 
during four observing runs from 1995 to 2003 with
the Swedish-ESO-Submillimetre Telescope SEST at
the La Silla observatory, Chile. The SEST 3 and 2 mm (SESIS) dual SIS
single sideband (SSB) receiver was used. The observations were made
in the dual beam switching mode (beam throw $11\farcm5$ in azimuth).
The SEST half-power beam width (HPBW) and the main beam efficiency, 
$\eta_{\rm MB}$, at frequencies 86 GHz, 115 GHz and 147 GHz are $57\arcsec$,
$45\arcsec$, $34\arcsec$ and 0.75, 0.70, 0.66, respectively. The
SEST high resolution 2000 channel acousto-optical spectrometer
(bandwidth 86 MHz, channel width 43 kHz) was split into two halves to
measure two receivers simultaneously. At the observed wavelengths, 2 mm
and 1 mm, the 43 kHz channel width corresponds to approx. 0.12 km s$^{-1}$ and
0.08 km s$^{-1}$, respectively. The observed molecular transitions, their rest
frequencies and the upper level energies are listed in Table \ref{table:lines}.

Calibration was achieved by the chopper wheel method. Pointing was
checked regularly towards known circumstellar SiO masers.
Pointing accuracy is estimated to be better than $5\arcsec$.

The $^{28}$SiO ($v=0$, $J=2-1$) and ($v=0$, $J=3-2$) observations
were carried out in October 1995 and April 1996 and
are described in detail in HLBZ98.
$^{29}$SiO($2-1$) and ($3-2$) observations were carried out in October 1998.
Typical values for the effective SSB system temperatures and
the RMS noise of the spectra were 140 K and 170 K, and 0.016 K and 0.021 K
at frequencies 86 GHz and 130 GHz, respectively.
Linear baselines were subtracted from the spectra.

The CH$_3$CCH ($J=5_K-4_K$) and ($J=6_K-5_K$) observations
were carried out in June 2003. At this time the electronics of
the SESIS 2 mm receiver had been decommissioned and only
the 3 mm receiver was available. These were the last spectral line observations
made with SEST. For each source typically six spectra, each with 2 minutes 
integration time, were obtained. The effective SSB system temperatures 
were 130 K for the CH$_3$CCH($5_K-4_K$) transition and
250 K for the CH$_3$CCH($6_K-5_K$). The RMS of the spectra were 0.017 K
and 0.024 K for the two transitions, respectively.
Because of the better performance of 
the receiver at the lower frequency, the $J=5_K-4_K$ transition was chosen 
for the survey. A third order baseline was subtracted from the spectra.
Four $K$-components ($K=0,1,2,3$) were detected towards all sources, and
Gaussian profiles were fitted to the lines.
The CLASS program of the GILDAS software package developed at IRAM
was used for the reductions.
In the fitting procedure the line separations between
the components were fixed and the line widths
for all $K$-components were assumed to be identical.

\subsection{Continuum}

The 1.2 mm continuum observations were carried out in June 2003
with the 37 channel SEST imaging bolometer array, SIMBA.
The SIMBA central frequency is 250 GHz and the bandwidth about 50 GHz.
The HPBW of a single bolometer element is $\sim24\arcsec$
and the separation between elements on the sky is $44\arcsec$.
The observations were conducted in stable weather.
Frequent skydips were used to
determine the atmospheric opacity and the values obtained varied
between 0.27 and 0.3. Pointing was controlled by observing sources
with known accurate coordinates and is estimated to be better than
5\arcsec. Uranus was used for flux density calibration.

The observations were done in the fastscanning mode with a scanning
speed of $80\arcsec\, {\rm s}^{-1}$. The typical map consisted
of 65 scans of $700\arcsec\,$ in length in azimuth and spaced by $8\arcsec\,$
in elevation. Each source was observed at least twice. The two  maps were done
at different LST values to reduce the possibility of strong artefacts
in the reduced maps. The data were reduced using the
MOPSI\footnote{MOPSI is a software package for infrared,
millimeter and radio data reduction developed and
constantly upgraded by R. Zylka.} software package according to
guidelines in the SIMBA Observers Handbook (2002) and Chini et al. (2003).

All the observed maps contained strong sources and therefore, as
suggested in the SEST handbook, no despiking was applied to the
data. The data reduction deviated from the suggestions in the SEST
handbook in the following details. The SEST dish shape (and the beam profile)
degrades as a function of the elevation as the dish deforms because of its own
weight. Gain-elevation correction is used to compensate the
decreasing of the observed source peak intensity as a function of the
telescope elevation. However, no gain-elevation correction to the
present data was applied because aperture photometry was used to
define the source flux. The source flux lost because of the decrease
of the telescope gain in the source position is recovered as the
source is mapped. The telescope gain is unity at elevation
of 74.6 degrees and decreases by 10 \% at 45 degrees.
Most of the present observation were obtained at elevations
50 degrees or higher where the said correction is small. Third order
baseline was applied to the data before correlated noise removal after
which all the bolometer elements with insufficient base range were
masked. A first order baseline was applied to individual scan lines
after this. The refinement of the SIMBA data reduction is discussed
in detail in Haikala (2006, in preparation).

\section{Observational results}

The observed sources are listed in Table \ref{table:sources}.
The columns of this table are: (1) source name; (2) and
(3) equatorial coordinates (J2000.0); (4) galactic
coordinates; (5) distance; (6) galactocentric distance;
(7) LSR velocity determined from the CH$_3$CCH line; 
(8) notes on association with molecular masers or an 
UC \ion{H}{ii} region. When no distance
reference is given in Col. 5, the value is the kinematic distance 
calculated from the CH$_3$CCH velocity using the rotation curve determined by
\cite{brand1993} and $R_0=8.5$ kpc (galactocentric distance of the Sun)
and $\Theta_0=220$ km s$^{-1}$
(circular velocity at a distance $R_0$).

\subsection{Dust emission}

The obtained SIMBA maps are presented in Fig. \ref{figure:simbamaps}.
The maps are plotted
to the same angular and intensity scales. In
Table \ref{table:dustmaxima} we list the coordinates
($\alpha_{2000.0}$, $\delta_{2000.0}$) and
the intensities ($I_{\nu}^{\rm max}$)
at the dust emission maxima identified on the maps. 
The SEST HPBW at the frequency of the continuum observations is 24\arcsec 
which is less than half of the HPBW at the frequency of the
observed SiO $J=2-1$ and the CH$_3$CCH lines. In order to estimate
the average dust and H$_2$ column densities within
the beam used for the 3 mm line observations, the SIMBA maps were
smoothed to the resolution of 57\arcsec. The resulting smoothed surface
brightness ($I_{\nu}^{\rm smo}$) towards the position of the line measurements
is listed in Col. 2 of Table \ref{table:smoothed_intensities}.
The 1.2 mm flux density ($S_{\rm 1.2\,mm}$) integrated over the source and
the angular size (FWHM) of the core ($\Theta_{\rm s}$) determined
from the original map is listed in Cols. 3 and 4 of this Table.
The angular sizes were estimated from
two-dimensional Gaussian fits to the surface brightness
distribution (geometric mean of the observed major and
minor axes of Gaussians). 
The source size was corrected for the broadening by 24\arcsec 
the telescope beam in accordance with the assumption of Gaussian
distributions. The flux density was calculated in an 72\arcsec diameter
circular aperture which correponds to three times the original beam size.

%1st SIMBA figure
\begin{figure*}
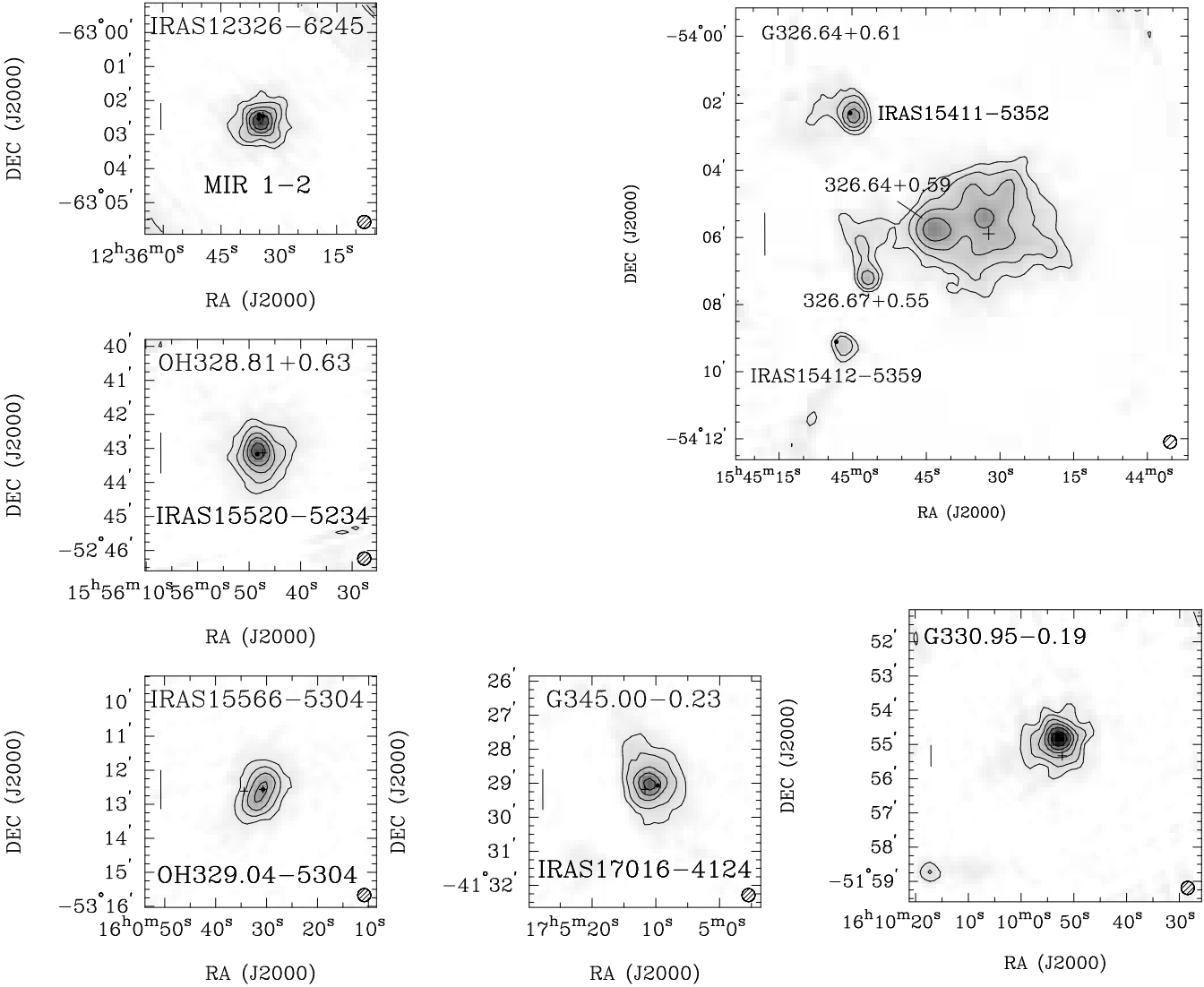

\unitlength=1mm
\begin{picture}(0,150)
%1
\put(-10,150){
\begin{picture}(0,0) \includegraphics{figures/iras12326-6245.ps} \end{picture}}
%2
\put(75,150){
\begin{picture}(0,0) \includegraphics{figures/g326.64+0.61.ps} \end{picture}}
%3
\put(-10,100){
\begin{picture}(0,0) \includegraphics{figures/oh328.81+0.63.ps} \end{picture}}
%4
\put(-10,50){
\begin{picture}(0,0) \includegraphics{figures/iras15566-5304.ps} \end{picture}}
%5
\put(103,60){
\begin{picture}(0,0) \includegraphics{figures/g330.95-0.19.ps} \end{picture}}
%8
\put(47,50){
\begin{picture}(0,0) \includegraphics{figures/g345.00-0.23.ps} \end{picture}}
\end{picture}
\caption{SIMBA images of the cores (1). The grey scales and contours represent
the intensity of 1.2 mm dust emission. The contour levels are 
0.25, 0.5, 1.0, 2.0, 4.0, 8.0 and 16.0 Jy/beam in all maps. The maps are 
plotted to the same angular scale.
The target positions of the line observations
are denoted with crosses. The filled circles indicate the locations of 
mid- and far-infrared sources.
The  scale bar on the left corresponds to 1 pc and
the beam HPBW is shown in the lower right corner.
The image borders have been masked out.}
\label{figure:simbamaps}
\end{figure*}

%2nd SIMBA figure
\begin{figure*}
\unitlength=1mm
\begin{picture}(0,70)
%6
\put(-5,70){
\begin{picture}(0,0) \includegraphics{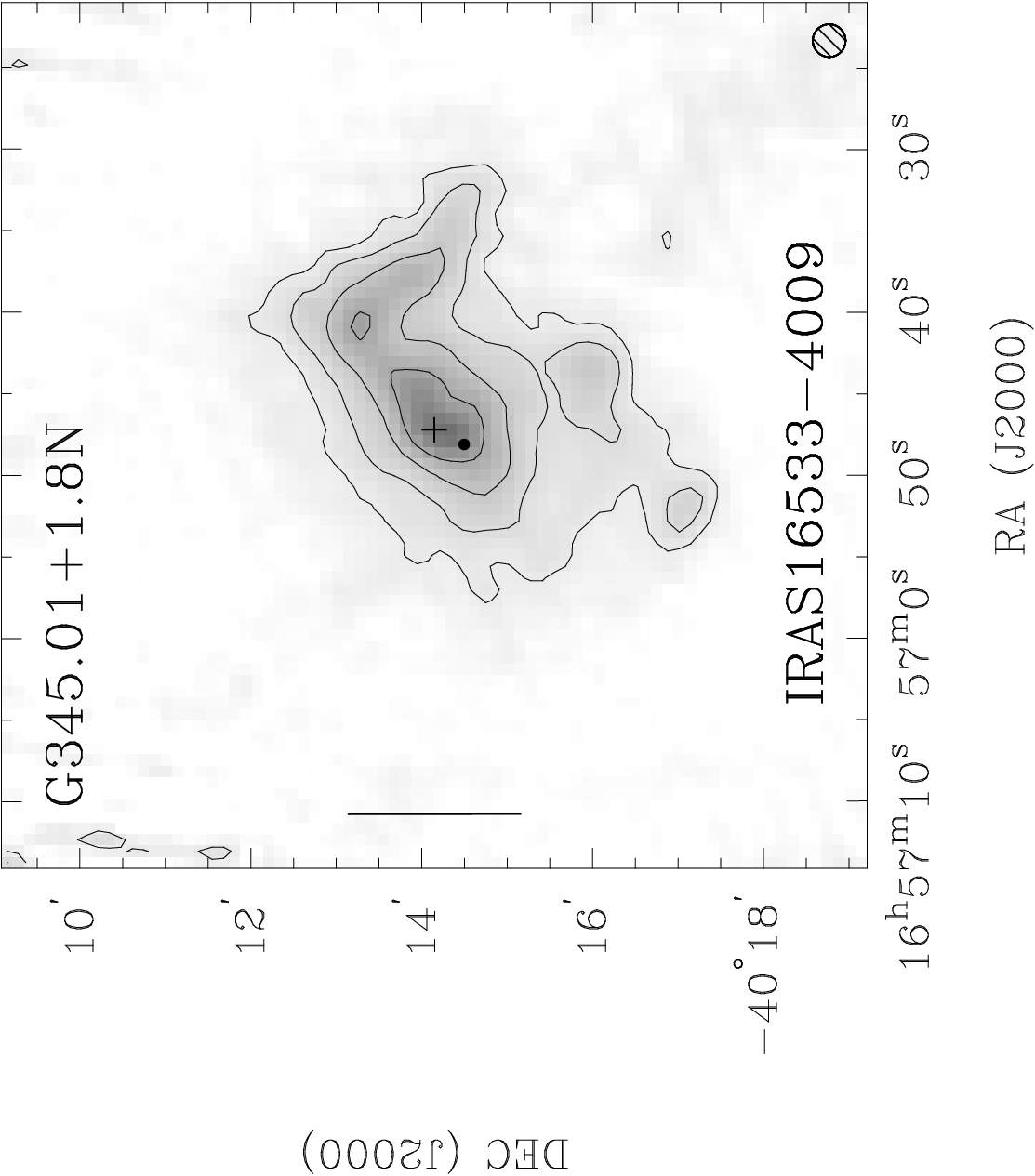} \end{picture}}
%7
\put(80,70){
\begin{picture}(0,0) \includegraphics{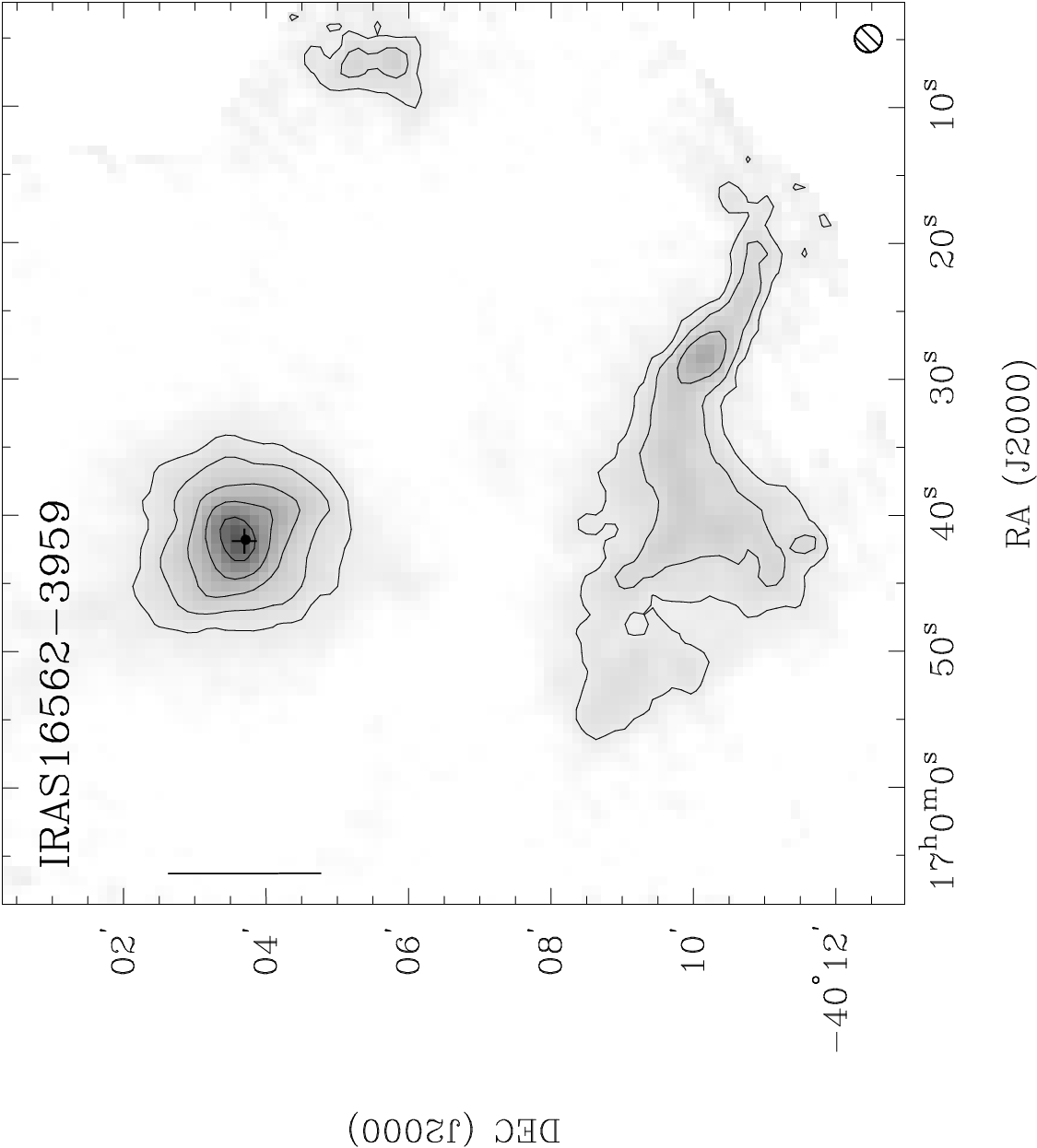} \end{picture}}
\end{picture}
\addtocounter{figure}{-1}
\caption{continued.}
\end{figure*}

%3rd SIMBA figure
\begin{figure*}
\unitlength=1mm
\begin{picture}(0,140)
%9
\put(-10,140){
\begin{picture}(0,0) \includegraphics{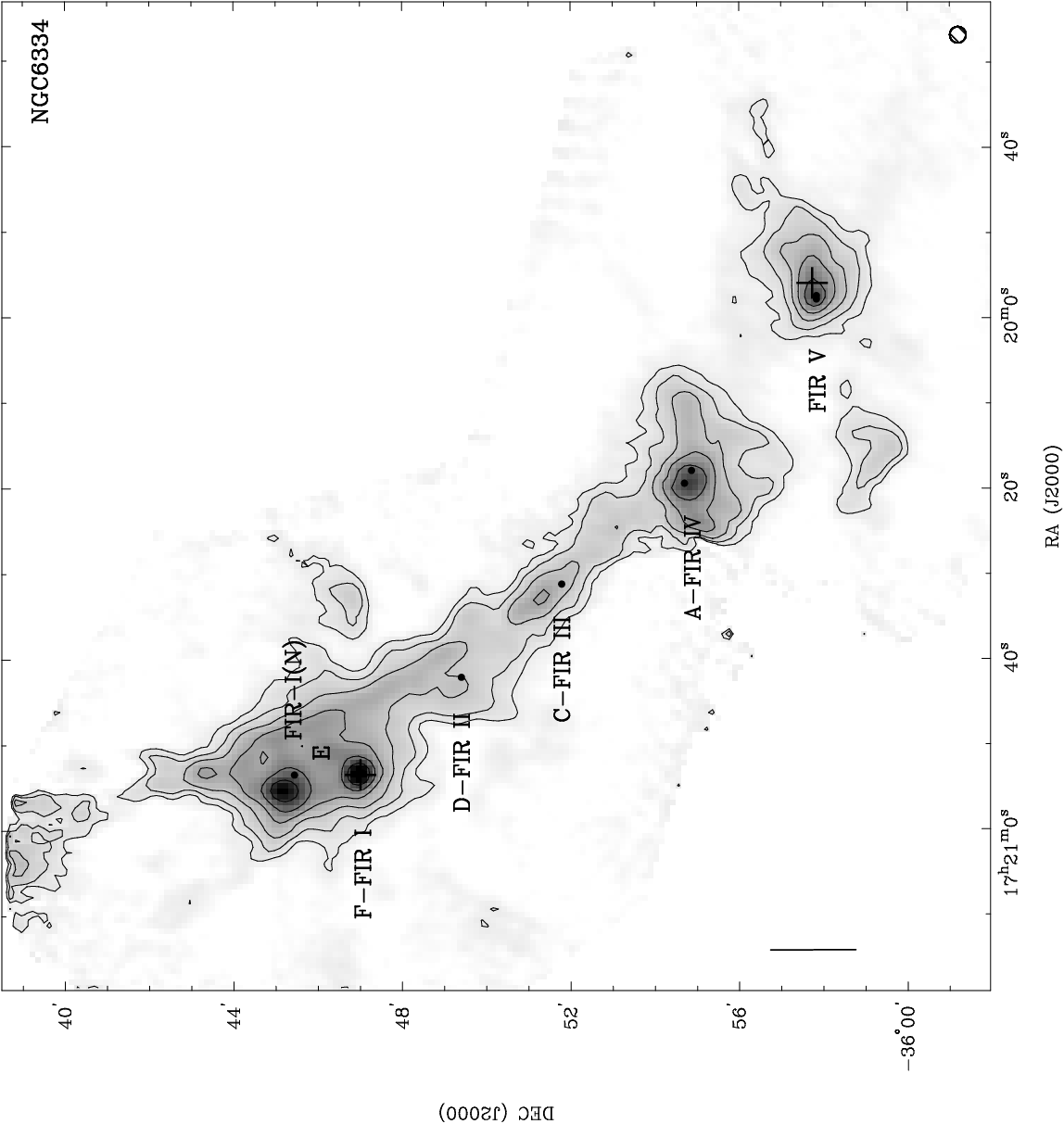} \end{picture}}
\end{picture}
\addtocounter{figure}{-1}
\caption{continued.}
\end{figure*}

%4th SIMBA figure

\begin{figure*}
\unitlength=1mm
\begin{picture}(0,150)
%10
\put(-10,150){
\begin{picture}(0,0) \includegraphics{figures/g351.77-0.54.ps} \end{picture}}
%11
\put(85,150){
\begin{picture}(0,0) \includegraphics{figures/g353.41-0.36.ps} \end{picture}}
%12
\put(74,80){
\begin{picture}(0,0) \includegraphics{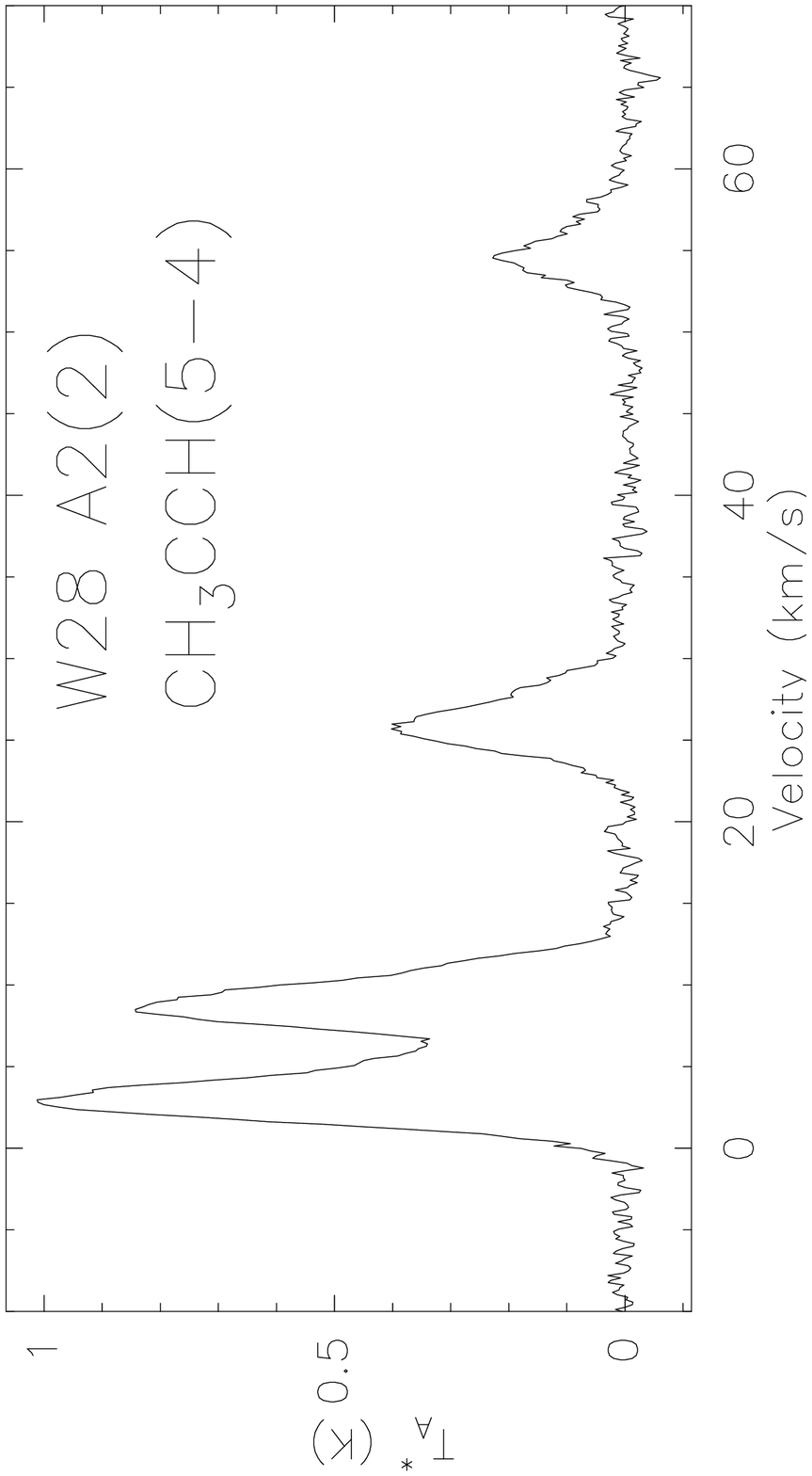} \end{picture}}
%13
\put(-10,80){
\begin{picture}(0,0) \includegraphics{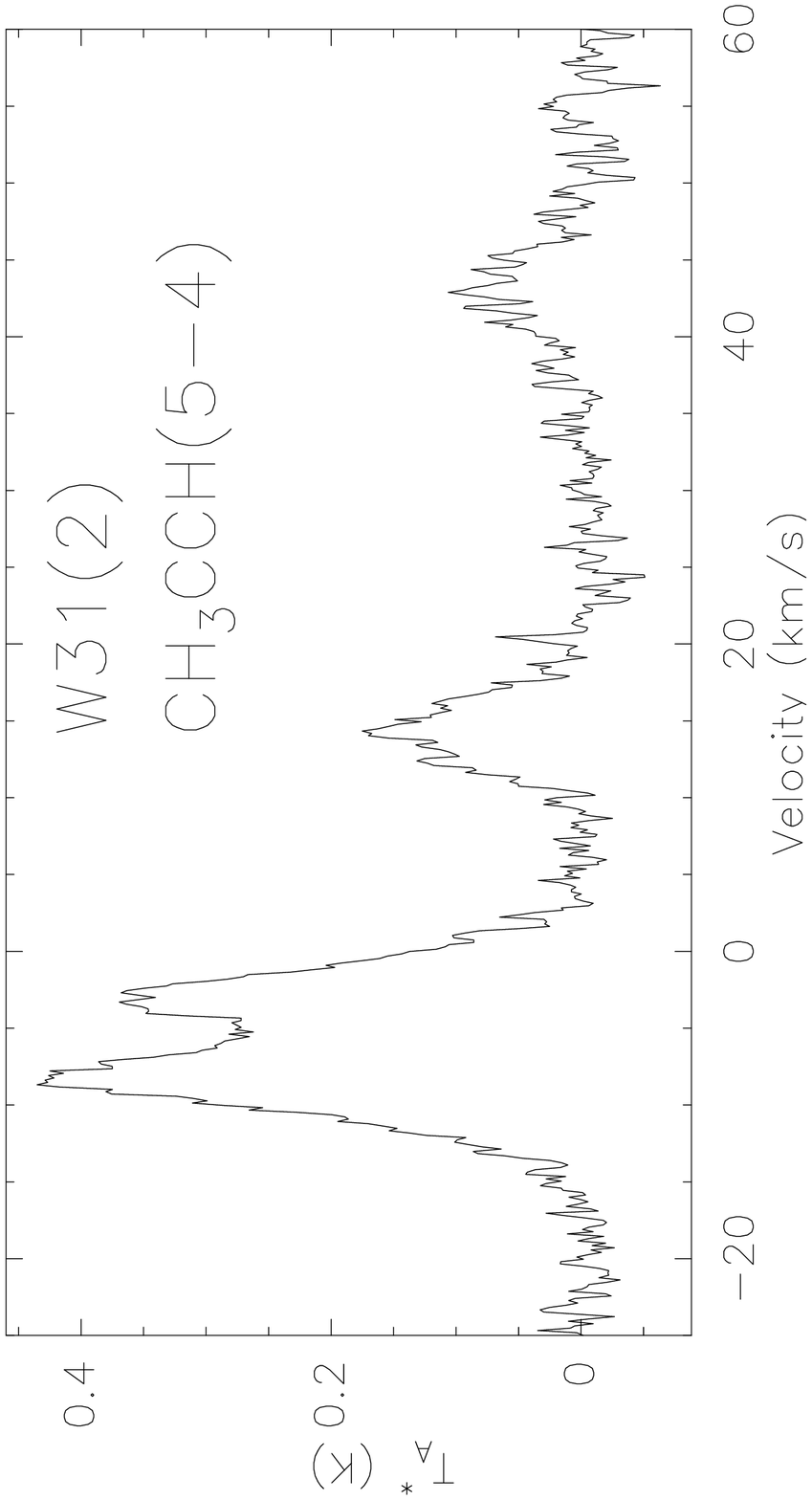} \end{picture}}
\end{picture}
\addtocounter{figure}{-1}
\caption{continued.}
\end{figure*}

%5th SIMBA figure
\begin{figure*}
\unitlength=1mm
\begin{picture}(0,90)
%14
\put(25,90){
\begin{picture}(0,0) \includegraphics{figures/w33.ps} \end{picture}}
\end{picture}
\addtocounter{figure}{-1}
\caption{continued.}
\end{figure*}

\subsection{SiO and CH$_3$CCH lines}

\begin{figure*}
\begin{center}
\includegraphics[width=4.0cm]{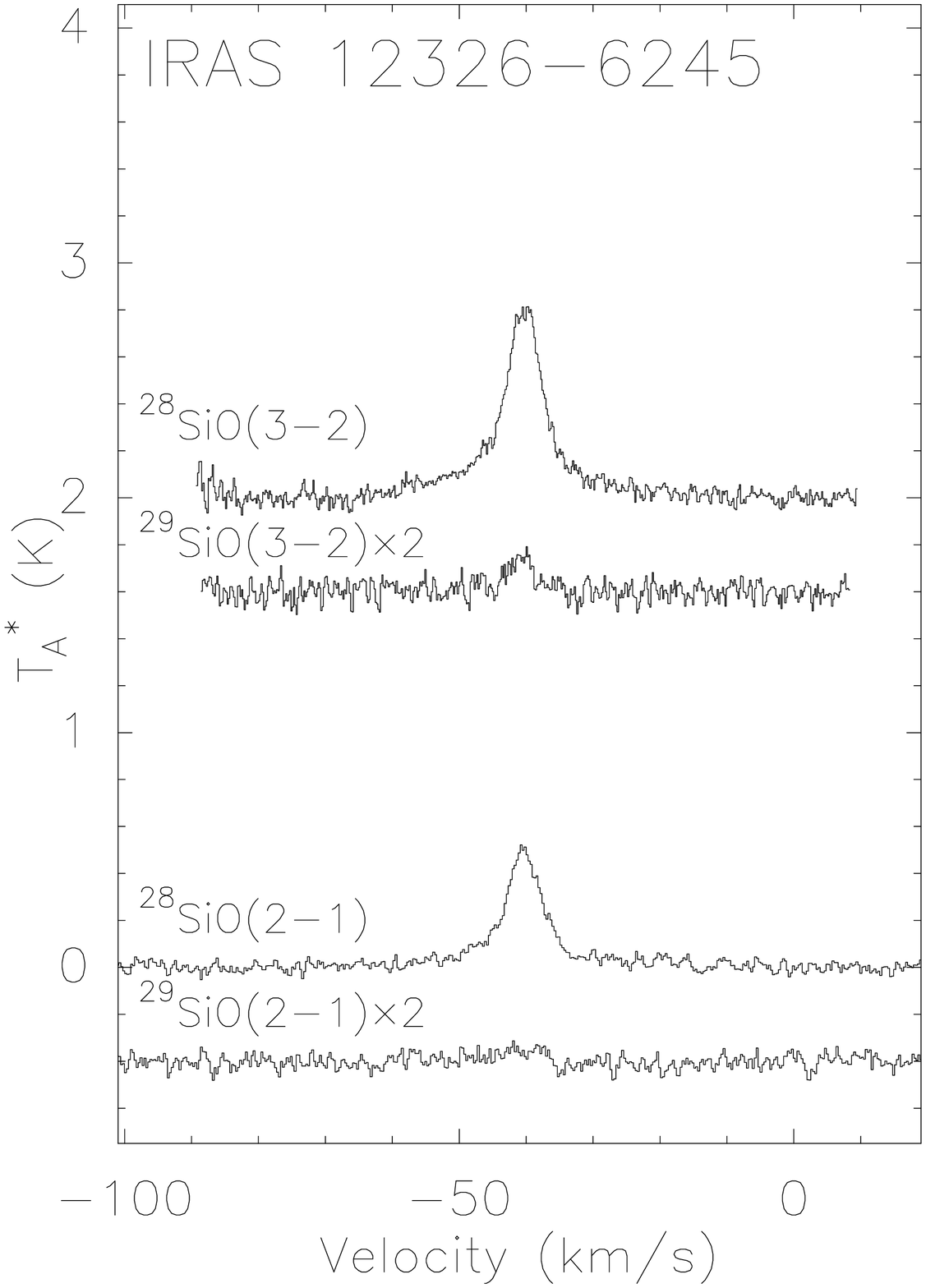}
\includegraphics[width=4.0cm]{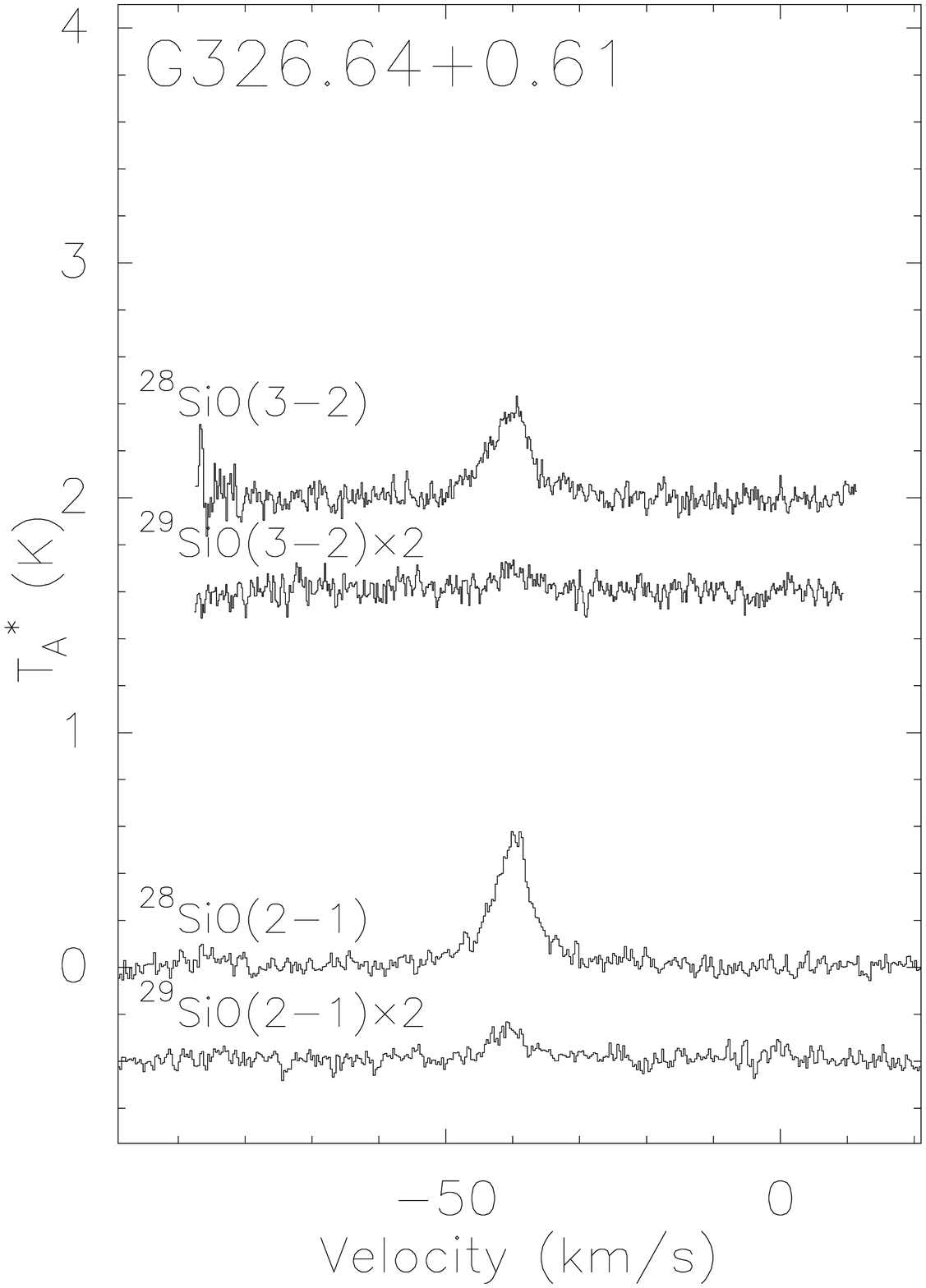}
\includegraphics[width=4.0cm]{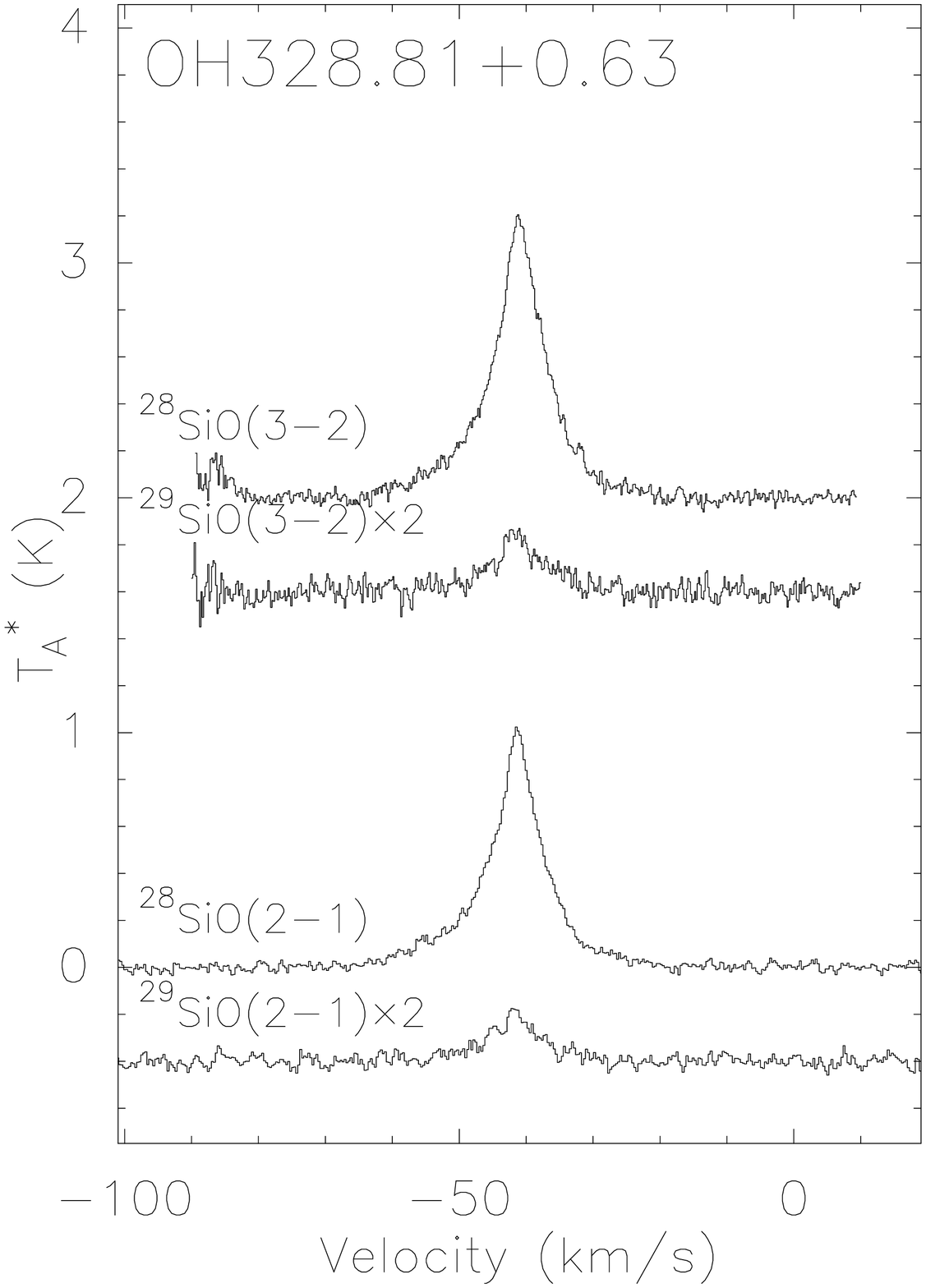}
\includegraphics[width=4.0cm]{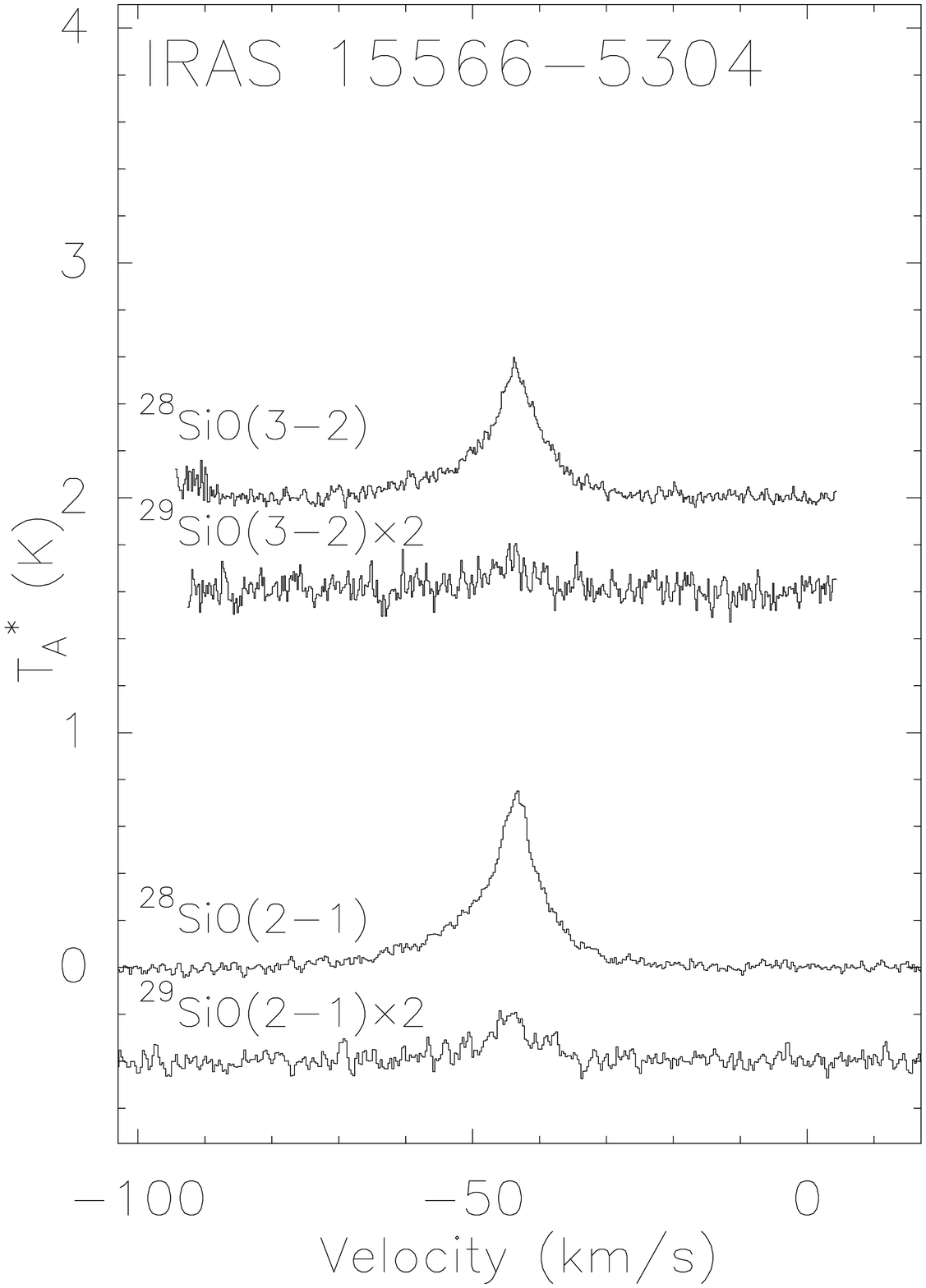}
\includegraphics[width=4.0cm]{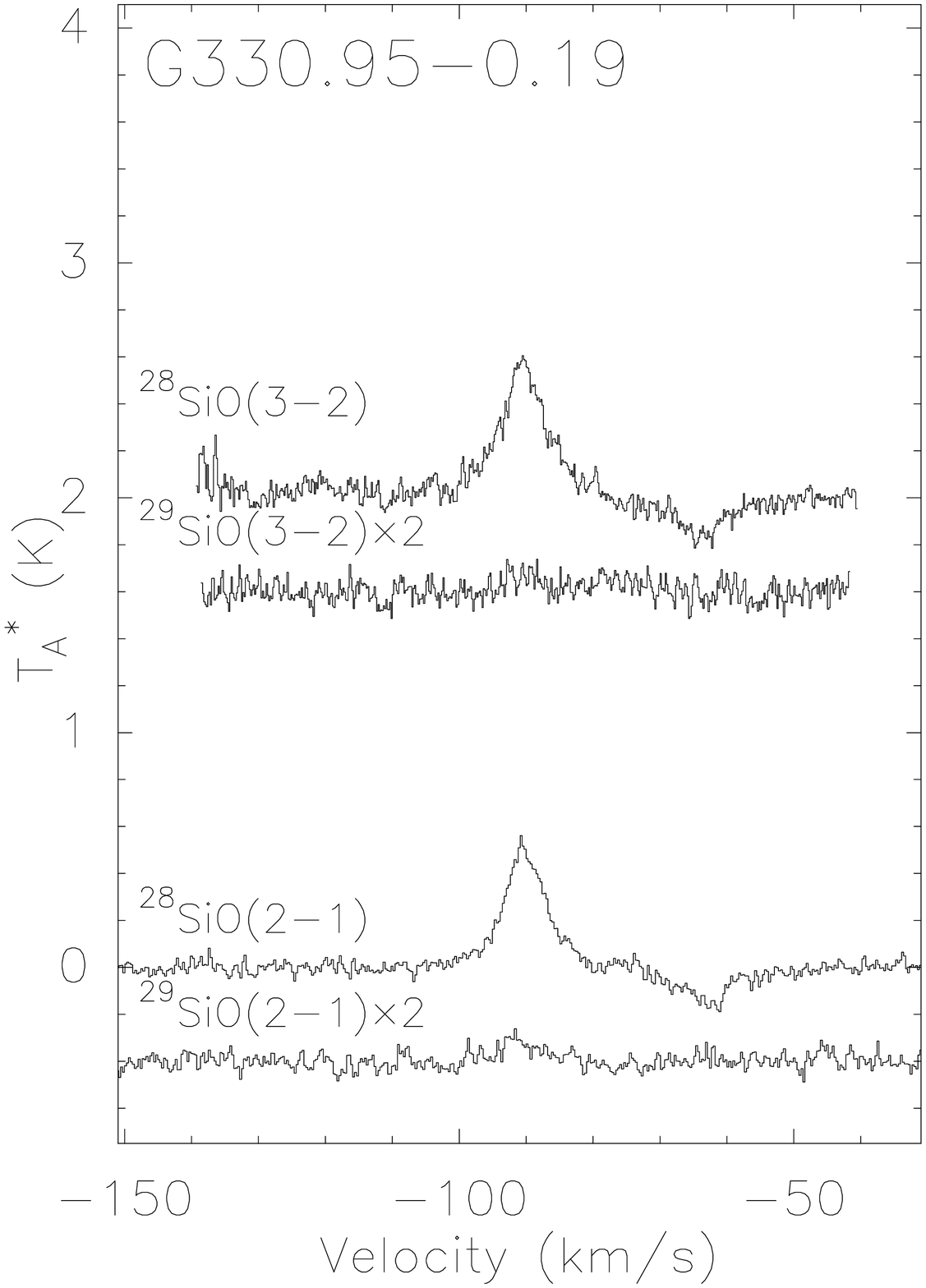}
\includegraphics[width=4.0cm]{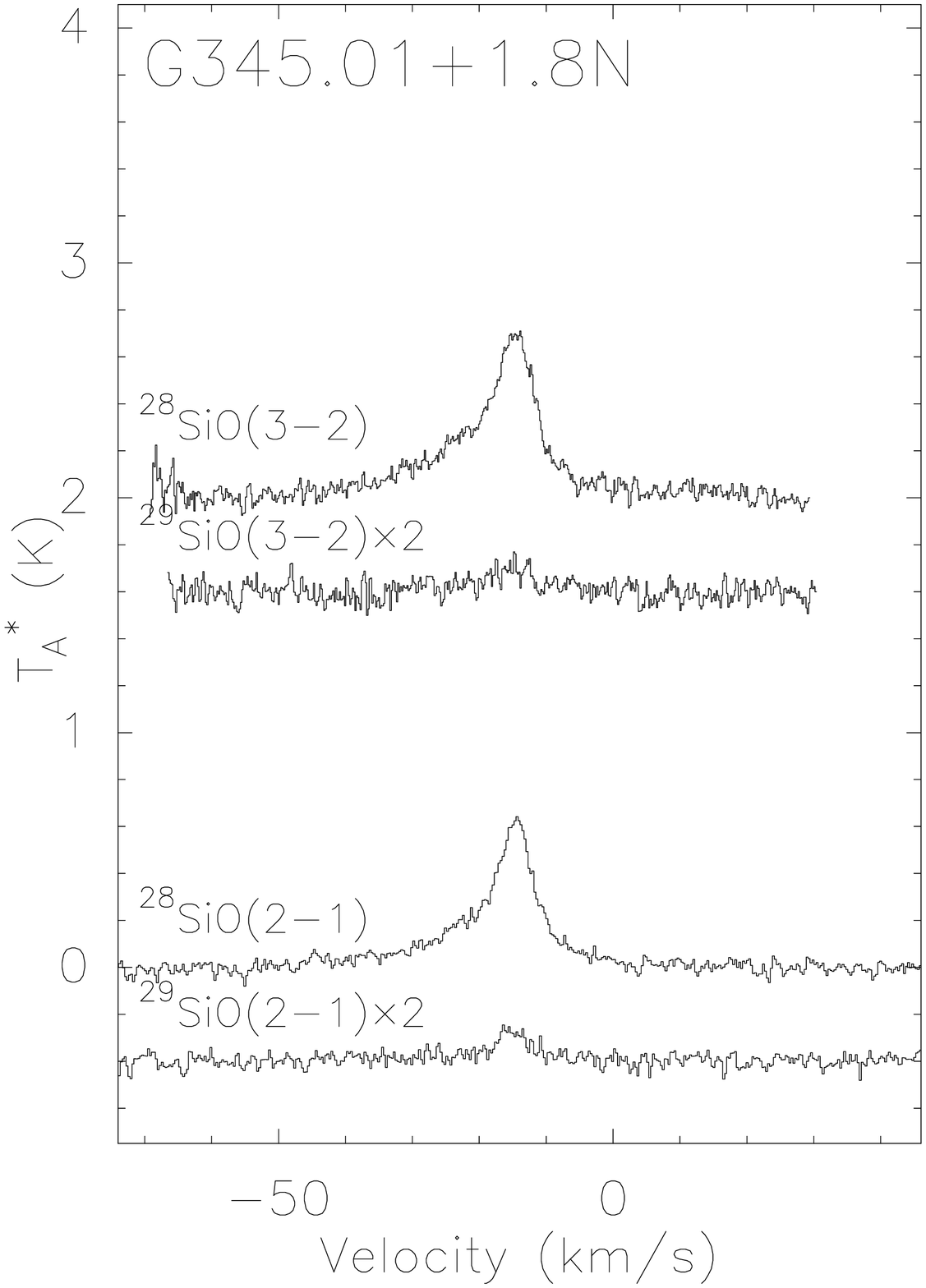}
\includegraphics[width=4.0cm]{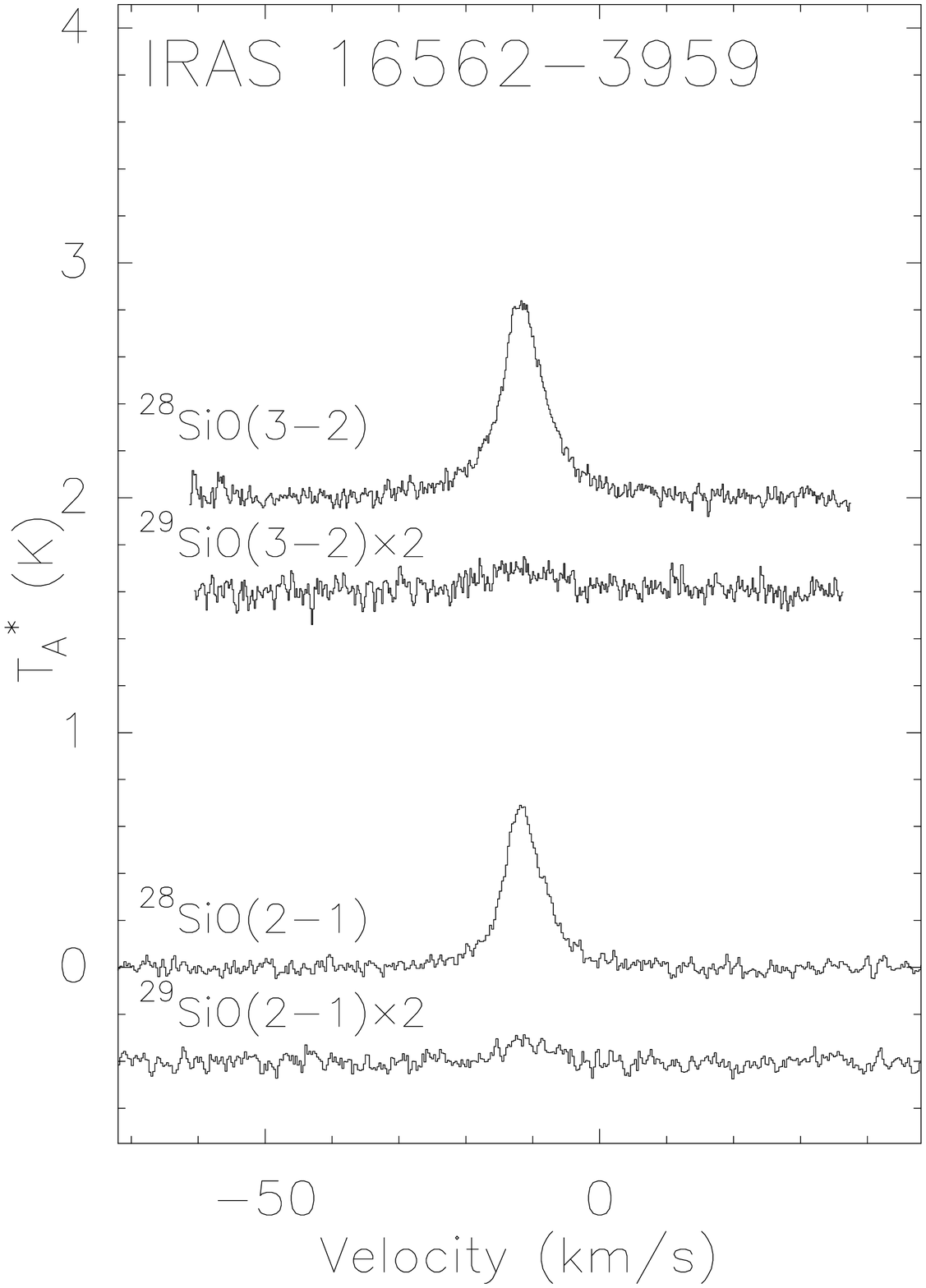}
\includegraphics[width=4.0cm]{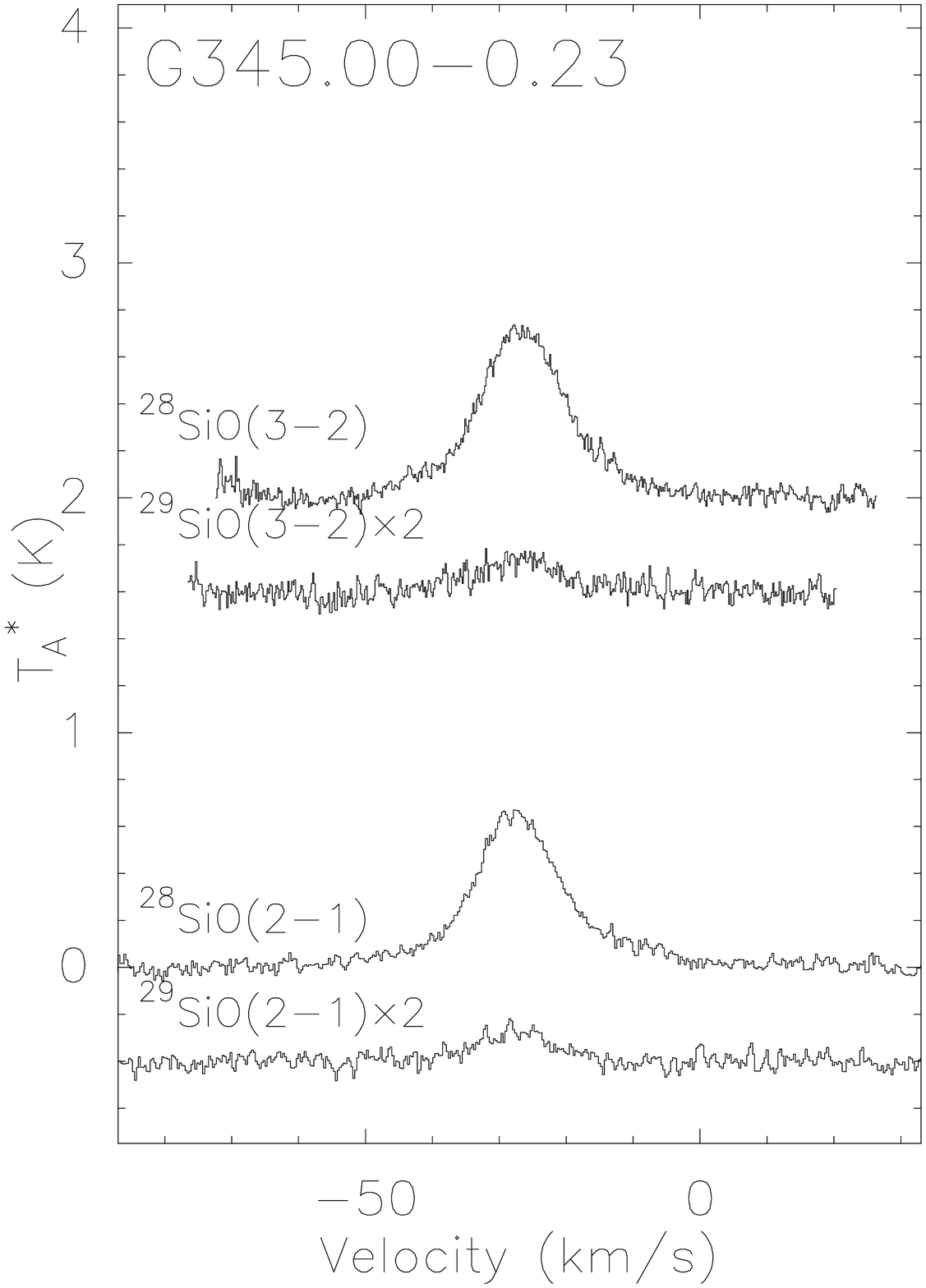}
\includegraphics[width=4.0cm]{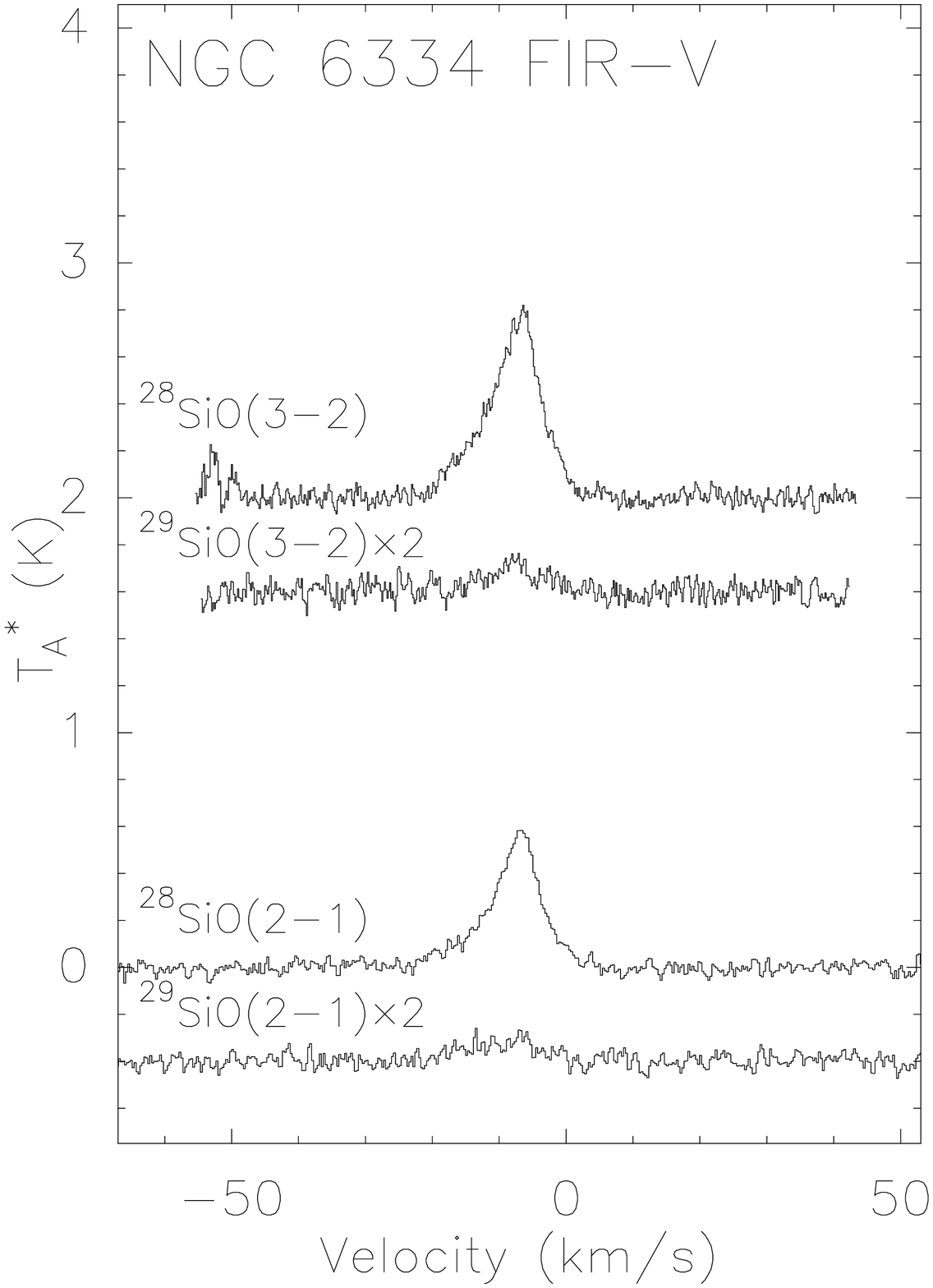}
\includegraphics[width=4.0cm]{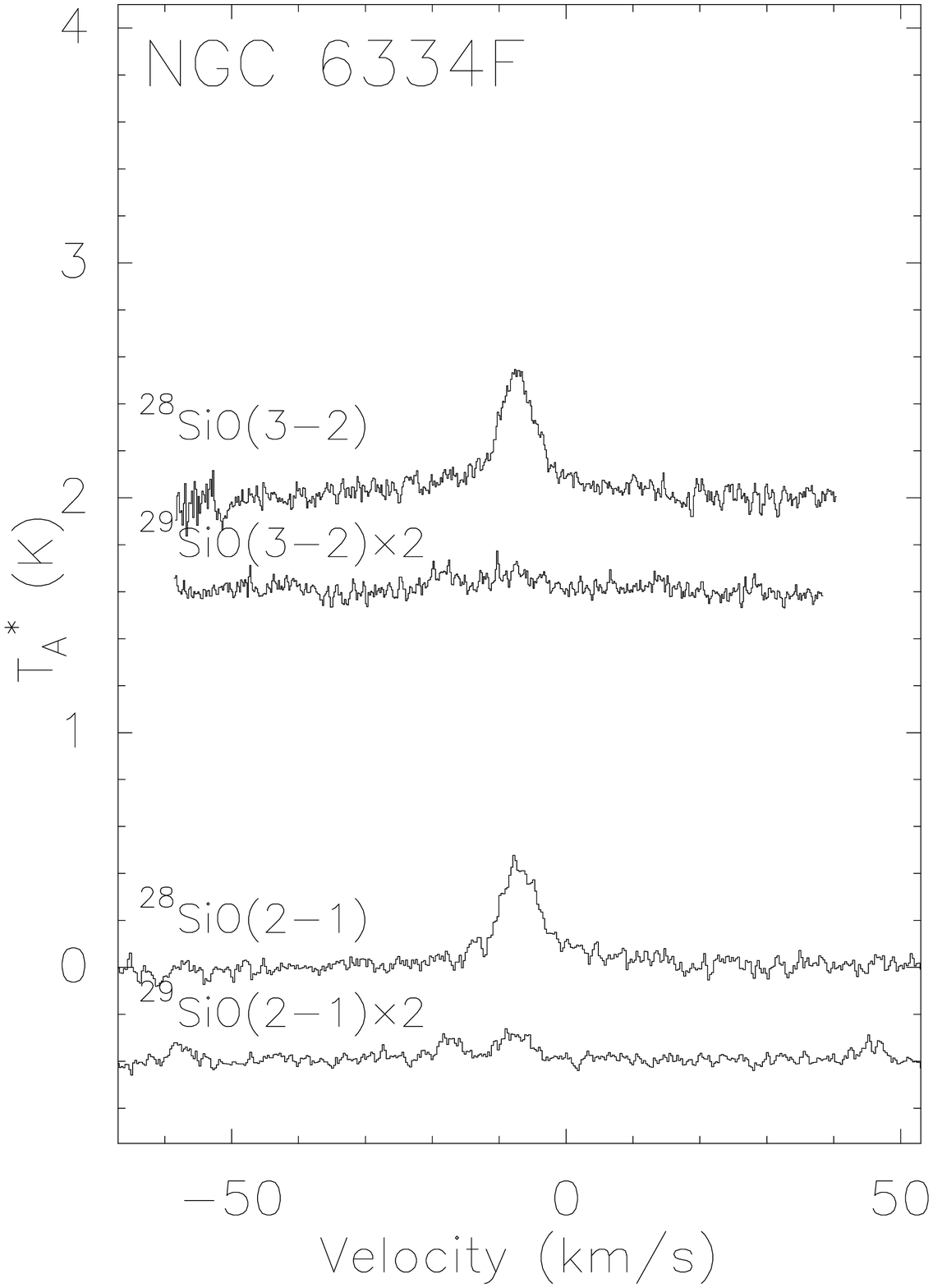}
\includegraphics[width=4.0cm]{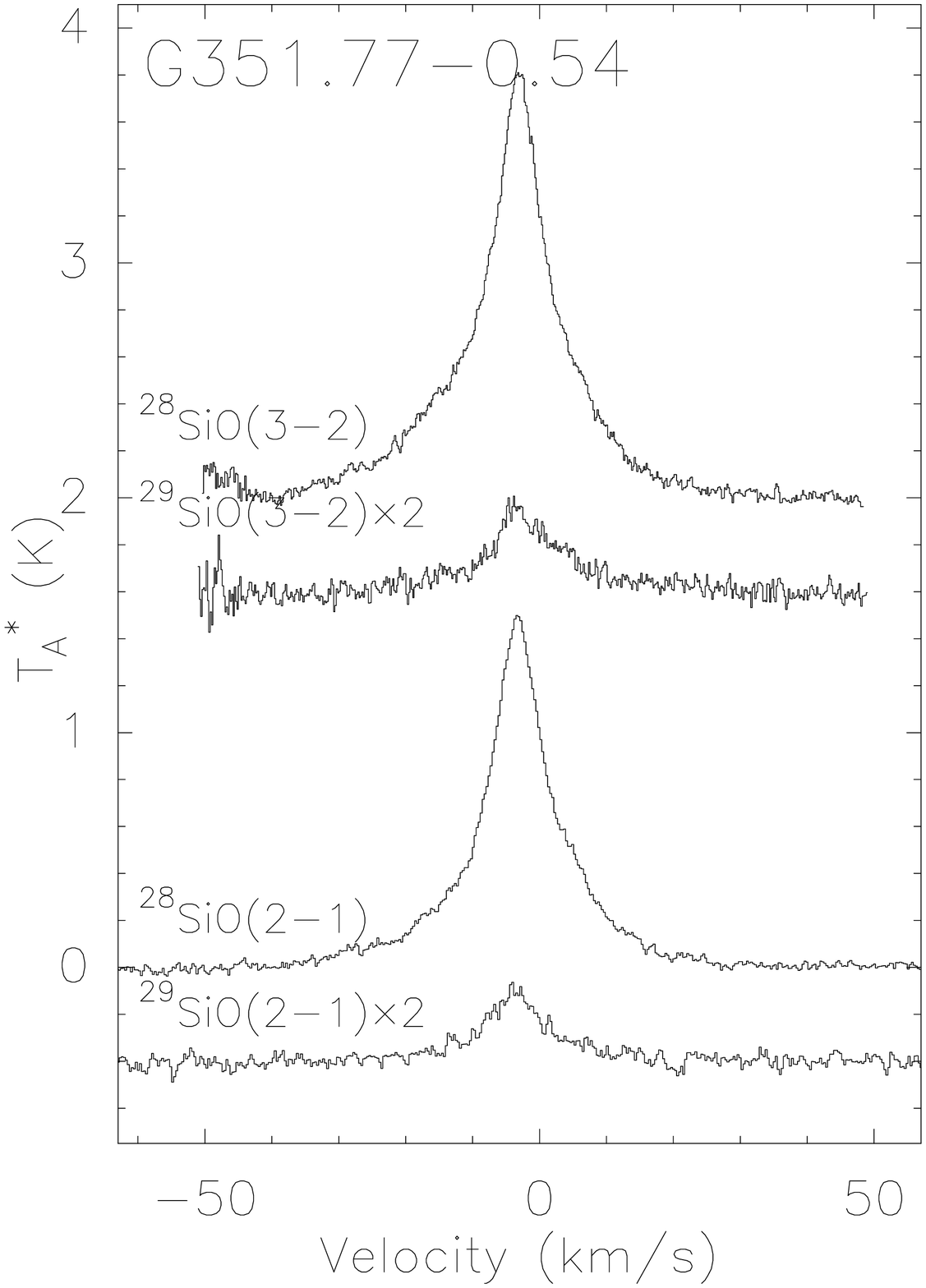}
\includegraphics[width=4.0cm]{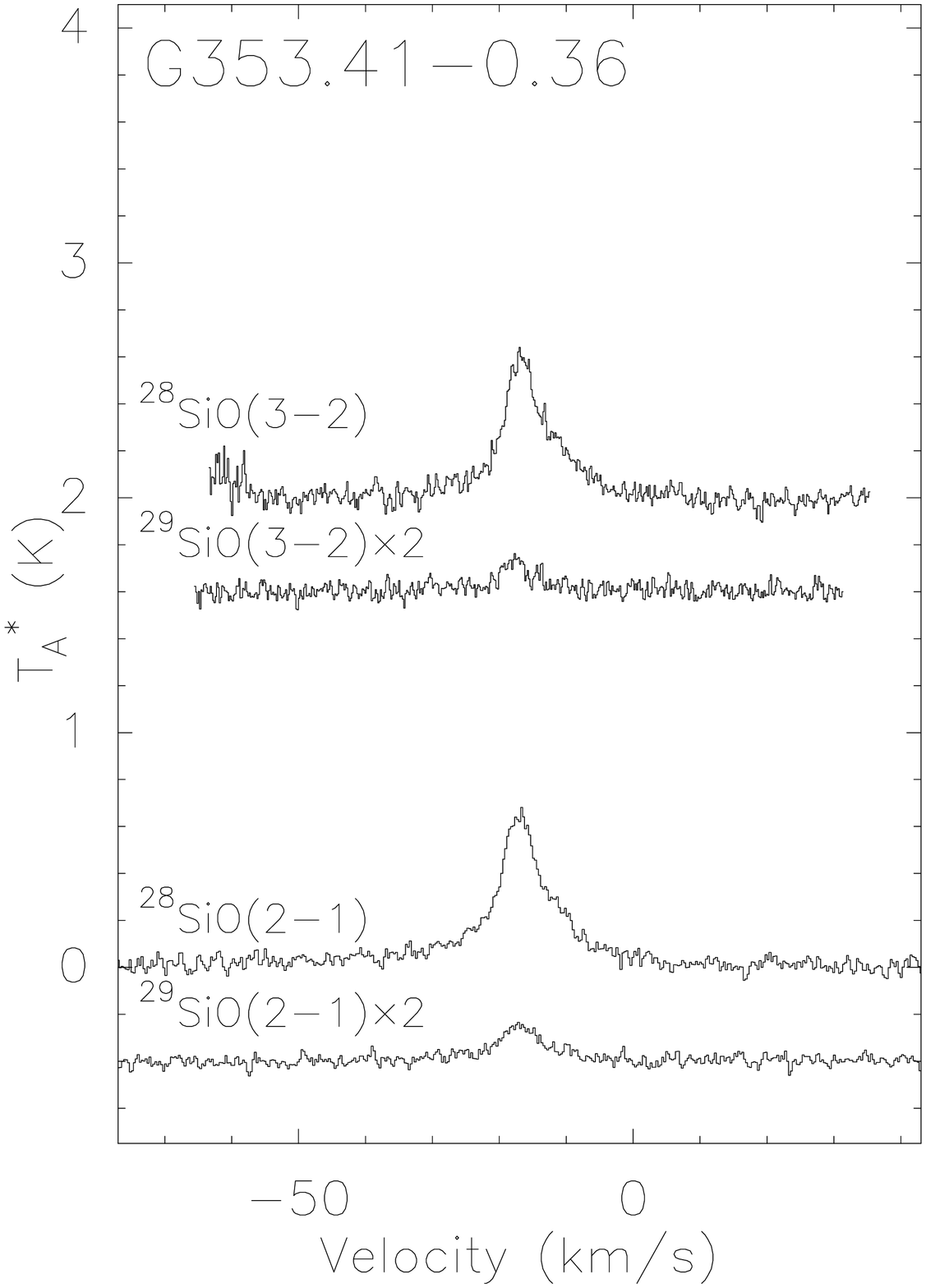}
\includegraphics[width=4.0cm]{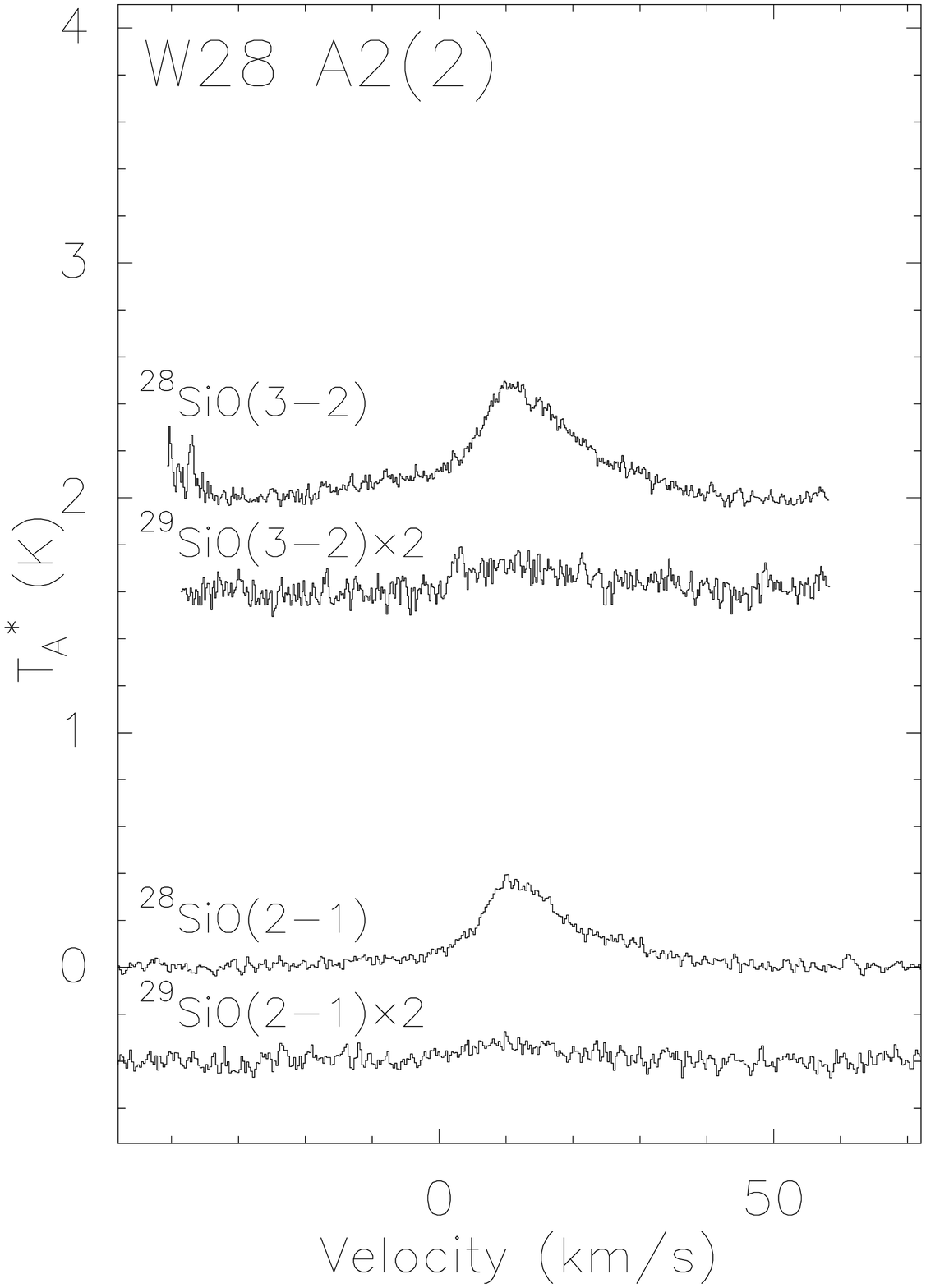}
\includegraphics[width=4.0cm]{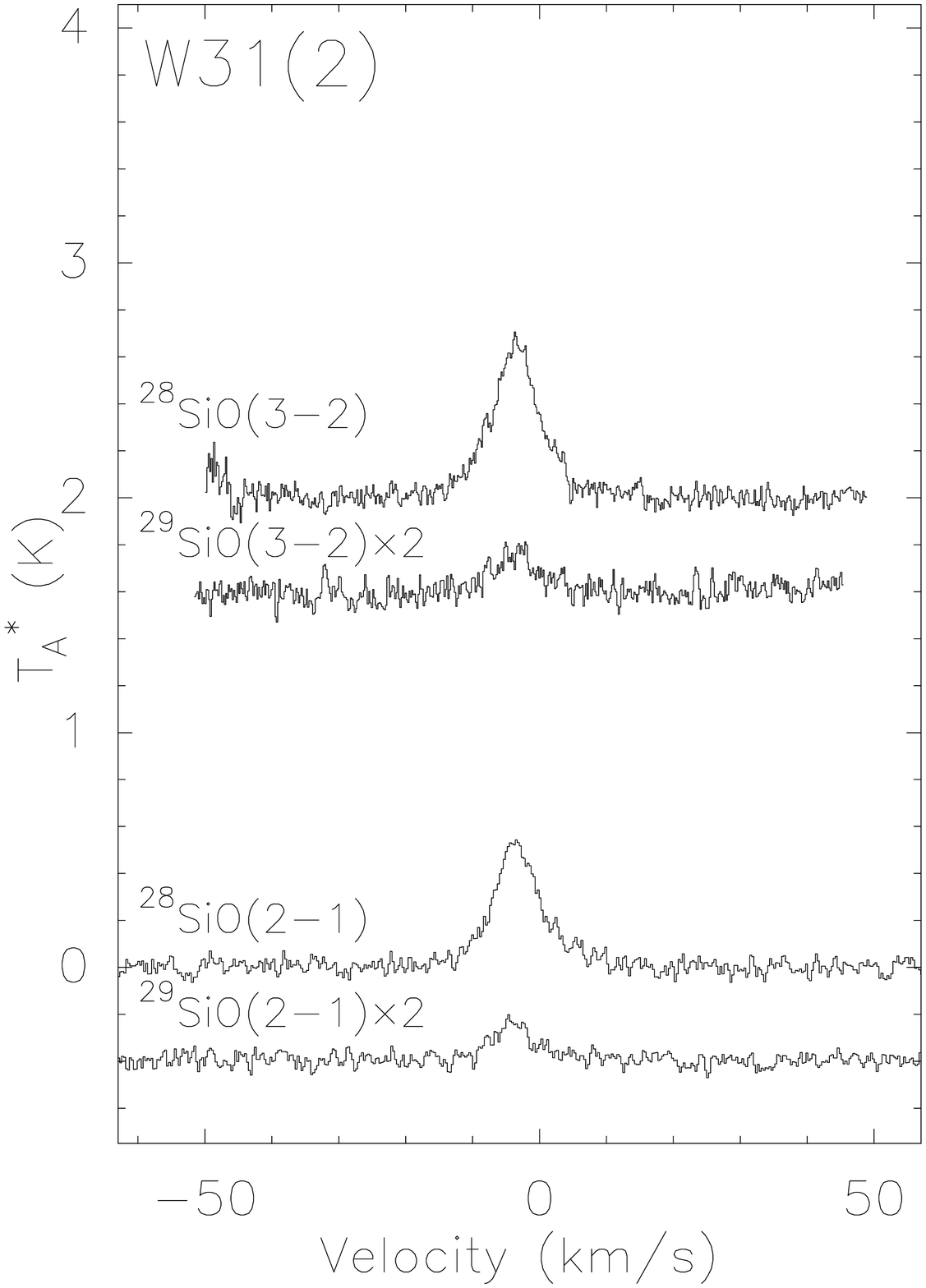}
\includegraphics[width=4.0cm]{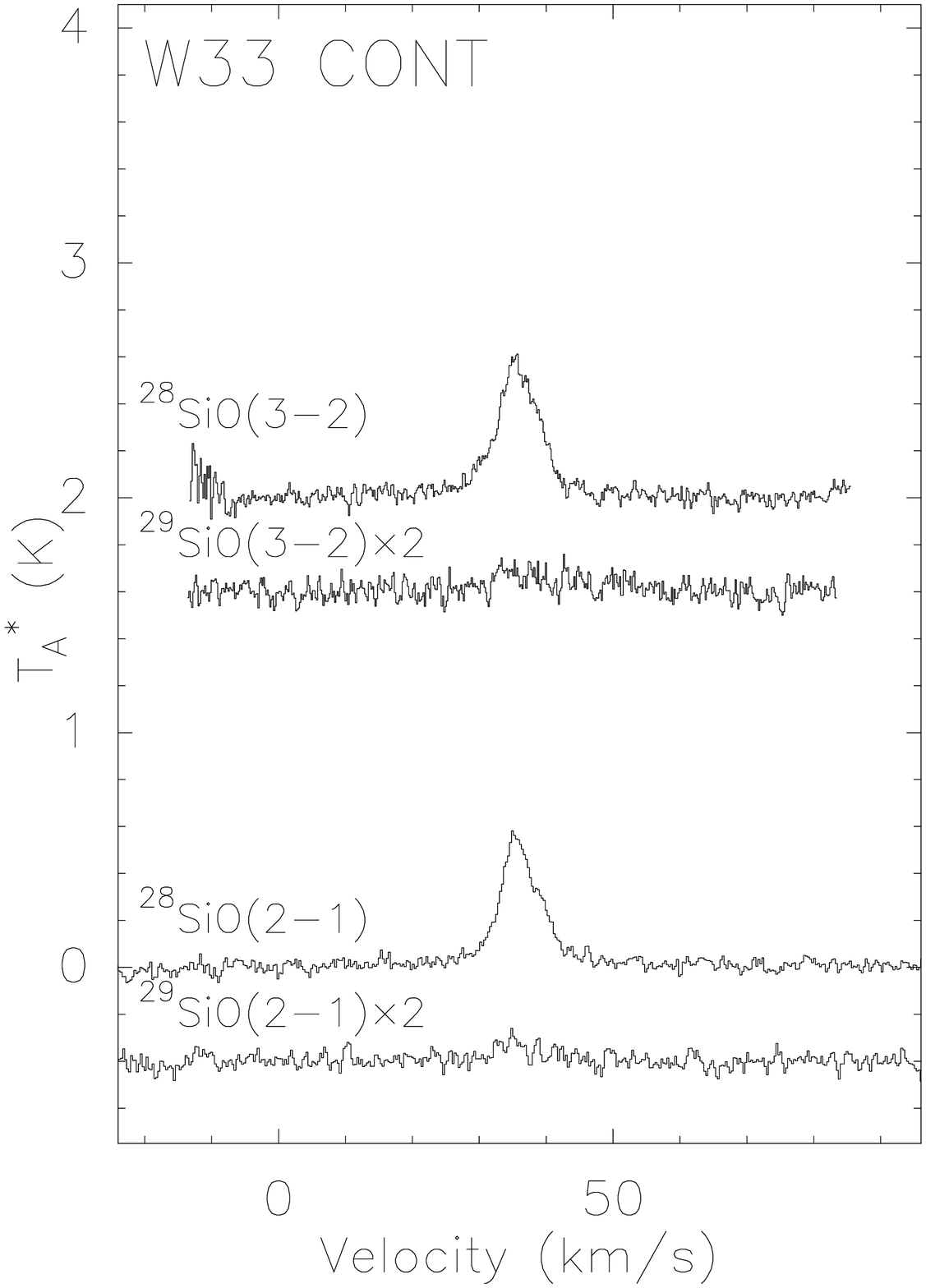}
\caption{The $v=0$, $J=3-2$ and $v=0$, $J=2-1$ lines of $^{28}$SiO and
$^{29}$SiO towards each observed source. The weaker $^{29}$SiO spectra
are multiplied by 2.}
\label{figure:SiO_spectra}
\end{center}
\end{figure*}

The obtained SiO spectra are presented in Fig. \ref{figure:SiO_spectra}.
The lines are nearly Gaussian in the central part and have broad,
asymmetric wings.
The line parameters are listed in Table \ref{table:29SiO_line_parameters}.
In this table we give the minimum, maximum, and peak velocities
($v_{\rm min}$, $v_{\rm max}$ and  $v_{\rm peak}$, respectively),
and the integrated intensities, $\int T_{\rm A}^*(v){\rm d}v$, 
over the given velocity range.

The spectra of the two SiO isotopologues were used to estimate
the optical thicknesses of the $J=2-1$ and $J=3-2$ lines
of $^{28}$SiO, $\tau_{2\rightarrow1}$ and $\tau_{3\rightarrow2}$.  
For this purpose all spectra were resampled to the
same LSR velocity grid with a channel width of 1 km s$^{-1}$. 
The smoothing of the spectra to this resolution was necessary to achieve
the sufficient signal to noise ratio. Using these resampled
spectra the optical thicknesses $\tau_{2\rightarrow1}$ and
$\tau_{3\rightarrow2}$ could be derived in 1--13 velocity channels near
the line peak for twelve sources. The terrestrial SiO isotopic abundance
ratio is $X_{\rm ter}(^{28}{\rm SiO}/^{29}{\rm SiO})=19.6$. In interstellar
medium values in the range 10--20 have been determined (\cite{penzias1981}).
Here we use the value $X=20$ when determining the optical thicknesses.
The $^{28}$SiO optical thicknesses $\tau_{2\rightarrow1}$ and
$\tau_{3\rightarrow2}$ are typically between 1 and 3 in the line
centres. 

The excitation temperatures, $T_{\rm ex}$, were estimated
from the optical thickness ratio by using Eq. (2) of
HLBZ98. These estimates are based on the assumption that 
$T_{\rm ex}(J=3-2) = T_{\rm ex}(J=2-1)$. They do not depend on the 
beam filling as the optical thicknesses were determined using pairs 
of lines having similar frequencies. The derived values of $T_{\rm ex}$ are
close to 5 K for all sources and velocity channels. 
The weighted average of $T_{\rm ex}$ and its standard deviation are
$T_{\rm ex}=4.3 \pm 0.1$ K (see Table \ref{table:sio_tex}).

From these results, the SiO column densities, $N$(SiO), were
estimated from the integrated intensities of the $^{29}$SiO spectra
by assuming optically thin emission and uniform excitation of
the rotational lines with $T_{\rm ex} = 5$ K for all sources.

\begin{figure*}
\begin{center}
\includegraphics[angle=270,width=4cm]{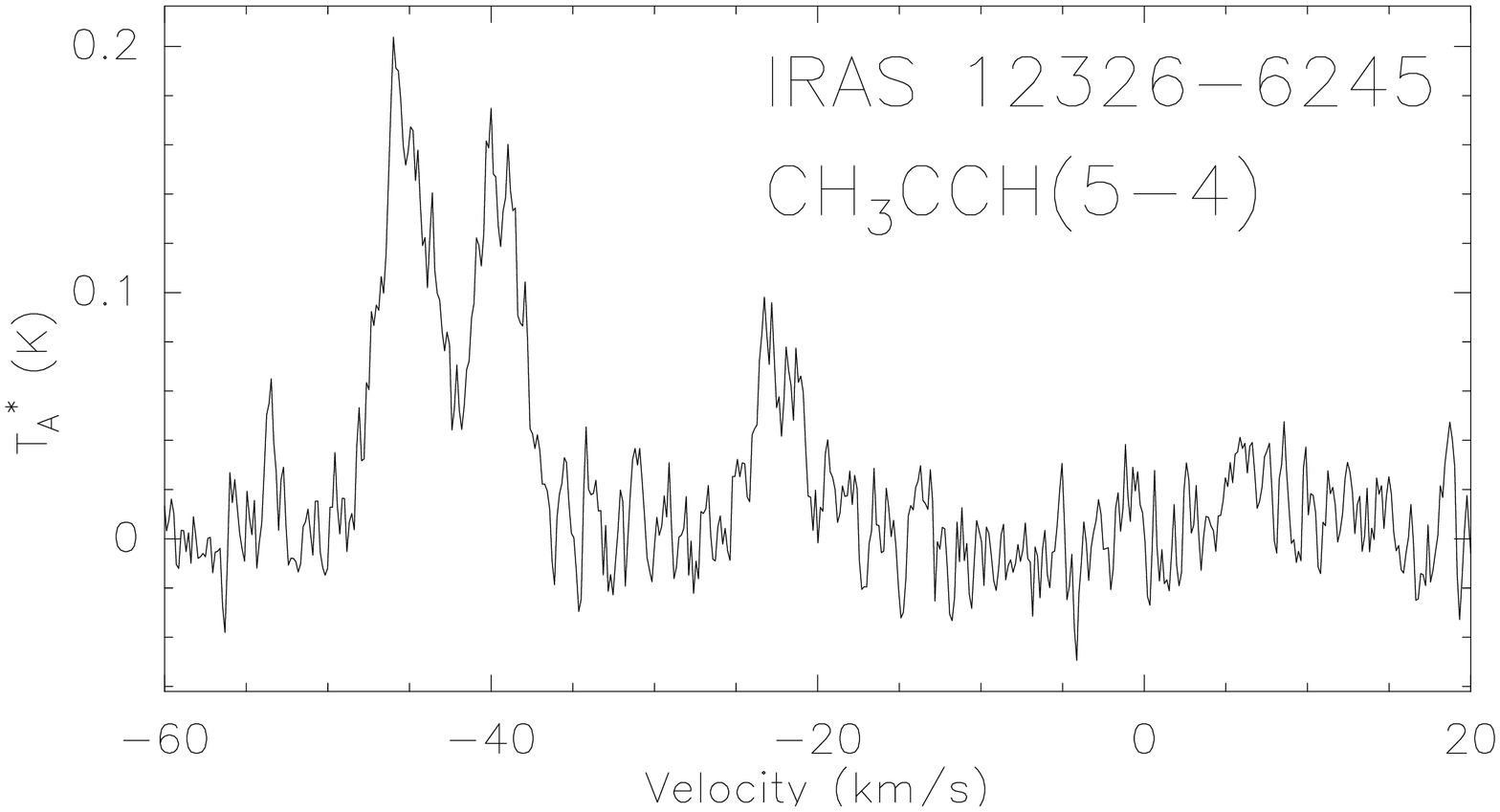}
\includegraphics[angle=270,width=4cm]{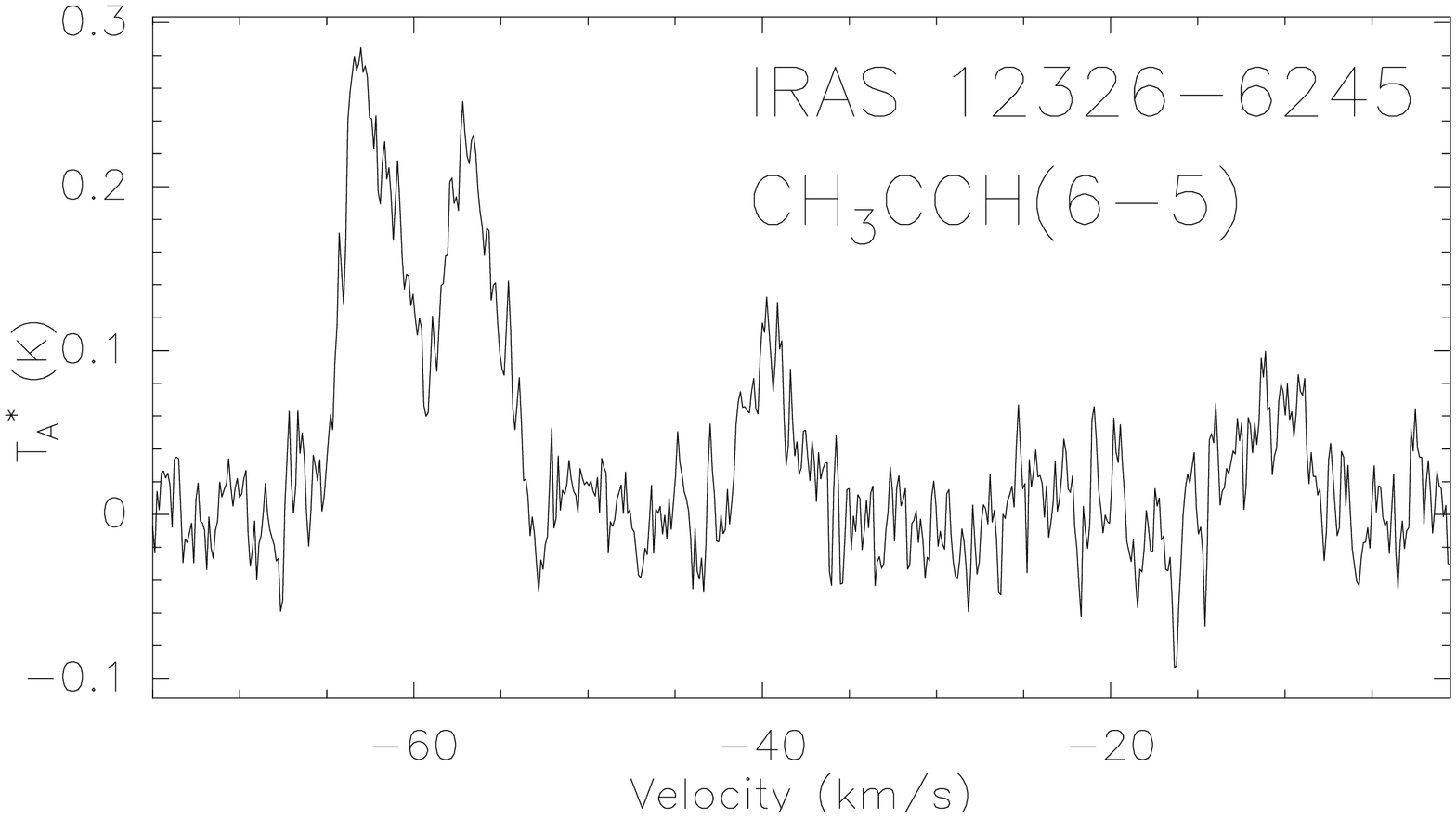}
\includegraphics[angle=270,width=4cm]{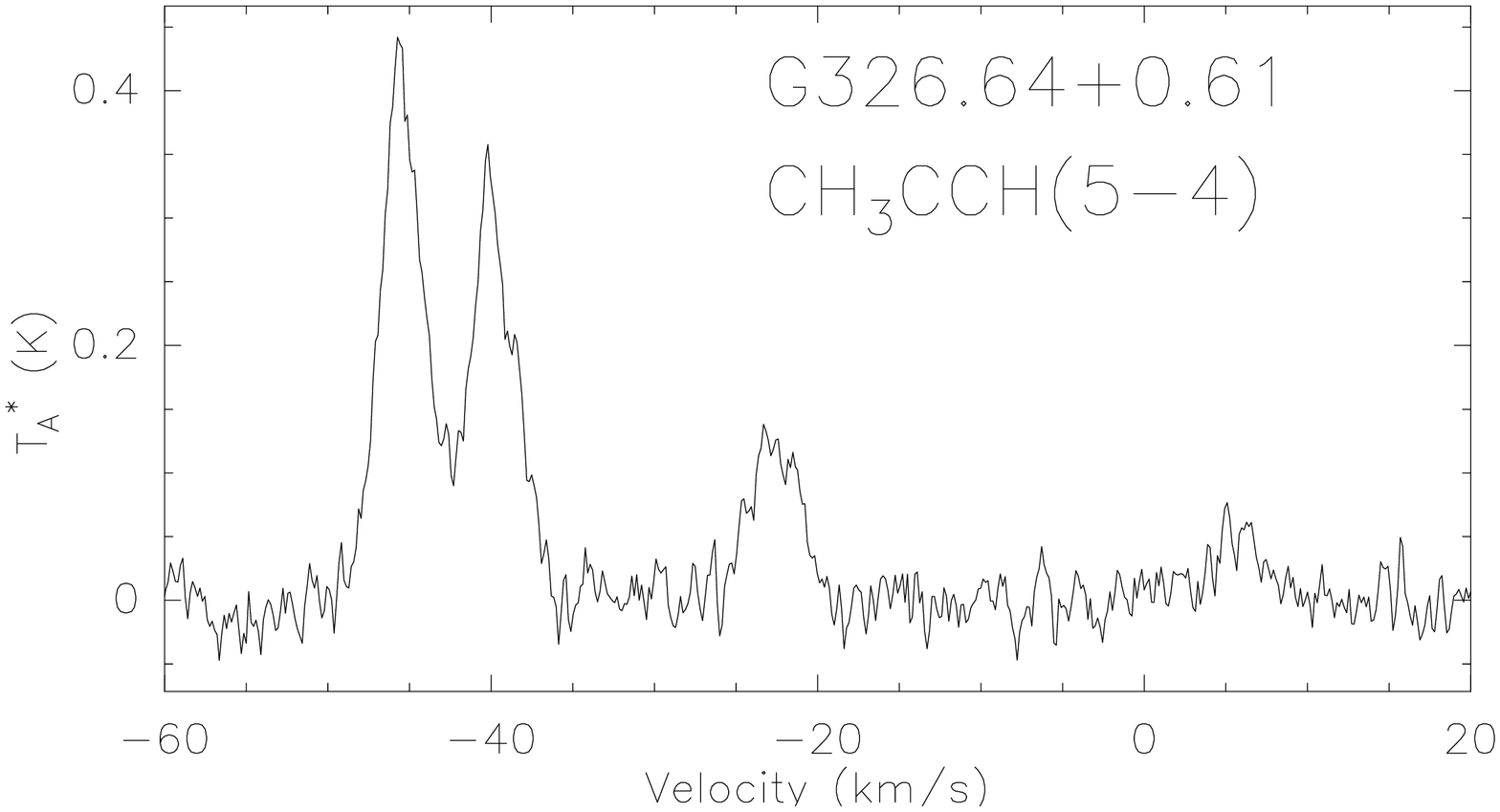}
\includegraphics[angle=270,width=4cm]{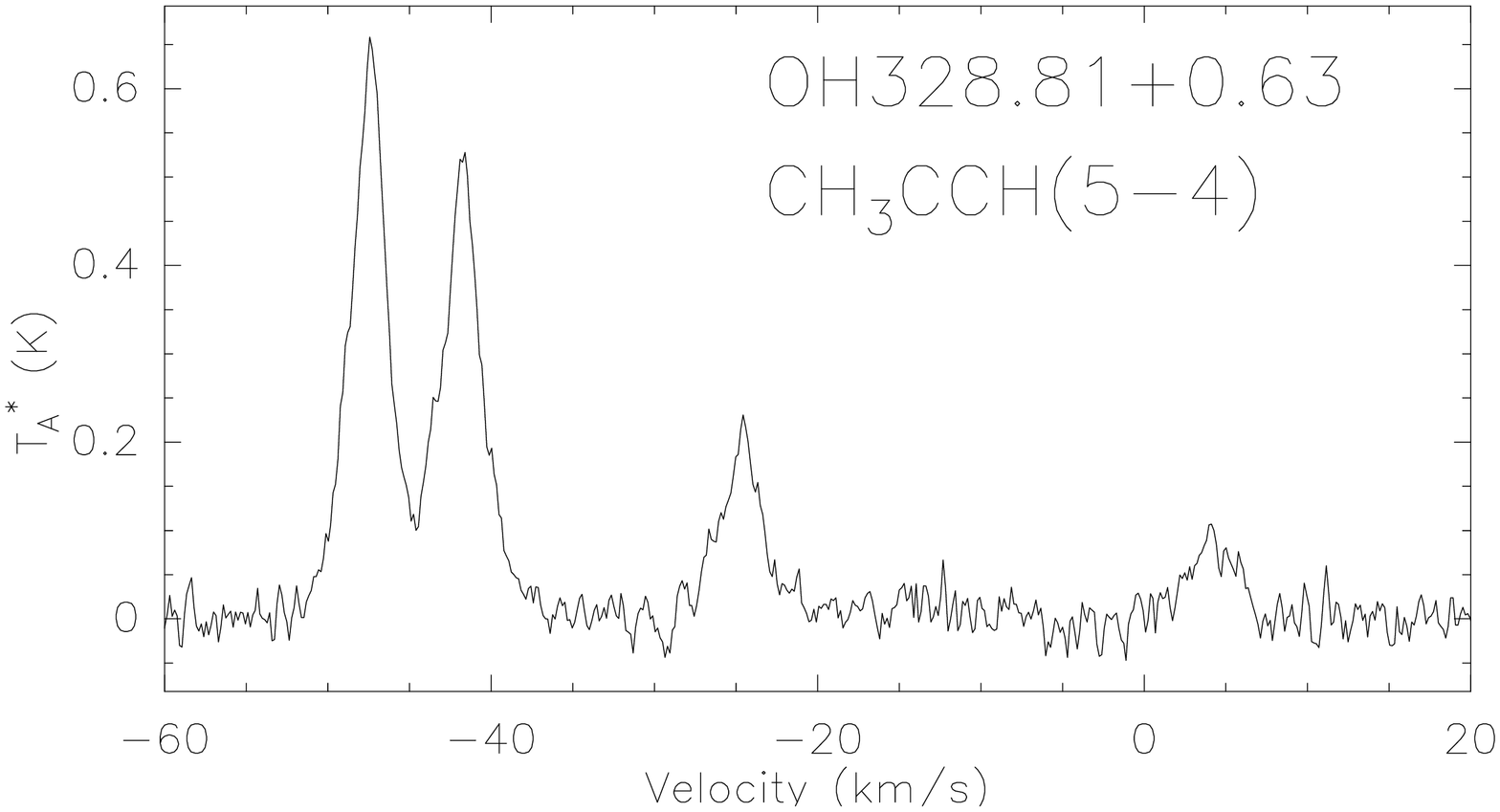}
\includegraphics[angle=270,width=4cm]{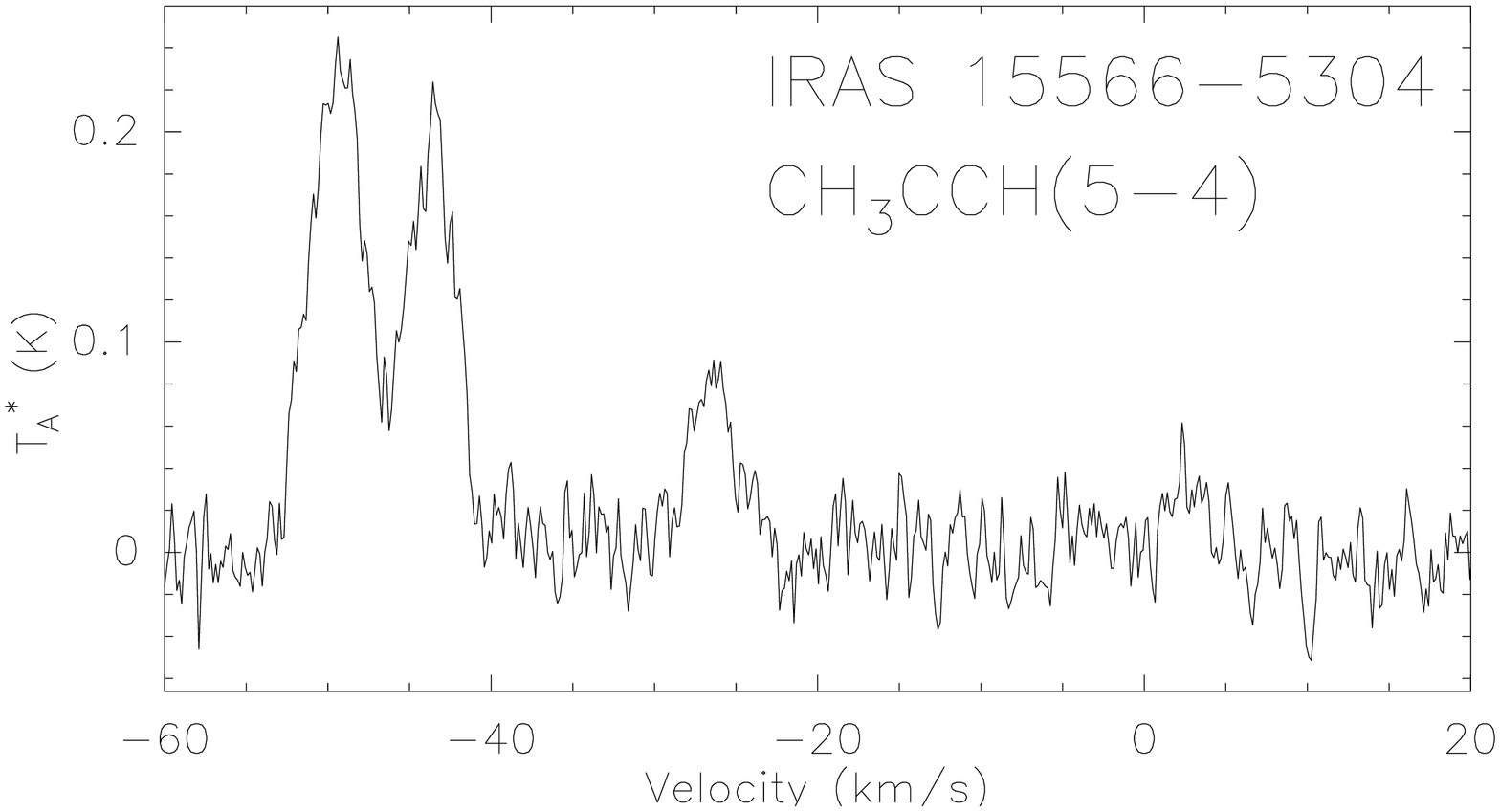}
\includegraphics[angle=270,width=4cm]{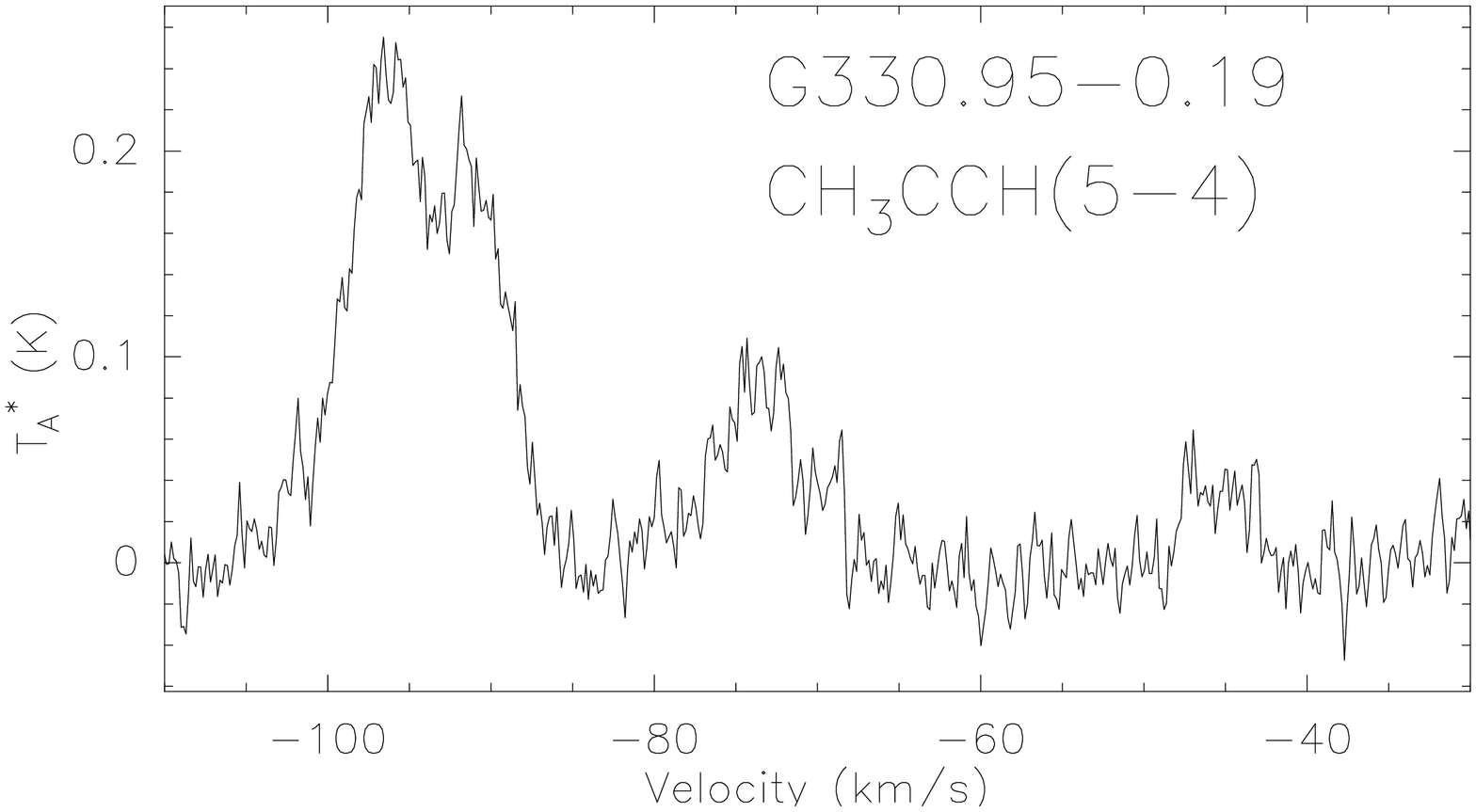}
\includegraphics[angle=270,width=4cm]{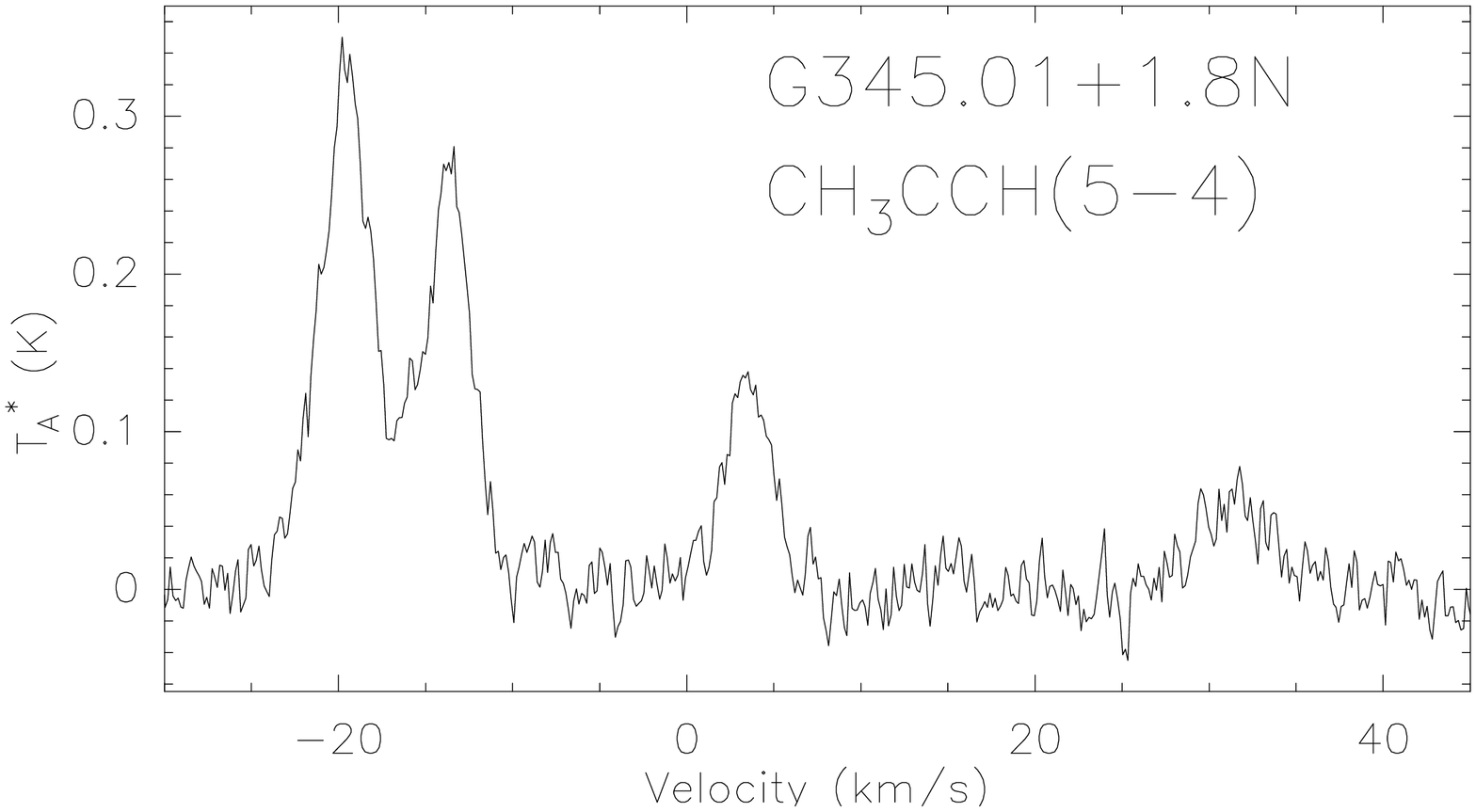}
\includegraphics[angle=270,width=4cm]{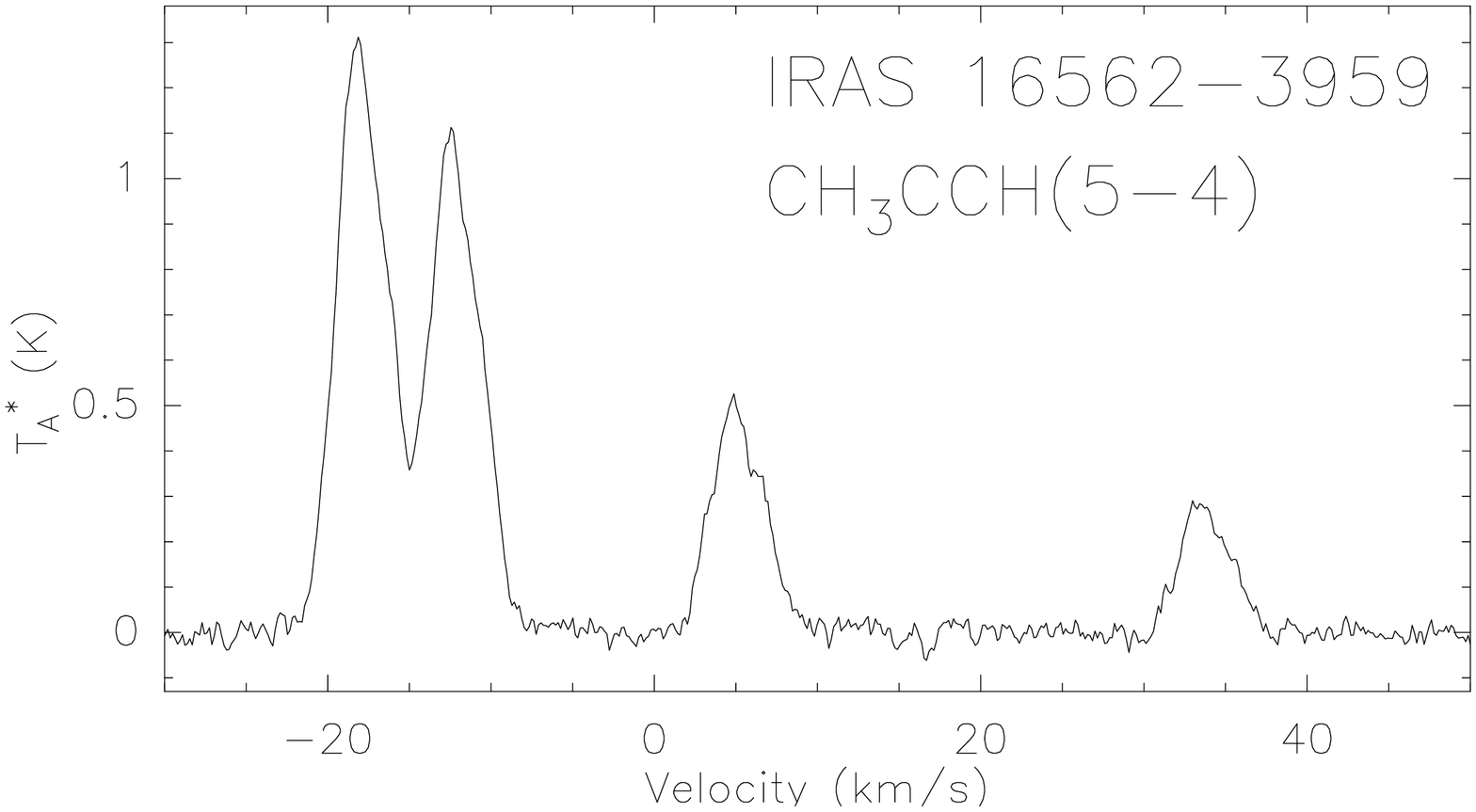}
\includegraphics[angle=270,width=4cm]{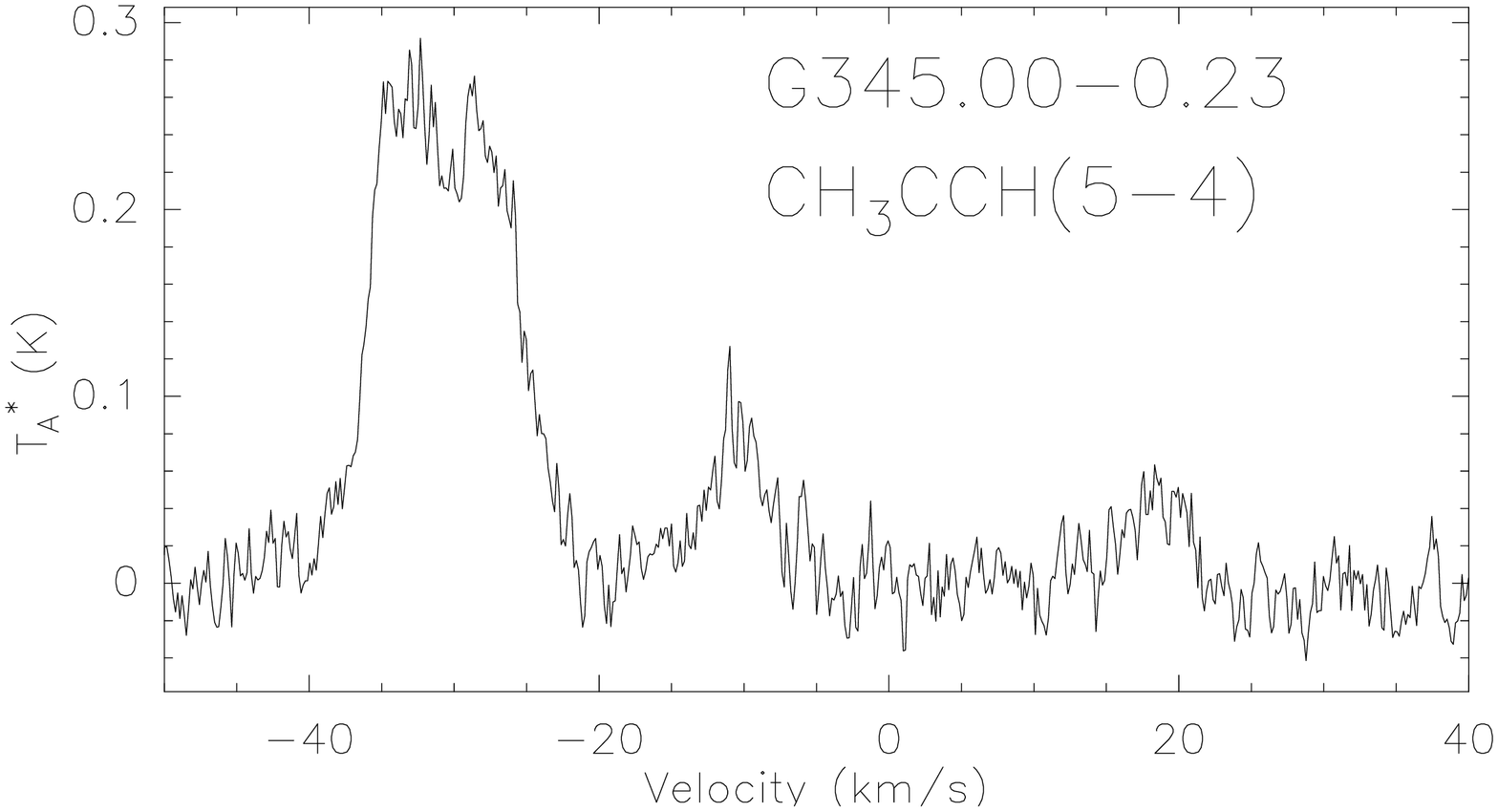}
\includegraphics[angle=270,width=4cm]{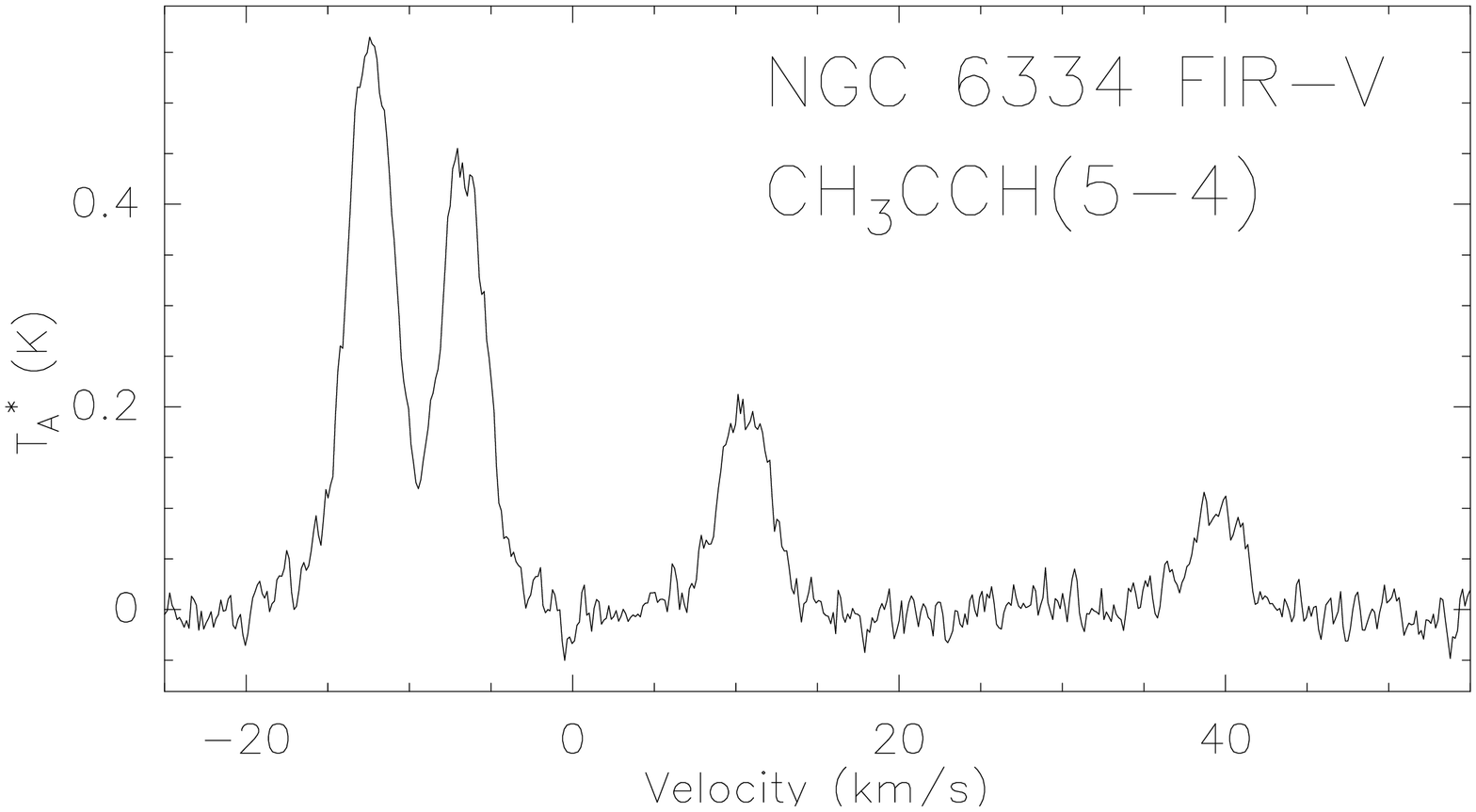}
\includegraphics[angle=270,width=4cm]{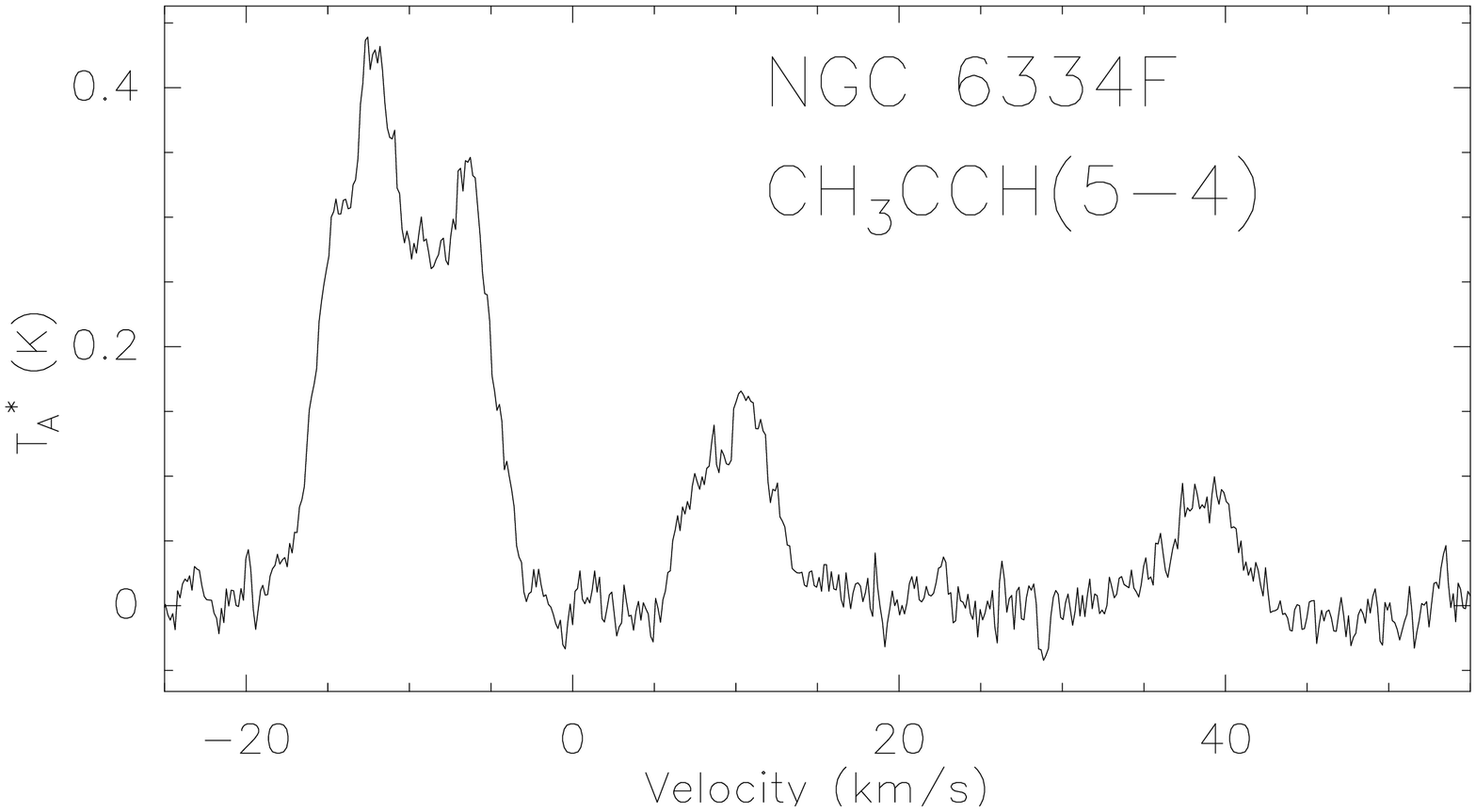}
\includegraphics[angle=270,width=4cm]{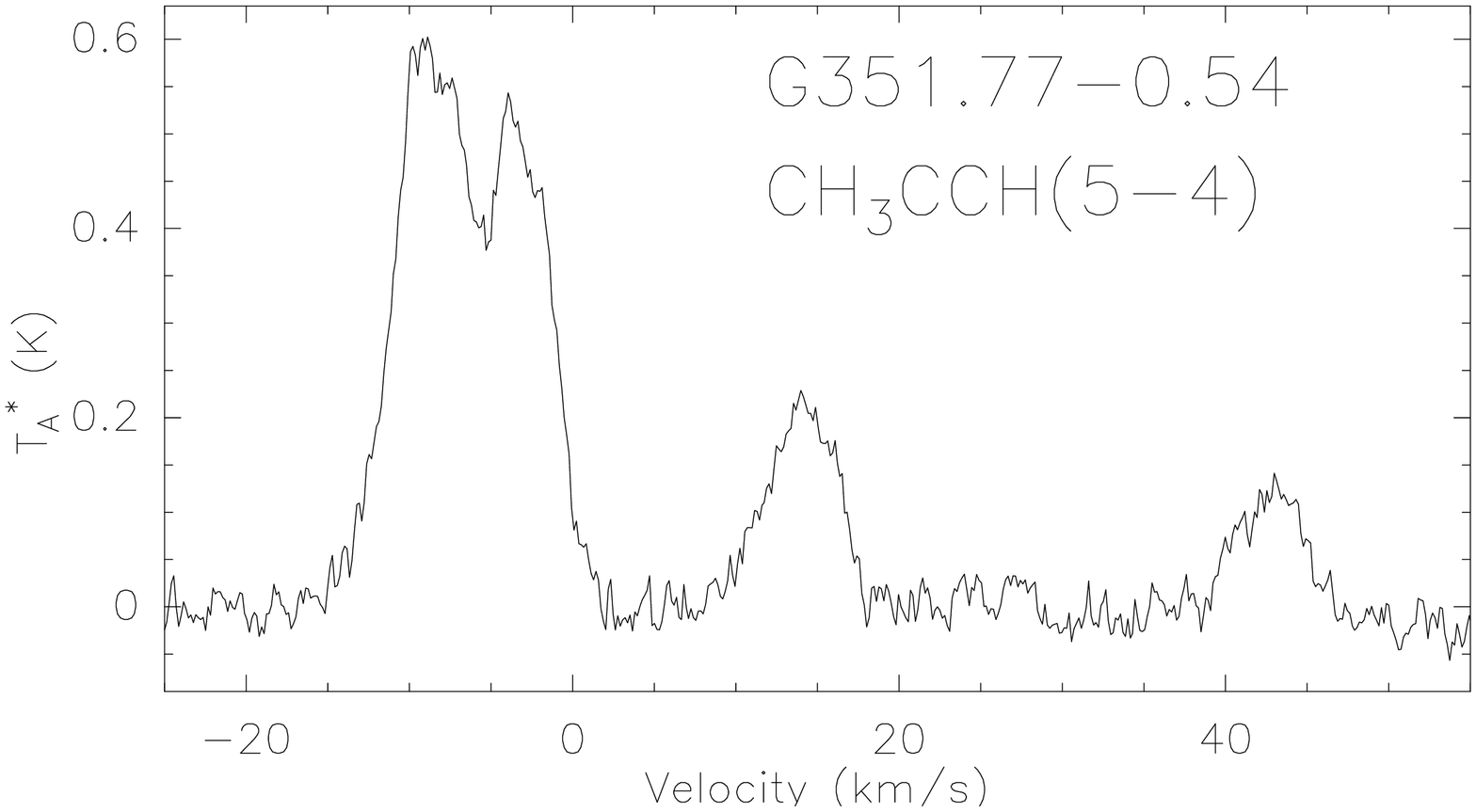}
\includegraphics[angle=270,width=4cm]{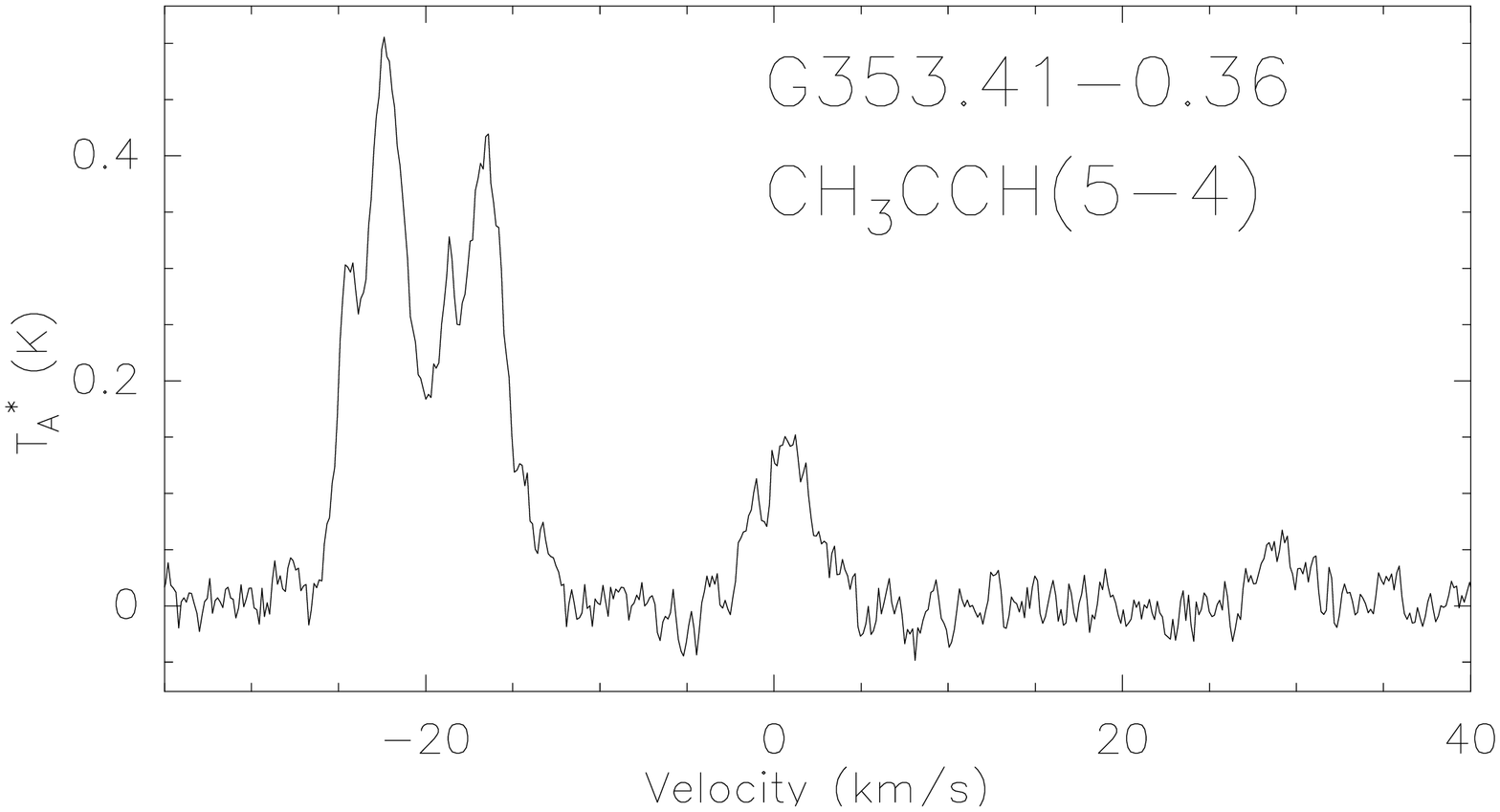}
\includegraphics[angle=270,width=4cm]{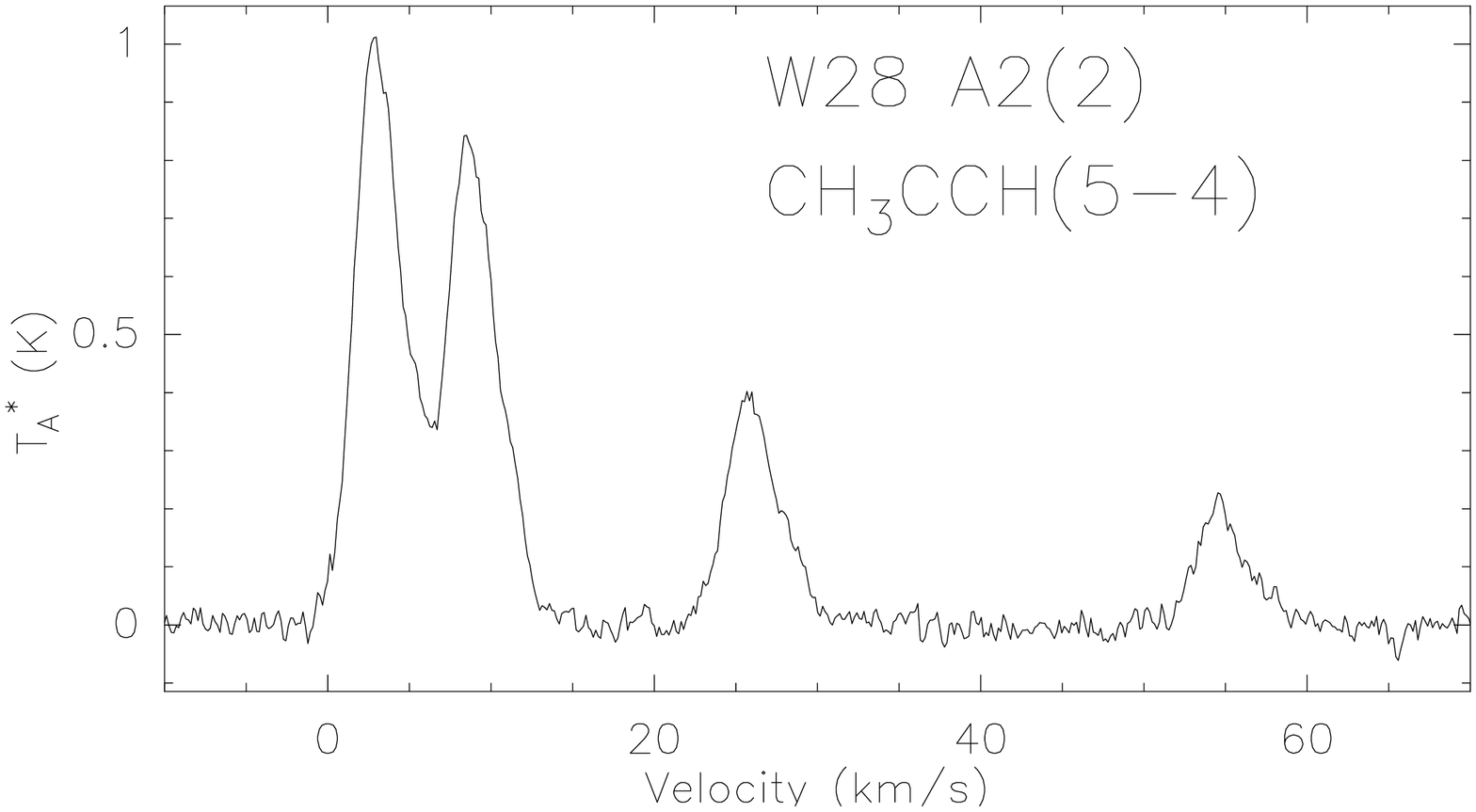}
\includegraphics[angle=270,width=4cm]{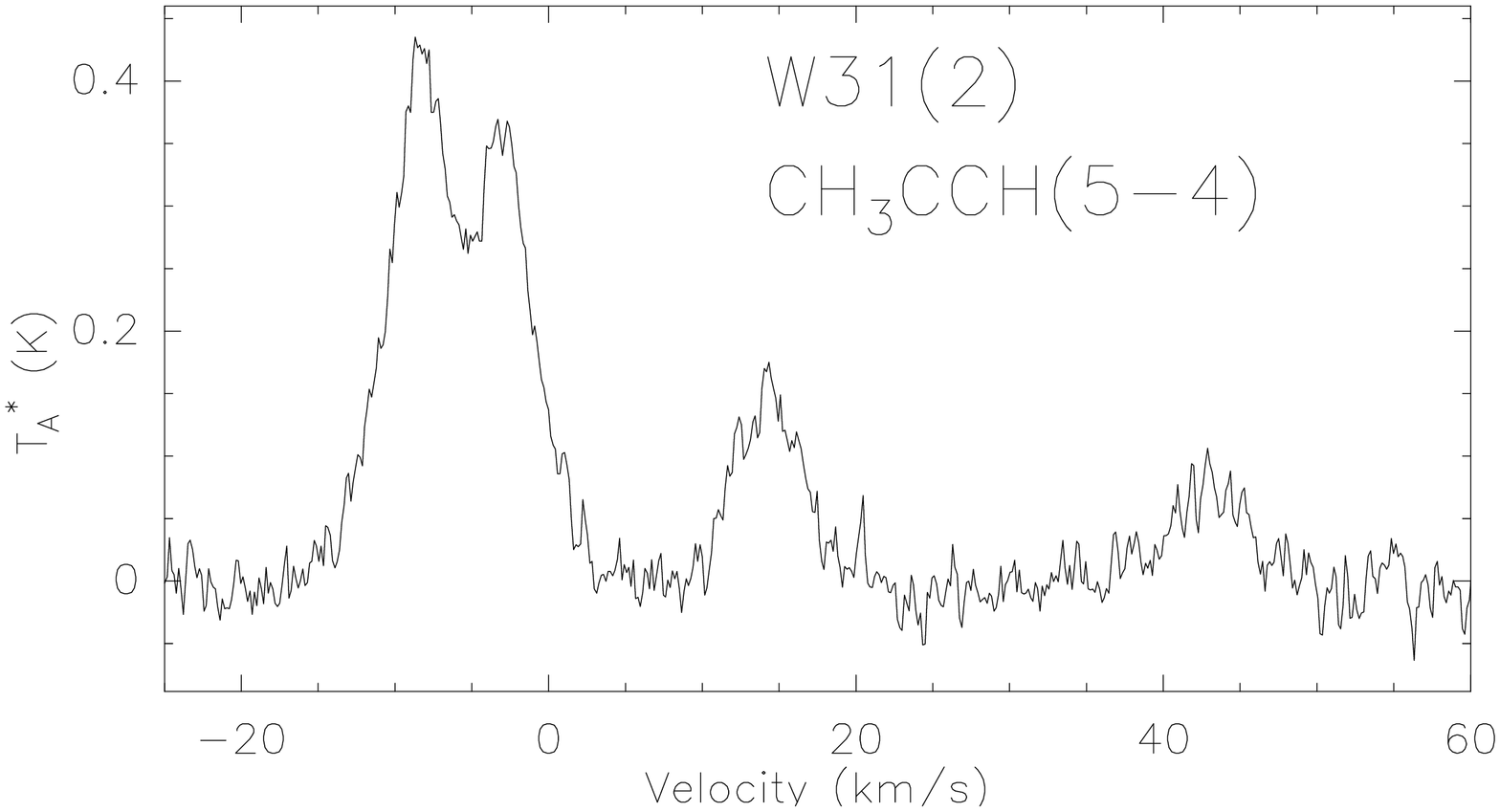}
\includegraphics[angle=270,width=4cm]{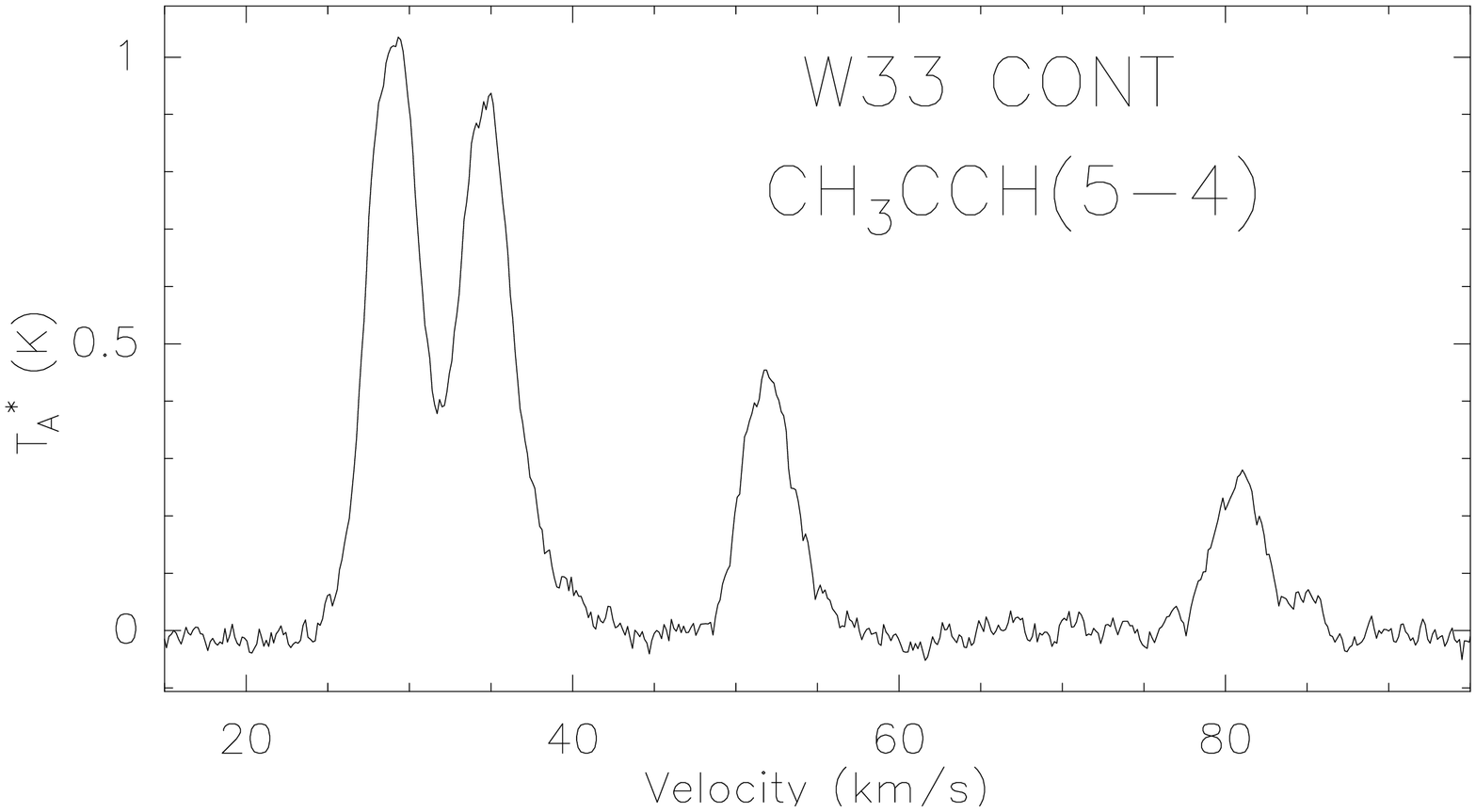}
\caption{Spectra of the CH$_3$CCH $J=5_K-4_K$ and $J=6_K-5_K$ rotational
transitions towards each observed source.
The X-axis represents the LSR velocity of the $5_1-4_1$
line ($6_2-5_2$ in the case of $6_K-5_K$) and the intensity scale
is antenna temperature, $T_{\rm A}^*$, in Kelvins.
The fairly low line intensities, $T_{\rm A}^{*} < 1 \; \rm K$,
suggest a low optical thickness.}
\label{figure:CH3CCH_spectra}
\end{center}
\end{figure*}

The CH$_3$CCH spectra are shown in Fig. \ref{figure:CH3CCH_spectra},
and the Gaussian parameters of the detected lines
are given in Table \ref{table:CH3CCH_line_parameters}.
The line peak velocities and
widths (FWHM) are listed in Cols. 2 and 3 of this table. The integrated
intensities of the components $K=0,1,2$ and 3 are given in Cols. 4 - 7.
The integrated intensities were used to derive the rotational
temperatures, $T_{\rm rot}$, and the CH$_3$CCH column densities,
$N({\rm CH_3CCH})$, with the population diagram method described and
discussed in, e.g., \cite{askne1984}, \cite{bergin1994}, and
\cite{goldsmith1999}.

\section{Physical/chemical parameters of the cores}

\subsection{Column densities and rotational temperatures}

In the case of uniform excitation and optically thin dipole transition,
the total column density of the emitting molecule is related 
to the integrated line intensity by the formula 

\begin{equation}
N_{\rm tot} =  \frac{3k\epsilon_0}{2\pi^2 \mu^2} \; \frac{1}{\nu S_u} \; 
\frac{e^{E_u/kT_{\rm ex}} \, Z(T_{\rm ex})}
     {1 - \frac{F(T_{\rm bg})}{F(T_{\rm ex})}} \; 
\frac{1}{\eta} \, \int \, T_{\rm A}^*(v) \, {\rm d}v \; ,
\label{eq01}
\end{equation}
where $k$ is the Boltzmann constant, $\epsilon_0$ is the vacuum
permittivity, $\mu$ is the permanent dipole moment, $\nu$ and $S_u$ are
the frequency of the transition and the line strength, $E_{\rm u}$
is the upper state energy, $Z$ is the rotational partition function, 
$\eta$ is the beam-source coupling efficiency,
and the function $F(T)$ is defined by

\begin{equation}
F(T) \equiv \frac{1}{e^{h\nu/kT}-1} \; .
\label{eq02}
\end{equation}
For the rotational transition $J_u \rightarrow J_u-1$ of a linear molecule, 
like SiO, $S_u=J_u$.
We have assumed that $T_{\rm ex}=5$ K for all sources
(see Table \ref{table:sio_tex}) and $\eta=\eta_{\rm MB}$ 
(0.75 at 86 GHz and 0.68 at 129 GHz), which means that
the source just fills the main beam. For the SiO isotopic abundance ratio 
we have used the value $X=20$.

The rotational temperatures, $T_{\rm rot}$, and the CH$_3$CCH column densities,
$N({\rm CH_3CCH})$, were derived by means of the population diagram method.
This method is based on comparison of intensities of spectral lines
which lie close in frequency although arising from rotational levels with 
different energies. In the case of CH$_3$CCH, the close lying transitions
represent different $K$-components.

Assuming uniform excitation and optically thin emission,
the 'rotational diagram equation', i.e. the equation relating
the integrated intensities of different $K$-components to
the rotational temperature, $T_{\rm rot}$, and
the total column density, $N_{\rm tot}$, can be written as

\begin{equation}
\ln\Bigg[\frac{\int T_{A}^{*}(v){\rm d}v}{\eta\nu S_{J,K}g_{K}g_{I}}\Bigg]=
\ln\Bigg(\frac{2\pi^{2}\mu^{2}}{3k\epsilon_{0}}\frac{N_{\rm tot}}{Z_{\rm rot}}\Bigg)-
\frac{1}{T_{\rm rot}}\frac{E_{u}}{k} \; ,
\label{eq03}
\end{equation}
where $g_K$ is the $K$ degeneracy and statistical weight $g_I$
takes the spins of hydrogen nuclei into account.
See, e.g., Appendix B of \cite{anderson1999},
for derivation of Eq. (\ref{eq03}).
As above, we have again assumed that $\eta=\eta_{\rm MB}$.

\begin{figure*}
\begin{center}
\includegraphics[width=8cm]{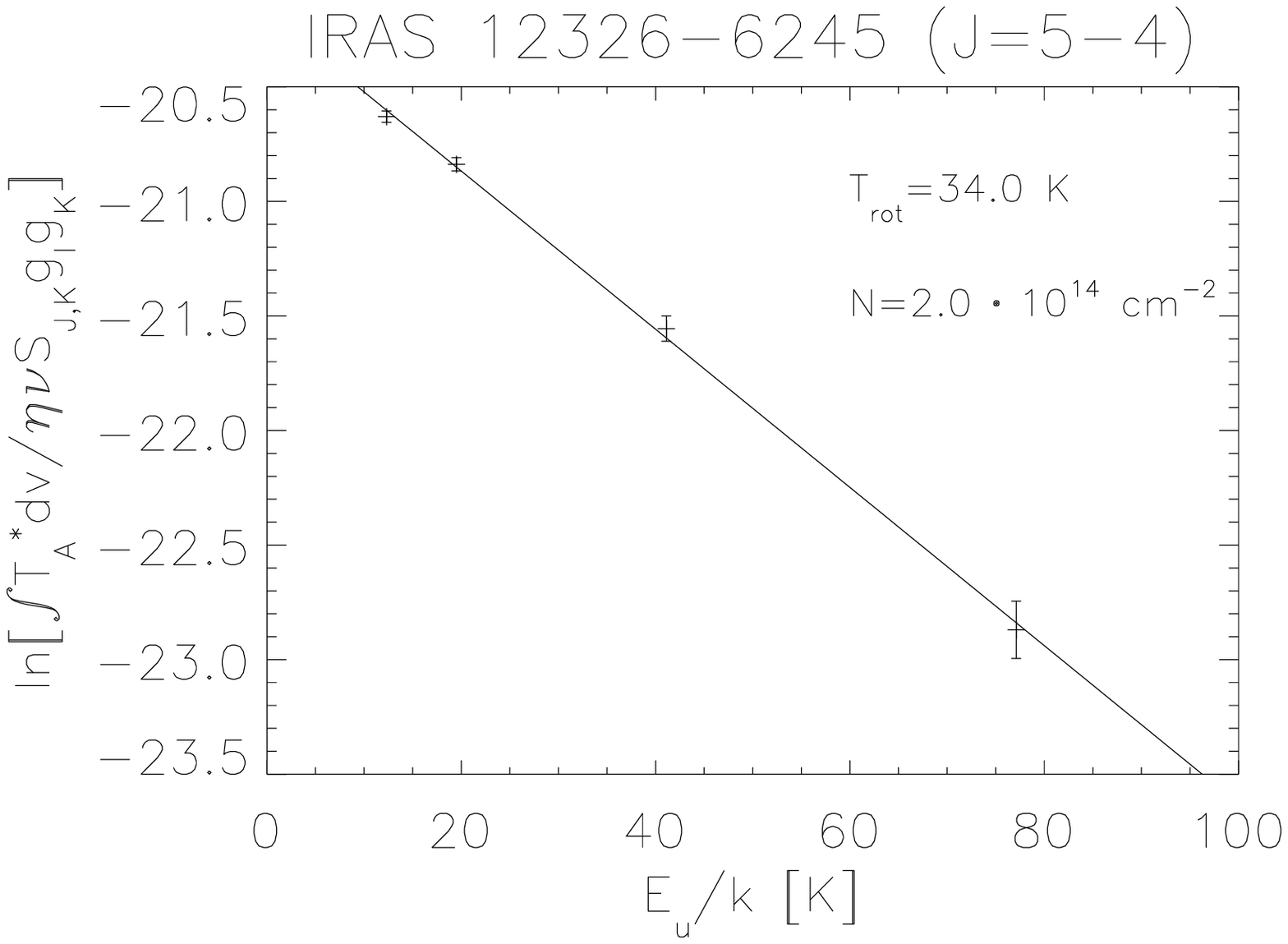}
\includegraphics[width=8cm]{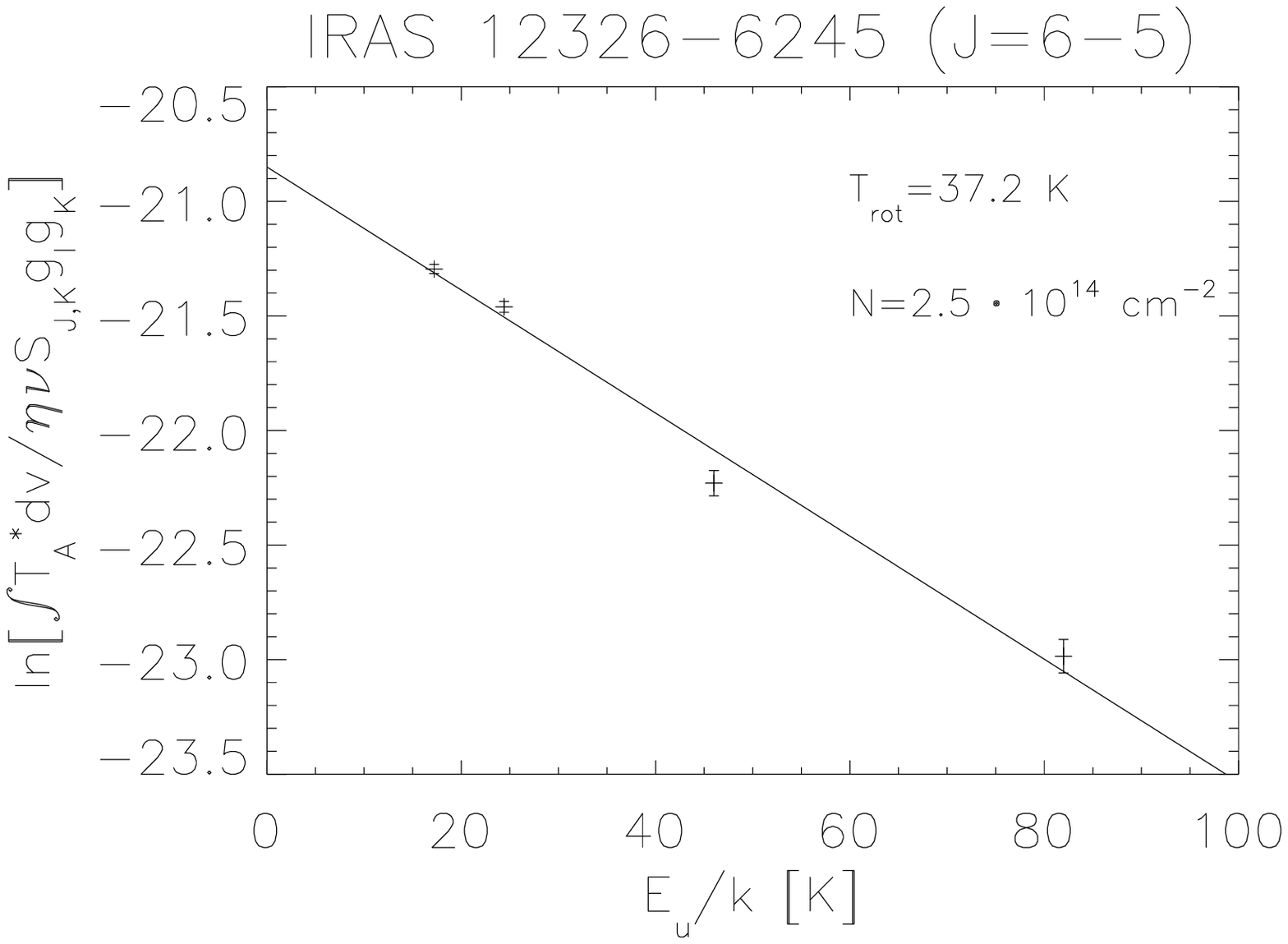}
\includegraphics[width=8cm]{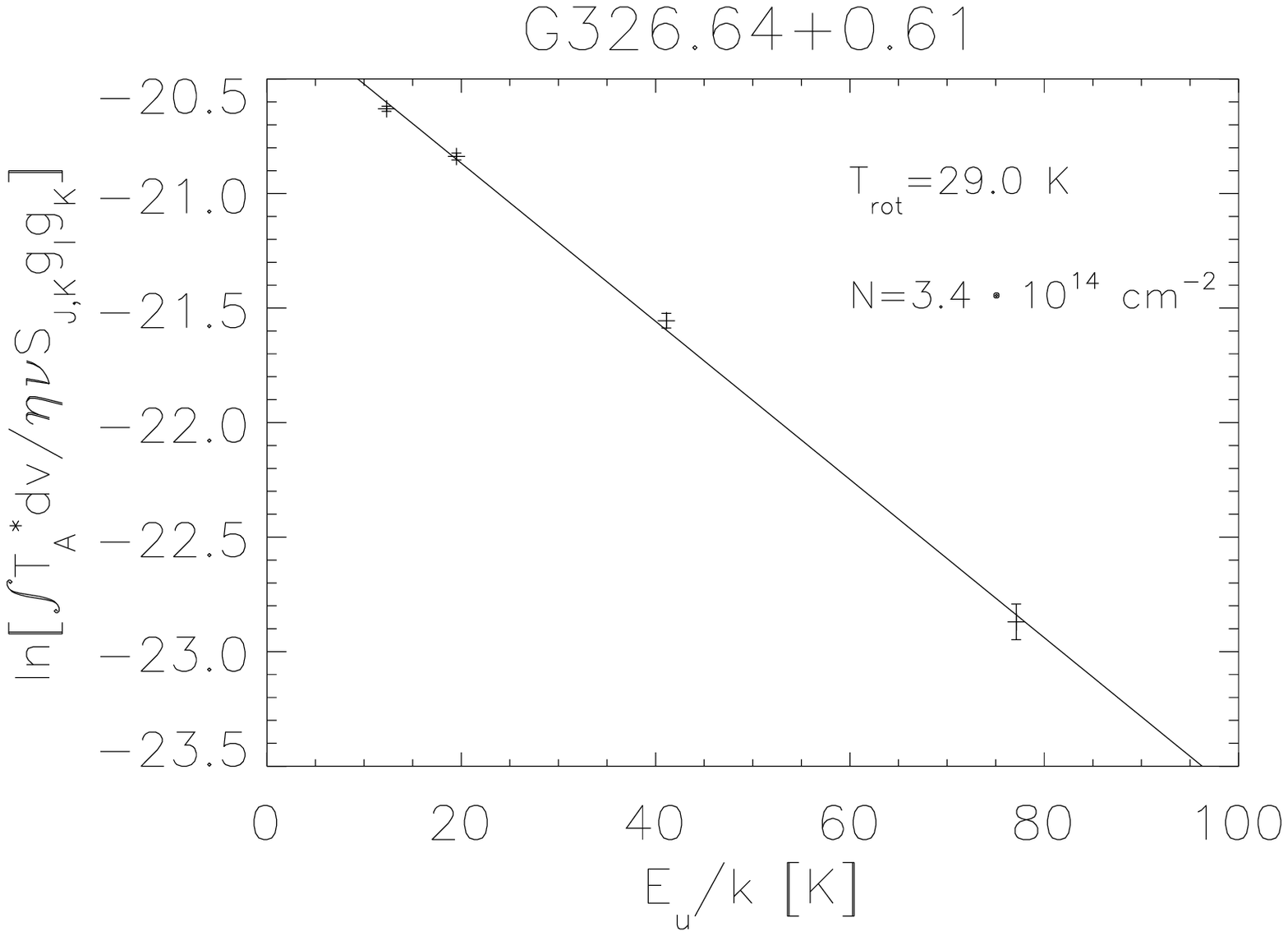}
\caption{Rotational diagrams of the CH$_3$CCH spectra for two of the observed
 sources. X-axis plots the upper transition state energies divided by
 the Boltzmann constant, $\frac{E_{\rm u}}{k}$, and Y-axis plots 
 the lefthand side of Eq. (\ref{eq03}). The lines represent linear
 least-squares fits to the data. The error bars correspond to
 the errors of integrated intensity according to the law of propagation
 of errors. The rotational temperature, $T_{\rm rot}$, and the total
 CH$_3$CCH column density, $N$, are given in the top right of each panel.}
\label{figure:diagrams}
\end{center}
\end{figure*}

Rotational diagrams for two sources are shown in
Fig. \ref{figure:diagrams}.
Straight lines were fitted to the data using a least-squares fitting technique.
The data are consistent with a single $T_{\rm rot}$. The goodness of
the fit substantiates the assumptions that the gas is in LTE with
$T_{\rm rot} \approx T_{\rm kin}$ and that the lines are optically
thin.

The H$_2$ column density, $N$(H$_2$), of each source
was derived from dust continuum emission using the
following equation:

\begin{equation}
N(\textrm{H$_2$})=\frac{I_{\nu}^{\rm dust}}{B_{\nu}(T_{\rm d})\mu m_{\rm H}\kappa_{\rm d}R_{\rm d}} \; ,
\label{eq04}
\end{equation}
where $I_{\nu}^{\rm dust}$ is the observed dust surface brightness
(see Table \ref{table:smoothed_intensities}),
$B_{\nu}(T_{\rm d})$ is the Planck function for a blackbody
of dust temperature $T_{\rm d}$, $\mu=2.29$ is the mean molecular weight
accounting for a 10 \% contribution of helium
(e.g., \cite{hill2005}), $m_{\rm H}=1.673534 \cdot 10^{-27}$ kg is the
mass of the hydrogen atom, $\kappa_{\rm d}$ is the mass absorption coefficient
per unit mass of dust, and $R_{\rm d}$ is the dust-to-gas mass ratio.

We assume that $T_{\rm d}$ is equal to the gas kinetic
temperature derived from CH$_3$CCH observations.
$T_{\rm kin}$ and $T_{\rm d}$ are believed to be well coupled deep
in the cloud, above a density of about $10^4$ cm$^{-3}$,
due to collisions between gas and dust grains
(see, e.g., \cite{takahashi1983}; \cite{lee2004}).
Values of 0.1  m$^2$ kg$^{-1}$ at $\lambda=$ 1.2 mm (\cite{ossenkopf1994})
and $\frac{1}{100}$ are adopted for $\kappa_{\rm d}$
and $R_{\rm d}$, respectively.

The rotational temperatures, the H$_2$, SiO and CH$_3$CCH column densities and
the fractional SiO and CH$_3$CCH abundances
($\chi(x) \equiv N(x)/N(\rm H_2)$) are listed
in Table \ref{table:abundances}.
The average value and the standard deviation of
$T_{\rm rot}$ is $32.7 \pm 3.6$ K. The $N$(H$_2$) values are found to be
$\sim 10^{22}-10^{23}$ cm$^{-2}$, and $N$(SiO) and $N$(CH$_3$CCH) lie
in the range $0.8-5.5 \cdot 10^{13}$ cm$^{-2}$ and
$2.0-15.0 \cdot 10^{14}$ cm$^{-2}$, respectively.
The values of $\chi({\rm SiO})$ and $\chi({\rm CH_3CCH})$ are
found to be $\sim 3.0 \pm 1.7 \cdot 10^{-10}$ and
$6.9 \pm 4.1 \cdot 10^{-9}$, respectively.

\subsection{Linear sizes, mass estimates and densities}

The linear sizes have been computed from the angular
diameters listed in Table \ref{table:smoothed_intensities}
using the distances listed in Table \ref{table:sources}.

The core masses have been estimated using the 1.2 mm continuum maps. 
We have also calculated the virial masses using the velocity
dispersions and kinetic temperatures from the CH$_3$CCH data. 
The mass of a core, $M_{\rm cont}$, has been calculated assuming that 
the dust emission is optically thin and
that the dust-to-gas ratio, $R_{\rm d}$, and the dust absorption coefficient
per unit mass, $\kappa_{\rm d}$,
are constant. These assumptions imply the following equation:

\begin{equation}
M_{\rm cont}=\frac{S_{\nu}d^2}{B_{\nu}(T_{\rm d})\kappa_{\rm d}R_{\rm d}} \; ,
\label{eq05}
\end{equation}
where $S_{\nu}$ is the 1.2 mm continuum integrated flux within
the selected aperture (see Table \ref{table:smoothed_intensities}).

The virial masses, $M_{\rm vir}$, have been estimated by approximating 
the mass distribution by a homogenous, isothermal sphere without magnetic 
support and external pressure, using the formula

\begin{equation}
M_{\mathrm{ vir}}[\mathrm{M_{\sun}}]=\big( 0.5 \, (\Delta v_{1/2}[\mathrm{km} \;
\mathrm{s}^{-1}])^2 + 0.01 \, T[\mathrm{K}]\big) d[\mathrm{kpc}] \Theta_{\mathrm s}[\arcsec]\; ,
\label{eq06}
\end{equation}
where $\Delta v_{1/2}$ is the FWHM of the CH$_3$CCH($5_K-4_K$) line, 
$T$ is the gas kinetic temperature derived from the CH$_3$CCH
observations, $d$ is the distance to the source, and $\Theta_{\rm s}=2R/d$ is
the angular diameter of the source as estimated from the SIMBA maps. 
For W28 A2(2) no virial mass estimate was obtained as its
(deconvolved) angular size is less the half the beam size.

The average H$_2$ number densities, $n(\rm H_2)$, were calculated using the 
masses, $M_{\rm cont}$, and radii, $R$, estimated from the dust 
continuum maps. The obtained radii, masses and average densities are listed
in Table \ref{table:properties}. The average radius of the sources
is $0.2 \pm 0.16$ pc, the average value of the masses estimated
from continuum emission is $1.8 \pm 2.1 \cdot 10^3$ M$_{\sun}$,
and the average density is $4.5 \pm 6.5\cdot 10^6$ cm$^{-3}$.

\section{Discussion}

\subsection{SiO and CH$_3$CCH abundances}

The 1.2 mm dust emission is likely to be dominated by the cool, $T\sim
30$~K, envelopes, although it has a contribution from hot cores ($T > 100$~K)
around newly born massive stars. The dust temperatures derived by
\cite{faundez2004} towards several of our sources using SIMBA and IRAS
100 and 60 $\mu$m data are comparable to the CH$_3$CCH rotational
temperatures, $T_{\rm rot}$, derived here. Taken the uncertainties due
to different beam sizes of the SIMBA and IRAS into account the
agreement is reasonable and substantiates the assumption $T_{\rm dust}
\approx T_{\rm rot}$ used in the total $N({\rm H_2})$ estimates.

While it can be assumed that the CH$_3$CCH lines and 1.2 mm dust
continuum emission trace to a large part the same material, this is
not obvious for dust and SiO. All SiO lines observed in this
survey show high-velocity wings, are single-peaked, and have their
maxima near the systemic velocity of the cloud. These characteristics
can be qualitatively explained with bow-shock models where the jet
generating the shock is seen in a small angle, i.e. either head on or
tail on. A large inclination should result in a double peaked profile
from a bow shock, and in a narrow line in the case where the emission
originates in a turbulent wake behind a bow shock (HLBZ98, Fig.~12). In
view of the distance to the sources and the angular resolution of the
present observations, it is clear that the telescope beam encompasses
entire star-forming regions with various gas components and young
stars at different evolutionary stages. The line profiles may
therefore have contribution from several outflows with different
inclinations.

In the light of previous SiO mapping observations it seems possible, however,
that part of the low-velocity emission originates in the quiescent envelope. 
This kind of component has been discovered in
high-resolution SiO mappings of star-forming regions 
(Lefloch et al. 1998; Codella et al 1999; Shepherd et al. 2004;
Fuente et al. 2005). In these studies SiO emission characterized by
narrow and broad lines have been found to come from separate
regions. The SiO abundances derived for the low-velocity component
are $\sim 10^{-10}$. This value is smaller than the abundances
derived towards high-velocity shocks ($\ga 10^{-8}$, e.g.,
\cite{bachiller1991}; \cite{gibb2004}), and larger than the upper
limits derived in cold, dark clouds and PDRs ($\sim 10^{-12}$, e.g.,
\cite{ziurys1989}; \cite{martin1992}; \cite{schilke2001}).

In the present single-pointing observations 
typically half of the integrated SiO intensity comes from the velocity
range with detectable CH$_3$CCH, which is supposed to trace the quiescent
gas component (see Table \ref{table:29SiO_line_parameters}).
However, the same radial velocity does
not necessarily mean spatial coincidence. Although SiO abundances
derived in outflows are based on comparison with CO in the same
velocity range, a similar spatial velocity correlation cannot be
assumed near the cloud's systemic velocity. Depending on the
structure and orientation of the shock, it is possible to find 
shock-produced SiO at low radial velocities. Therefore the column density
of the low-velocity SiO gives only upper limit to the SiO column density
in the quiesent envelope, and the same is probably true for the fractional
SiO abundance using the H$_2$ column density determined by dust. 

A comparison of the SiO main beam brightness temperatures, $T_{\rm
MB}$, with the derived optical thicknesses, $\tau$, and excitation
temperatures, $T_{\rm ex}$, suggests that the beam filling factors
are close to unity, and the SiO emission regions are extended. We find
that the upper limits for the fractional SiO abundance in the
low-velocity gas are on the order of $10^{-10}$, i.e. similar to the
SiO abundances derived previously for quiescent gas component in some
star-forming regions. It should be noted, however, that the same
result would be achieved if SiO is present only in shocked layers
filling 1\% of the gas volume. This is not necessarily in
contradiction with nearly uniform beam filling.

The model for greatly enhanced SiO production in powerful shocks by
silicate grain destruction and subsequent high-temperature gas-phase
chemistry is well established (\cite{schilke1997}; \cite{caselli1997};
\cite{pineau1997}). The possible occurrence of SiO in the quiescent
gas component is not as well understood. In the postshock gas SiO
should convert in a reaction with OH into SiO$_2$ (SiO + OH
$\longrightarrow$ SiO$_2$ + H), which is eventually removed from the
gas phase by accretion onto the dust grains.
According to \cite{codella1999}, the times needed to remove SiO from
the gas phase and to slow down speeding SiO 'blobs' are similar, $\sim
10^4$ yr, and this would support the outflow remnant scenario. In
another, perhaps more feasible model high-velocity SiO emission comes
from bow-shocks generated by high-velocity jets. In this case
low-velocity SiO can originate in turbulent wakes behind
bow-shocks (see \cite{raga1993} and the model profiles in Fig. 12 of HLBZ98).

\begin{figure}
\resizebox{\hsize}{!}{\includegraphics{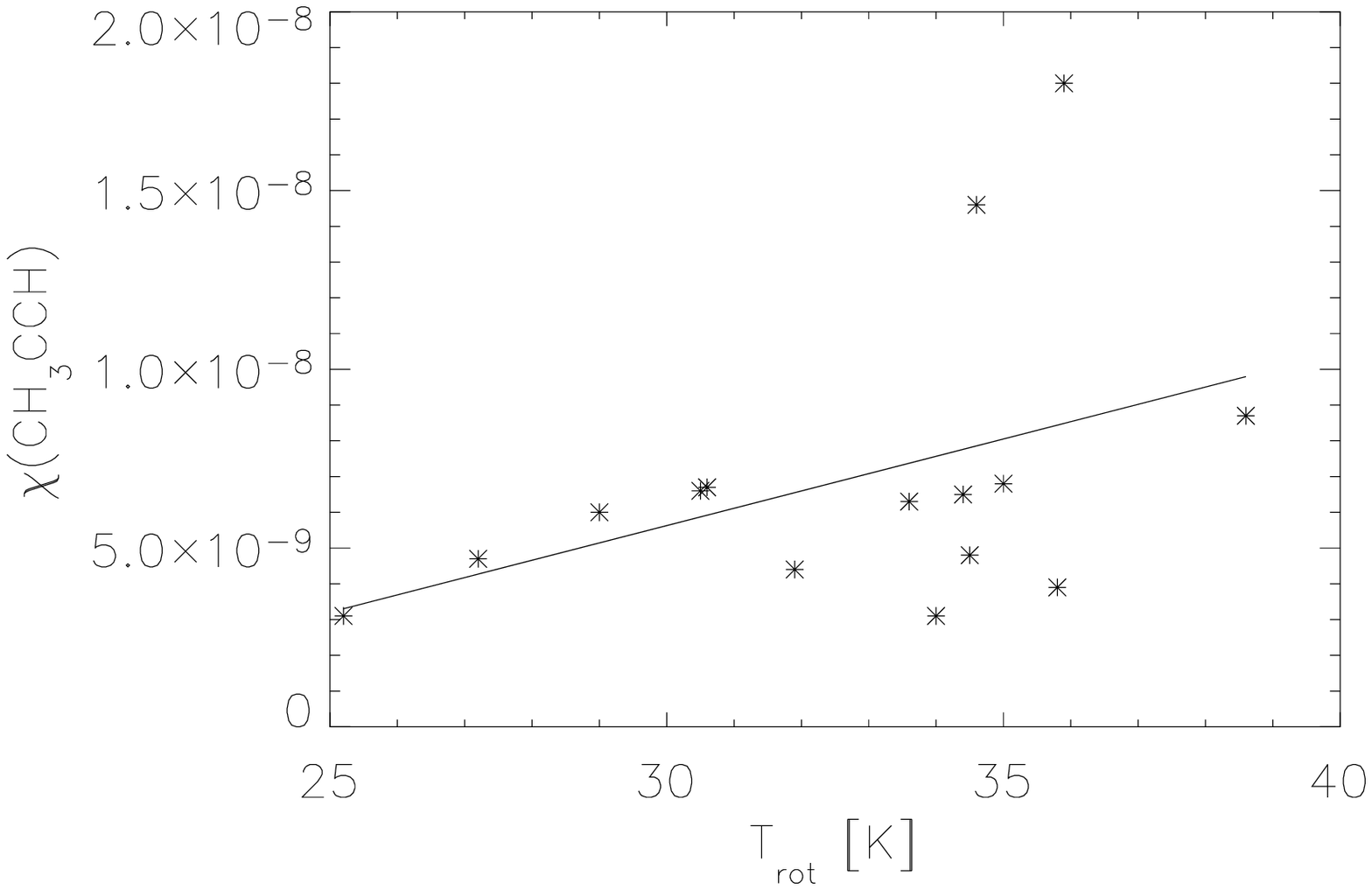}}
\resizebox{\hsize}{!}{\includegraphics{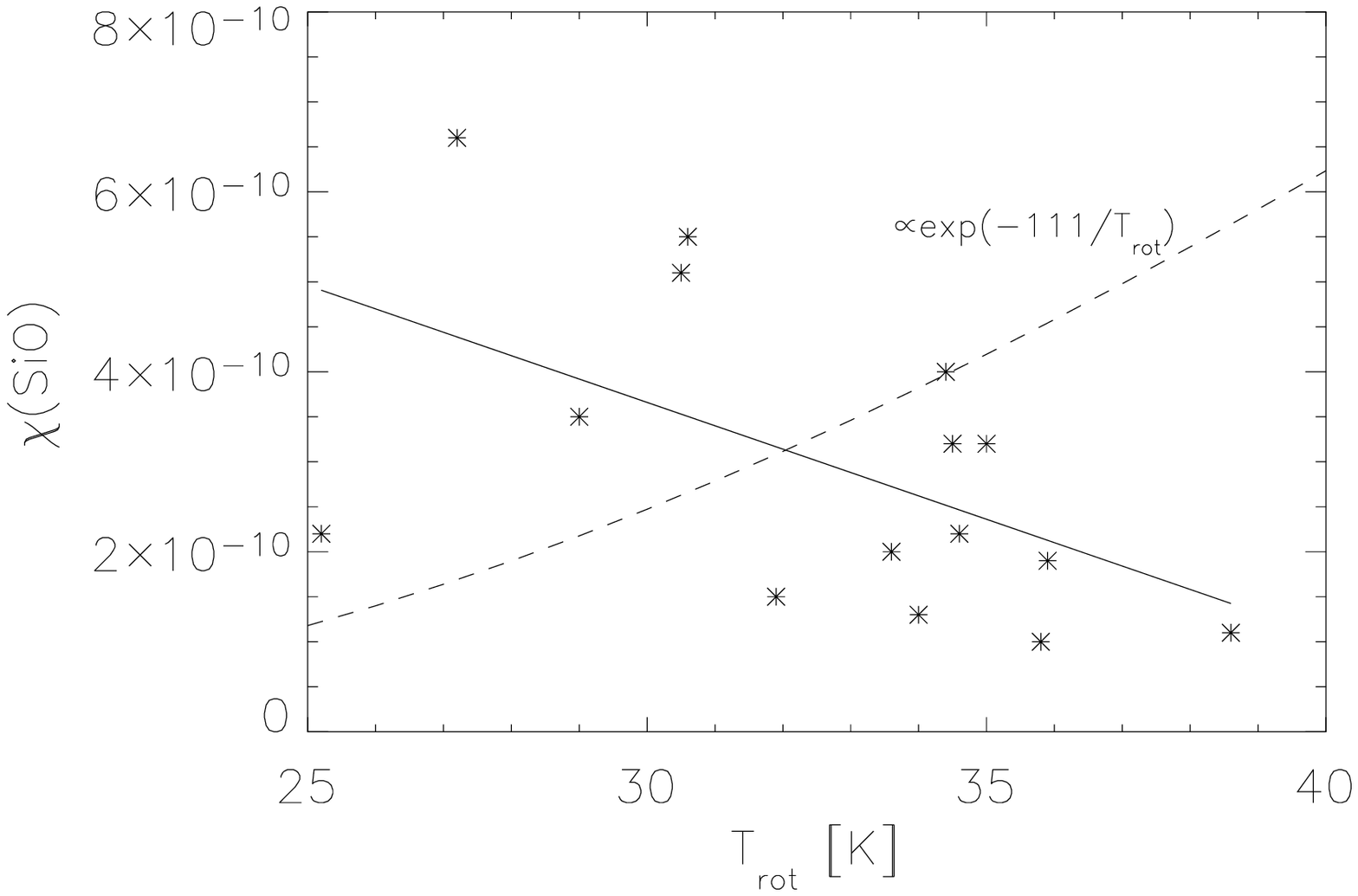}}
\caption{Fractional CH$_3$CCH and SiO abundances as functions
of the rotational temperature, $T_{\rm rot}$, derived from CH$_3$CCH.
The solid lines represent linear least-squares fits to the data.
The dashed line in the bottom panel represents a change
in the fractional SiO abundance proportional to $\exp(-111{\rm K}/T)$
as suggested by \cite{langer1990}.}
\label{figure:correlations}
\end{figure}

It has been suggested that SiO can be produced also in warm, quiescent
gas via neutral or ion-neutral reactions, possibly preceded by
evaporation of Si-bearing molecules from the icy mantles of dust
grains. The suggestion of \cite{ziurys1989} and \cite{langer1990}
that neutral reactions with activation energies are important for the
SiO production in warm gas, would imply a strong correlation between
the SiO abundance and the average kinetic temperature. As shown in
Fig. \ref{figure:correlations}, bottom panel, even in the rather
narrow temperature range covered, $25-39$ K, a substantial change in
the fractional SiO abundance would be expected if it were proportional
to $\exp(-111{\rm K}/T)$ as suggested by \cite{langer1990}. In the
model of MacKay (\cite{mackay1995}; \cite{mackay1996}) icy silicon is
mainly in the form of SiH$_4$, which can evaporate when the dust is
warmed up. Also in this model one could expect the warmer cores have
larger SiO abundances.

The abundances derived for the low-velocity SiO do not
show, however, any positive correlation with the average kinetic
temperature. Instead, they even seem to decrease slightly when the
temperature rises (Fig. \ref{figure:correlations}, bottom panel).

Also the CH$_3$CCH production can be enhanced by intensified
desorption.  CH$_3$CCH is one of the lowest energy products in
reaction between the methylidine radical (CH) and ethylene
(C$_2$H$_4$, \cite{canosa1997}) which is formed on dust grains. The
diagram presented in Fig. \ref{figure:correlations}, top panel,
suggests a positive correlation between the fractional CH$_3$CCH
abundance and $T_{\rm kin}$, which would conform with the desorption
scenario. On the other hand, this tendency does not agree with the
prediction of the pure gas-phase model of \cite{lee1996} (see also
\cite{alakoz2002}), which predict a strong anticorrelation between the
CH$_3$CCH abundance and the kinetic temperature. The CH$_3$CCH
abundances derived here are similar to those found in the Orion
Ridge, M17 and Cepheus A (\cite{ungerechts1997}; \cite{bergin1997}),
and substantially larger than those determined towards cold, starless
cores (\cite{kontinen2000}; \cite{markwick2005}).

These findings lend support to the possibility that indeed
grain-surface reactions are important for the production of CH$_3$CCH
and its gas-phase abundance depends on temperature via the thermal
desorption.  On the other hand, no evidence for such behaviour is
found in the case of SiO. This complies with the shock origin of
SiO. The diminishing tendency towards warmer cores suggested by
Fig.~\ref{figure:correlations} may reflect an evolutionary effect.

According to the picture presented by \cite{fuente2005} based on a
study of the environments of two young stars in the region of the
nebula NGC 7129, protostellar envelopes are dispersed and warmed up
during the early stellar evolution, at the same time as shocks
associated with outflows get less energetic. For SiO this means
decreasing abundance with time. The abundance variation from core to
core is not very large in our sample. This is probably related to the
fact that all the studied objects are associated with powerful masers
and/or UC \ion{H}{ii} regions, and therefore the dynamical
and chemical age variations are likely to be rather limited
(\cite{garay1999}; \cite{bergin1997}).

\subsection{Dynamical states}

Most of the cores studied here appear to be gravitationally bound
which is indicated by the fact that $M_{\rm cont} \ga M_{\rm vir}$.
The $M_{\rm cont}/M_{\rm vir}$ ratios range from 0.7 to 6.8, the
average being 2.5. For five cores $M_{\rm cont}$ is substantially 
larger than $M_{\rm vir}$, whereas the rest appear to be close to the
virial equilibrium ($M_{\rm vir} \sim M_{\rm cont}$, within a factor of two).
The virial masses, $M_{\rm vir}$, were
calculated assuming homogenous density distributions. A more
realistic distribution described by a powerlaw of the
type $n(r) \propto r^{-p}$, where $p$ is
typically $\sim 1.5-2.5$, would make the
virial mass estimates smaller (see, e.g., \cite{fontani2005}).
On the other hand, the effect of the
density gradient can be compensated by the increase of the kinetic
temperature towards the core centre. All the cores studied here
contain newly born massive stars which are known to give rise to
expanding UC \ion{H}{ii} regions and hot cores.
Magnetic fields are likely to provide
additional support against the gravitational collapse (e.g.,
\cite{garay1999}; \cite{fontani2002}).
For example, Sarma et al. (2000) detected magnetic fields toward sources
A, E, and D in the NGC 6334 complex, which are of the same order as
the critical fields required to support the cores against
gravitational collapse.

\subsection{Comments on individual sources}

Ten of our fourteen targets have been included in recently published, 
extensive millimetre continuum surveys with SIMBA. \cite{faundez2004} have
mapped about 150 massive cores around IRAS point sources with FIR
colors typical of UC \ion{H}{ii} regions, and discuss the general
characteristics of these objects. They present the maps
of IRAS 12326-6245, G326.64+0.61 (IRAS 15048-5356), OH328.81+0.63
(IRAS 15520-5234), G330.95-0.19 (IRAS 16060-5146), IRAS 16562-3959,
G345.00-0.23 (IRAS 17016-4124), NGC 6334F (IRAS 17175-3544), and
G351.77-0.54 (IRAS 17233-3606). The survey of \cite{hill2005} covering
about 130 regions include IRAS 12326-6245 (G301.14-0.2), G330.95-0.18,
W28 A2 (G5.89-0.39), and W31 (2) (G10.62-0.38). 

The structure of the sources in Fig. \ref{figure:simbamaps}
agrees well with those presented in \cite{faundez2004} and \cite{hill2005}.
The peak fluxes in \cite{hill2005} for the sources in common with this work
agree also well. The integrated flux densities for the sources are however
higher in \cite{hill2005} and especially in \cite{faundez2004} than
in the present paper. Hill et al. calculated the integrated flux
density inside a polygon containing all the emission around the source
whereas in the present paper the integration was done inside a circle
corresponding to three times the telescope beam width.
Contrary to the present paper \cite{hill2005} also applied the gain-
elevation correction to the SIMBA data. The different integration aperture
and the gain-elevation correction can well explain the Hill et al. (2005)
$\sim30-50 \, \%$ higher flux densities. \cite{faundez2004} give no details on
how the integrated fluxes were calculated so it can not be decided
if the higher values are due to e.g. the integration aperture.

The kinetic temperatures from CH$_3$CCH are lower than
the dust temperatures derived by \cite{faundez2004} from the spectral
energy distributions (SEDs) using the flux densities at 1.2 mm and in
the four IRAS bands. However, as our 1.2 mm flux densities are, in general,
lower than those derived by Fa{\'u}ndez et al.,
we end up with comparable masses.
The masses derived by \cite{hill2005} are on the average larger 
than those derived by us, mainly because they have obtained 
slightly higher flux densities and assumed that $T_{\rm d}=20$ K. 
Finally, the mass estimates are proportional to the distance squared, and 
there are some differences in the derived kinematic distances depending on the 
Galactic rotation curve and $v_{\rm LSR}$ used. 
 
Next we discuss six of our sources, IRAS 12326-6245, NGC 6334,
G351.77-0.54, W28 A2(2), W31 and W33 in more detail.
In general, the sources are typical high-mass star-forming cores associated
with several molecular masers, UC \ion{H}{ii} regions and infrared sources
(see Table \ref{table:sources} and  Fig. \ref{figure:simbamaps}).   

\subsubsection{IRAS 12326-6245}

IRAS 12326-6245 is a luminous FIR source ($L_{\rm bol}=3.8 \cdot 10^5$
L$_{\sun}$) which is located at a kinematically estimated distance
of 4.4 kpc (\cite{zinchenko1995}; \cite{osterloh1997}).
This object is associated with two UC \ion{H}{ii} regions,
both of which are associated with mid-infrared sources (\cite{henning2000}). 
Also several molecular masers including H$_2$O, OH and CH$_3$OH have been
identified at the position of IRAS 12326-6245
(see \cite{henning2000} for references).

The molecular line maps (see \cite{henning2000} and references therein) 
shows that one of the most energetic and massive bipolar molecular outflow
(mass outflow rate $\dot{M} \sim 0.02$ M$_{\sun}$ yr$^{-1}$)
in the southern sky is originated close to IRAS 12326-6245.

The core 1.3 mm flux density and the gas mass obtained by
Henning et al. (12.0 Jy and 2400 M$_{\sun}$, respectively)
are very close to the values in the present study
(12.8 Jy and 2200 M$_{\sun}$, respectively).

\subsubsection{NGC 6334}

The star-forming molecular ridge associated with
the giant \ion{H}{ii} region
NGC 6334 is the best studied of our targets. For recent, comprehensive
molecular line and continuum studies see Kraemer et al. 1999,
Kraemer \& Jackson 1999, Jackson \& Kraemer 1999, Sandell 2000, 
McCutcheon et al. 2000 and Brooks \& Whiteoak 2001. 
These authors summarize the accumulated
knowledge of the region, and include keys of the
nomenclature. Sandell (2000) mapped the northern part of the
molecular ridge with UKT14 bolometer at the JCMT using four millimetre
and submm wavelengths in the range 1.1 -- 0.35 mm.

The six brightest 1.2 mm peaks listed in Table \ref{table:dustmaxima}
can be associated with star-forming cores detected as far-infrared sources
or UC \ion{H}{ii} regions
as follows: 1) FIR-I/UC \ion{H}{ii} F;
2) I(N); 3) FIR-IV/UC \ion{H}{ii} A; 4) FIR-V;
5) FIR-III/UC \ion{H}{ii} C;
6) FIR-II/UC \ion{H}{ii} D (\cite{mcbreen1979};
\cite{rodriguez1982}; \cite{gezari1982}).
The 1.2 mm peak No. 7 is located on the southeastern side
of radio shell between FIR-IV and V, in the region of
the PDR G351.2+0.70 (\cite{jackson1999}).
It is associated with a CO clump in
the survey of \cite{kraemer1999b} (their Fig. 5).
The dust emission peak No. 8 on the northwestern side of
the dense molecular ridge is associated with
the object called 'dust-cloudlet' by \cite{sandell2000}.

Assuming a uniform dust temperature of 30 K, we obtain from the SIMBA
map a mass of 11\,400 M$_{\sun}$ for the entire ridge. The share of its
northern end with the cores I, I(N) and II is 5400 M$_{\sun}$, and for
the regions III, IV, and V we get 900 M$_{\sun}$, 2200 M$_{\sun}$, and
1300 M$_{\sun}$, respectively.

The main features of the 1.1 mm map of \cite{sandell2000} can be
recognized on the present SIMBA map, despite a lower S/N available 
here. A careful inspection of the image reveals 
filaments reaching out from the dense ridge, bearing much likeness to 
the structures associated with OMC-1 discovered by \cite{johstone1999} 
on a JCMT/SCUBA map. The suggestion of \cite{johstone1999} that these features 
could represent the cavity walls of past outflows is appropriate 
also in this case. Filamentary structures are rather common features in 
GMCs forming high mass clusters, so there is no reason to consider them as
artefacs of the data reduction.

\begin{figure}
\resizebox{\hsize}{!}{\includegraphics[angle=270]{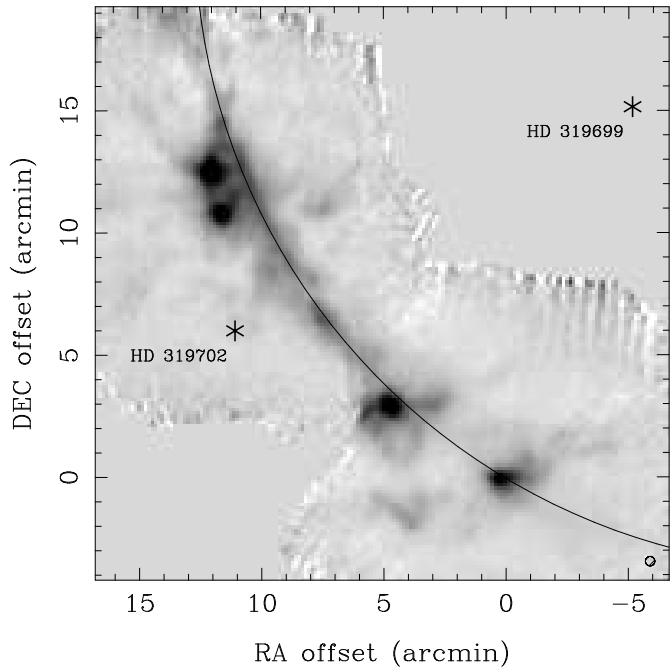}}
\caption{The SIMBA dust continuum map of the star forming ridge 
in the region of NGC 6334. The ridge seems to follow an arc of a 
circle described in the text. The locations of two nearby O-type 
stars are indicated.}
\label{figure:ngc6334_circle}
\end{figure}

The 'linear filament' discussed by \cite{sandell2000} continues to the
north and southwest and in fact the ridge is arc-shaped.  In
Fig. \ref{figure:ngc6334_circle} an arc of a circle of radius
$27\arcmin$ (13 pc) and centred at R.A. $17^{\rm h}18^{\rm m}50\fs0$,
Dec. $-35\degr35\arcmin30\arcsec$ is superposed on the SIMBA map. The
alignment evident in the figure suggests that the ridge has been
formed by an expanding \ion{H}{ii} region with its centre
near the given position. This \ion{H}{ii} region has 
probably been ionized by one or more of
the matured OB stars in the region. The closest O-type star on the
western side of the ridge is HD~319\,699 (O7, distance 1.6 kpc).
The UV radiation from this star is likely to dominate the excitation of the
extended emission nebula in this direction. It seems, however,
that the star has  not been in the centre of the expansion,
because it lies about $11\arcmin$
southeast of the centre of the circle, and its proper motion vector
(although very uncertain) points northwest (The Tycho-2 Catalogue;
\cite{hog2000}). The O-stars lying southeast and south of the ridge,
HD~319\,702 (O9, $d=1.4$ kpc), HD~319\,703 (O6, $d=2.0$ kpc), and
HD~156\,738 (O7, $d=1.3$ kpc) are surrounded by separate, roundish
ionization regions which probably have not contributed to shaping the
ridge (see, e.g., Fig. 5 of
\cite{kraemer1999b} or {\it SkyView}, http://skyview.gsfc.nasa.gov/).

\subsubsection{G351.77-0.54}

G351.77-0.54 is the brightest SiO line source of the sample
(see Fig. \ref{figure:SiO_spectra}).
It is associated with the luminous steep-spectrum far-infared source
IRAS 17233-3606, the strongest known ground-state OH maser source at
1665 MHz, other OH masers, and H$_2$O and CH$_3$OH masers (e.g.,
\cite{caswell1995a}; \cite{macleod1998}; \cite{valtts2000}).
The kinematic distance derived using the LSR velocity from
CH$_3$CCH is 700 pc, which would make this source the nearest
in our sample. The distance is smaller than the usually
adopted value ($\sim 1.5-2.2$ kpc), but it is close to the distance
derived by \cite{macleod1998} (1.0 kpc) from the LSR velocity of the
H$_2$CO $1_{10} - 1_{11}$ absorption. The neighbourhood of the source
direction to that of the Galactic centre causes a large uncertainty to the
kinematic distance estimate. Nevertheless, the proximity is supported by the
intensity of the SiO emission, suggesting a large beam-filling factor.

\subsubsection{W28 A2}

G5.89-0.39 (also known as W28 A2) is one of the best-studied
UC \ion{H}{ii} regions, but probably it does not belong to the
best understood.
W28 A2 is associated with a highly energetic outflow, one of
the most powerful ones in the Galaxy. Actually, in W28 A2 there might be 
multiple outflows and driving sources.

A bipolar outflow in the east-west direction with a mass of
$\sim 70$ M$_{\sun}$ was identified by Harvey \& Forveille (1988)
in the $J=1-0$ lines of CO, $^{13}$CO, and C$^{18}$O toward this source.
C$^{34}$S observations shows a north-south oriented outflow
(\cite{cesaroni1991}). There is growing evidence that the O5 star in G5.89
produced the N-S molecular outflow and hence is forming through
accretion process (see \cite{arce2006} and references therein).
The SiO ($v=0$, $J=1-0$) observations made with the VLA by Acord et al. (1997)
shows a northeast-southwest bipolarity, which is confirmed by
1.3 mm continuum and several molecular line observations by
Sollins et al. (2004). The 1.3 mm continuum source peaks
in the centre of the \ion{H}{ii} region (\cite{sollins2004}).
Moreover, NIR observations of the H$_2$ $v=1-0$ $S(1)$ rovibrational line and
the Br${\gamma}$ emission show evidence of three outflows with
distinct orientations and driving sources (\cite{puga2006}). 
Puga et al. suggest that the northwest-southeast oriented outflow is possibly 
connected to the H$_2$ knots seen in that direction. 
One of them ('knot B'), NW of W28 A2, can be identified in our SIMBA map.
The two other features seen in our map, NW of knot B, at
$\alpha_{2000}=18^{\rm h}00^{\rm m}15^{\rm s}$,
$\delta_{2000}=-24\degr01\arcmin30\arcsec$, and
$\alpha_{2000}=18^{\rm h}00^{\rm m}08^{\rm s}$, 
$\delta_{2000}=-24\degr00\arcmin30\arcsec$,
could be related to NW-SE oriented outflow.  
                           
Thus it is unlikely that there exist a single central high-mass star
in W28 A2 as was initially thought to be the case due to the
symmetric morphology of \ion{H}{ii} region.

The kinematic distance derived from \ion{H}{i} observations towards
the supernova remnant W28 by Vel\'azquez et al. (2002) is 1.9 kpc.
Adopting this value instead of 2.7 kpc derived by us, the mass of the
cloud would decrease from 970M$_{\sun}$  to 480 M$_{\sun}$.

Thompson \& Macdonald (1999) found rotational temperature of $\sim 70$ K
from CH$_3$CCH $J=20-19$ and $J=21-20$ observations and
G\'omez et al. (1991) derived the temperature of $\sim 90$ K from
NH$_3$(2, 2) and (3, 3) inversion transitions for the molecular envelope
surrounding the UC \ion{H}{ii} region. 
We have observed much lower CH$_3$CCH transition $J=5-4$, which trace
the gas in outer layers and thus giving lower rotational temperature (34.6 K).
The CH$_3$CCH abundance estimate of $2.4 \cdot10^{-7}$ by Thompson \&
Macdonald is an order of magnitude higher than our value of $1.5\cdot10^{-8}$.
NH$_3$(3, 3), (4, 4), and (5, 5) observations by Acord et al. (1997) imply
the kinetic temperature of around 100 K for the outflowing gas and
they found the SiO abundance of $1-3 \cdot 10^{-9}$ in this gas,
which is also an order of magnitude higher than the abundance estimate
in the present study in the position of W28 A2 (2). These results support
the idea that the CH$_3$CCH and SiO abundances are are clearly larger
in hot core than in the warm envelope.

\subsubsection{W31}

W31 is one of the largest \ion{H}{ii} complexes in our Galaxy. 
G10.6-0.4 lying in the centre of our SIMBA map is one of the main extended 
\ion{H}{ii} regions in this complex along with G10.2-0.3 and G10.3-0.1 
(see, e.g., \cite{corbel2004} and references therein).
NH$_3$(1, 1) and (3, 3) observations show that the molecular core
in G10.6-0.4 is rotating and collapsing inward toward the UC \ion{H}{ii}
region where molecular masers are seen distributed in the following way:
H$_2$O and CH$_3$OH masers linearly in the plane of rotation, and OH masers 
along the rotation axis (see, e.g., \cite{sollins2005} and references therein).
Further evidence for collapse comes from two-peaked profile of CO(4-3) line 
observed with the APEX (\cite{wyrowski2006}).
Observations made with the VLA by Sollins \& Ho (2005) suggest that there is
a rotating molecular accretion flow which flattens but does not form
an accretion disk.
Velocities of molecular lines observed from the surrounding neutral gas in
G10.6-0.4 suggest that accretion flow passes through the \ion{H}{ii} region
boundary and continues inward as an ionized flow (\cite{keto2002}).
The associated infrared source (see Fig. \ref{figure:simbamaps}),
IRAS 18075-1956, has a luminosity of $9.2\cdot10^5$ L$_{\sun}$
(\cite{casoli1986}). 

The distance of W31 is not well known. However, spectral types of O
stars identified in W31 and the NIR photometry made by Blum et al. (2001) show
that the distance to W31 must be $\la 3.7$ kpc. Corbel and Eikenberry (2004)
suggest that G10.6-0.4 is located on the $-30$ km s$^{-1}$ spiral arm at a
distance from the Sun of $4.5\pm0.6$ kpc, which is consistent with
the spectrophotometric distance obtained by Blum et al.  
We have used the value of 3.1 kpc which was derived by Blum et al. assuming
that O star spectra are consistent with ZAMS. Note that Sollins \& Ho (2005)
and Sollins et al. (2005) adopt the kinematic distance of 6.0 kpc from
\cite{downes1980}.

\subsubsection{W33}

The W33 complex is heavily obscured in the visual. It contains
a compact radio core G12.8-0.2. The FIR sources W33 A (G12.91-0.26)
and W33 B (G12.70-0.17) associated with OH/H$_2$O masers lie on the
opposite sides of G12.8-0.2 (\cite{capps1978}; \cite{stier1984}).

IR and radio continuum observations have revealed a cluster of compact sources
in the region of G12.8-0.2 (\cite{dyck1977}; \cite{haschick1983}).
The total IR luminosity of G12.81-0.19 assuming a distance of 3.7 kpc is
$1.4 \cdot 10^6$ L$_{\sun}$ (\cite{stier1984}). According to the results
of Stier et al. the minimum mass of the star associated with
the radio core is 20 M$_{\sun}$. G12.8-0.2 is considered as a good
candidate for a very young OB cluster (see, e.g., \cite{beck1998}).

\section{Summary and conclusions}

Spectral line observations at 3 and 2 mm, and dust continuum
observations at 1.2 mm performed with the SEST were used to derive
physical characteristics and fractional SiO and CH$_3$CCH abundances in
15 high-mass star-forming cores. The sample represents typical GMC
cores with the following average properties $<T_{\rm kin}> = 33$ K,
$<n(\rm H_2)> = 4.5 \cdot 10^6 \; {\rm cm}^{-3}$,
and $<M> = 1800$ $ \rm M_{\sun}$.
While most cores seem to be near the virial equilibrium,
for five of them the gravitational potential energy
is large compared with the kinetic energy. 

The fractional SiO abundances determined from the velocity range with
detectable CH$_3$CCH emission were found to be $\sim 1-7 \cdot 10^{-10}$. 
CH$_3$CCH abundances were found to be $3-18 \cdot 10^{-9}$. 
In consistence with earlier results towards GMC cores
(e.g., \cite{ungerechts1997}; \cite{bergin1997}) 
the CH$_3$CCH abundance is found to be clearly larger than in the dense cores
of dark clouds. It also shows a slight increase as
a function of $T_{\rm kin}$.
As this tendency does not agree with the predictions of gas-phase chemistry
models (e.g. \cite{lee1996}), a possible explanation is that warm
conditions lead to an intensified desorption of the precursor
molecules of CH$_3$CCH from the icy mantles of dust grains. This suggestion 
is supported by the chemistry model of \cite{canosa1997}.

The SiO abundances are midway between the upper limit from dark clouds
and PDRs ($\sim 10^{-12}$) and the values derived for powerful shocks
($\sim 10^{-8}$). The abundance seem to decrease with
rising temperature. This finding contradicts with the models where
SiO production is dominated by neutral reactions with activation energies
(\cite{ziurys1989}; \cite{langer1990}), and with models where
desorption of Si-containing species from grain mantles are significant
for the SiO production (\cite{mackay1996}). As suggested by 
the recent observational results of \cite{fuente2005}, warmer cores
represent more evolved objects where highly
energetic protostellar outlows releasing SiO into the gas phase are less
frequent; this together with the rapid post-shock processing
have lead to a diminished SiO abundance.

\begin{acknowledgements}
We thank the referee for a careful and critical 
reading of the manuscript and for very helpful comments and suggestions.
The work has been supported by the Academy of Finland through grants
Nos. 1201269 and 1210518. We acknowledge the use of the SIMBAD database,
operated at CDS, Strasbourg, France, and the NASA's SkyView facility
(http://skyview.gsfc.nasa.gov) located at NASA Goddard Space Flight Center.
\end{acknowledgements}

\clearpage

\begin{table}
\caption{The rest frequencies and the upper level energies of
the observed lines.}
\centering
\label{table:lines}
\normalsize
\begin{tabular}{c c c c}
\hline\hline 
Molecule & Transition & $\nu$ [MHz] & $E_{\rm u}/k$ [K]  \\
\hline
$^{29}$SiO & $J=2-1$       &  85759.0000 &  6.195  \\
$^{28}$SiO & $J=2-1$       &  86846.9600 &  6.252  \\
$^{29}$SiO & $J=3-2$       & 128636.7064 & 12.390  \\
$^{28}$SiO & $J=3-2$       & 130268.6100 & 12.504  \\
 CH$_3$CCH & $J_K=5_0-4_0$ &  85457.2720 & 12.304  \\
 CH$_3$CCH & $J_K=5_1-4_1$ &  85455.6220 & 19.505  \\
 CH$_3$CCH & $J_K=5_2-4_2$ &  85450.7300 & 41.107  \\
 CH$_3$CCH & $J_K=5_3-4_3$ &  85442.5280 & 77.112  \\
 CH$_3$CCH & $J_K=6_0-5_0$ & 102547.9842 & 17.225  \\
 CH$_3$CCH & $J_K=6_1-5_1$ & 102546.0241 & 24.426  \\
 CH$_3$CCH & $J_K=6_2-5_2$ & 102540.1447 & 46.029  \\
 CH$_3$CCH & $J_K=6_3-5_3$ & 102530.3487 & 82.033  \\
\hline 
\end{tabular} 
\end{table}

\begin{table*}
\caption{Source list. J2000.0 equatorial coordinates ($\alpha$,
$\delta$), galactic coordinates ($l$, $b$), distance ($d$),
galactocentric distance ($R_{\rm GC}$), LSR velocity of
the CH$_3$CCH line ($v_{\rm LSR}$) and notes on associated
masers and UC \ion{H}{ii} regions.}
\centering
\label{table:sources}
\footnotesize
\begin{tabular}{l c c r c c r l}
\hline\hline 
Source & $\alpha_{2000.0}$ & $\delta_{2000.0}$
  & \multicolumn{1}{c}{$l$, $b$} & $d$ & $R_{\rm GC}$ & $v_{\rm LSR}$ 
  & notes  \\
 & [h:m:s] & [$\degr$:$\arcmin$:$\arcsec$] & 
   \multicolumn{1}{c}{[$\degr$]} & [kpc] & [kpc] & [km s$^{-1}$] & \\
\hline
IRAS 12326-6245 & 12:35:34.1 & -63:02:28 & 301.13-0.22 & 4.4$^a$ & 7.1 & -39.5 & OH$^A$/H$_2$O$^B$/CH$_3$OH$^C$/UC \ion{H}{ii}$^{D,E}$\\
G326.64+0.61 & 15:44:32.3 & -54:05:54 & 326.64+0.61 & 2.7 & 6.4 & -39.8 & H$_2$O$^F$\\ 
OH328.81+0.63 & 15:55:47.3 & -52:43:08 & 328.81+0.63 & 2.9 & 6.2 & -41.7 & OH$^F$/CH$_3$OH$^G$/UC \ion{H}{ii}$^H$\\
IRAS 15566-5304 & 16:00:30.7 & -53:12:34 & 329.03-0.20 & 3.0 & 6.2 & -43.6 & OH$^I$/H$_2$O$^J$/CH$_3$OH$^G$\\
G330.95-0.19 & 16:09:52.2 & -51:55:20 & 330.95-0.19 & 6.1 & 4.3 & -100.0 & OH$^H$/H$_2$O$^F$\\
G345.01+1.8N & 16:56:47.2 & -40:14:09 & 345.01+1.80 & 1.7 & 6.9 & -13.8 & OH$^K$/H$_2$O$^F$/CH$_3$OH$^G$/UC \ion{H}{ii}$^H$\\
IRAS 16562-3959 & 16:59:41.9 & -40:03:42 & 345.49+1.47 & 1.6 & 7.0 & -12.2 & OH$^H$/UC \ion{H}{ii}$^D$\\ 
G345.00-0.23 & 17:05:11.7 & -41:29:11 & 345.00-0.23 & 2.9 & 5.7 & -27.7 & OH$^K$/H$_2$O$^F$/CH$_3$OH$^G$\\
NGC 6334 FIR-V & 17:19:55.9 & -35:57:45 & 351.16+0.70 & 1.7$^b$ & 7.2 & -6.6 & OH$^K$/H$_2$O$^F$\\
NGC 6334F & 17:20:53.5 & -35:47:01 & 351.42+0.65 & 1.7$^b$ & 7.1 & -7.0 & OH$^L$/H$_2$O$^F$/CH$_3$OH$^G$/UC \ion{H}{ii} $^H$\\ 
G351.77-0.54 & 17:26:42.6 & -36:09:17 & 351.77-0.54 & 0.7 & 7.8 & -3.2 & OH$^K$/H$_2$O$^F$/CH$_3$OH$^G$/UC \ion{H}{ii}$^H$\\ 
G353.41-0.36 & 17:30:26.0 & -34:41:57 & 353.41-0.36 & 3.5 & 5.1 & -16.8 & OH$^{M,K}$/H$_2$O$^F$/CH$_3$OH$^F$/UC \ion{H}{ii}$^H$\\ 
W28 A2(2) & 18:00:30.4 & -24:04:00 & 5.88-0.39 & 2.7 & 2.4 & 6.1 & OH$^H$/H$_2$O$^N$/UC \ion{H}{ii}$^H$\\ 
W31 (2) & 18:10:28.7 & -19:55:50 & 10.62-0.38 & 3.1$^c$ & 5.5 & -2.8 & OH$^H$/H$_2$O$^N$/UC \ion{H}{ii}$^H$\\ 
W33 CONT & 18:14:13.6 & -17:55:25 & 12.81-0.20 & 3.7 & 4.9 & 34.8 & H$_2$O$^N$\\ 
\hline 
\end{tabular} 

\vspace{0.5cm} \footnotesize \textbf{References:} $^A$ \cite{cohen1988}; $^B$ \cite{caswell1989};
$^C$ \cite{caswell1995}; $^D$ \cite{bronfman1996}; $^E$ \cite{zinchenko1995}; 
$^F$ The catalog of non-stellar H$_2$O/OH masers (\cite{braz1983});
$^G$ \cite{caswell2000}; $^H$ \cite{caswell1998}; $^I$ \cite{caswell1995};
$^J$ \cite{scalise1989}; $^K$ \cite{caswell1983}; 
$^L$ \cite{moran1980}; \cite{gaume1987}; $^M$ \cite{caswell1981};
$^N$ Arcetri Atlas of galactic H$_2$O masers (\cite{comoretto1990}; \cite{palagi1993};
\cite{brand1994}). \textbf{Distance references:} $^a$ \cite{zinchenko1995}; \cite{osterloh1997};
$^b$ \cite{neckel1978}; $^c$ \cite{blum2001}. When no reference is given, the distance is a kinematic distance
calculated from the CH$_3$CCH velocities.
\end{table*}

\begin{table}
\caption{Dust emission maxima indentified on the original maps
($24^{\prime\prime}$ beam).}  \centering
\label{table:dustmaxima}
\small
\begin{tabular}{c c c c c c}
\hline\hline 
Source & peak No. & $\alpha_{2000.0}$ & $\delta_{2000.0}$ & $I_{\nu}$ [MJy/sterad]\\
\hline
IRAS 12326-6245 & & 12:35:34.7 &-63:02:40 & 546\\
G326.64+0.61    &1& 15:44:43.2 &-54:05:46 & 226\\
                &2& 15:44:33.2 &-54:05:22 & 201\\
                &3& 15:44:59.5 &-54:02:26 & 197\\
                &4& 15:44:56.9 &-54:07:14 & 101\\
                &5& 15:45:01.4 &-54:09:14 &  63\\
OH328.81+0.63   & & 15:55:48.6 &-52:43:04 & 419\\
IRAS 15566-5304 & & 16:00:31.1 &-53:12:38 & 166\\
G330.95-0.19    &1& 16:09:53.1 &-51:54:52 & 995\\
                &2& 16:10:17.3 &-51:58:44 &  36\\
G345.01+1.8N    &1& 16:56:47.2 &-40:14:21 & 303\\
                &2& 16:56:40.9 &-40:13:17 & 151\\
                &3& 16:56:43.7 &-40:15:57 &  65\\ 
                &4& 16:56:52.1 &-40:17:01 &  44\\
IRAS 16562-3959 &1& 16:59:41.7 &-40:03:34 & 401\\
                &2& 16:59:28.5 &-40:10:06 & 131\\
                &3& 16:59:06.9 &-40:05:50 &  49\\
G345.00-0.23    & & 17:05:11.0 &-41:28:59 & 372\\
NGC 6334        &1& 17:20:53.4 &-35:46:58 & 1150\\
                &2& 17:20:55.4 &-35:45:06 & 850\\
                &3& 17:20:19.3 &-35:54:51 & 519\\
                &4& 17:19:57.5 &-35:57:47 & 447\\
                &5& 17:20:32.4 &-35:51:23 & 147\\
                &6& 17:20:42.9 &-35:49:14 &  81\\
                &7& 17:20:15.3 &-35:59:39 &  61\\
                &8& 17:20:33.0 &-35:46:51 &  59\\
G351.77-0.54    &1& 17:26:42.9 &-36:09:17 &1007\\
                &2& 17:26:39.0 &-36:08:05 & 114\\
                &3& 17:26:47.6 &-36:12:05 &  82\\
                &4& 17:26:25.8 &-36:04:45 &  55\\
G353.41-0.36    & & 17:30:26.7 &-34:41:45 & 487\\
W28 A2(2)       &1& 18:00:30.4 &-24:04:00 & 626\\
                &2& 18:00:40.9 &-24:04:16 & 205\\
                &3& 18:00:15.2 &-24:01:20 &  40\\
W31 (2)         &1& 18:10:28.7 &-19:55:50 & 690\\
                &2& 18:10:18.5 &-19:54:22 &  60\\
W33 CONT        &1& 18:14:13.9 &-17:55:41 &1071\\
                &2& 18:14:39.7 &-17:52:05 & 172\\
                &3& 18:13:54.8 &-18:01:49 & 119\\
                &4& 18:14:36.3 &-17:55:01 &  57\\
                &5& 18:14:25.7 &-17:53:57 &  53\\
                &6& 18:14:07.7 &-18:00:37 &  48\\
\hline 
\end{tabular} 
\end{table}

\begin{table}
\caption{Intensities in selected positions on the maps smoothed to 
$57\arcsec$, the total 1.2-mm flux density and the angular 
size (FWHM) of the source determined from the original maps ($24\arcsec \;$beam).}
\centering
\label{table:smoothed_intensities}
\normalsize
\begin{tabular}{c c c c}
\hline\hline 
Source & $I_{\nu}$ [MJy/sterad]& 
$S_{1.2 \; \rm mm}$ [Jy] & $\Theta_{\rm s}$ [$\arcsec$]\\
\hline
IRAS 12326-6245 &  134 & 12.8 & 13\\
G326.64+0.61    &   98 & 10.9 & 27\\
OH328.81+0.63   &  136 & 14.6 & 22\\
IRAS 15566-5304 &   74 &  7.8 & 34\\
G330.95-0.19    &  154 & 22.8 & 13\\
G345.01+1.8N    &  154 & 16.4 & 47\\
IRAS 16562-3959 &  185 & 19.1 & 36\\
G345.00-0.23    &  110 & 11.4 & 19\\
NGC 6334 FIR-V  &  189 & 19.4 & 39\\
NGC 6334F       &  360 & 34.7 & 16\\
G351.77-0.54    &  289 & 27.5 & 17\\
G353.41-0.36    &  197 & 22.6 & 33\\
W28 A2(2)       &  163 & 15.5 & 15\\
W31 (2)         &  198 & 19.1 & 89\\
W33 CONT        &  395 & 41.1 & 32\\
\hline 
\end{tabular} 
\end{table}

\begin{table}
\caption{Optical thicknesses of the $^{28}$SiO($2-1$) and ($3-2$)
lines and their excitation temperatures. In case the optical thickness ratios
have been determined in several velocity channels (column 5),
the tabulated values are the averages over these channels.}
\label{table:sio_tex}
\normalsize
\begin{tabular}{ccccc}
\hline\hline
Source & $\tau_{2\rightarrow1}$  & $\tau_{3\rightarrow2}$ & $T_{\rm ex}$ [K] & 
$N_{\rm chan}$ \\ \hline

G326.64+0.61    & 3.0 $\pm$ 0.4 & 2.5 $\pm$ 0.7 & 5.2 $\pm$ 1.7 & 2\\
OH 328.81+0.63  & 2.1 $\pm$ 0.2 & 2.1 $\pm$ 0.2 & 4.8 $\pm$ 0.4 & 10\\
IRAS 15566-5304 & 3.8 $\pm$ 0.4 & 2.6 $\pm$ 0.6 & 4.7 $\pm$ 0.6 & 5\\
G345.01+1.8N    & 1.5 $\pm$ 0.3 & 1.3 $\pm$ 0.4 & 5.3 $\pm$ 2.7 & 1\\
G345.00-0.23    & 2.2 $\pm$ 0.3 & 1.5 $\pm$ 0.3 & 4.6 $\pm$ 0.7 & 6\\ 
NGC 6334 FIR-V  & 1.7 $\pm$ 0.4 & 1.1 $\pm$ 0.3 & 4.1 $\pm$ 1.6 & 1\\
NGC 6334F       & 2.3 $\pm$ 0.3 & 1.2 $\pm$ 0.4 & 3.4 $\pm$ 0.7 & 3\\
G351.78-0.54    & 1.9 $\pm$ 0.2 & 1.7 $\pm$ 0.1 & 4.2 $\pm$ 0.2 & 13\\
G353.41-0.36    & 2.2 $\pm$ 0.2 & 2.0 $\pm$ 0.3 & 6.1 $\pm$ 1.1 & 5\\
W28 A2(2)       & 2.4 $\pm$ 0.6 & 2.0 $\pm$ 0.5 & 5.2 $\pm$ 1.7 & 3\\
W31 (2)         & 2.9 $\pm$ 0.4 & 2.9 $\pm$ 0.4 & 4.0 $\pm$ 0.4 & 6\\ 
W33 CONT        & 2.7 $\pm$ 0.7 & 3.0 $\pm$ 0.6 & 7.2 $\pm$ 5.1 & 1\\ \hline
\end{tabular}         
\end{table}

\begin{table*}
\caption{$^{29}$SiO integrated intensities for the line centres, i.e., for the
velocity intervals with detectable CH$_3$CCH emission,
and for the whole line with wings.} 
\centering
\label{table:29SiO_line_parameters}
\small
\begin{tabular}{l c c c c c c c c c}
\hline\hline 
   ~   &  ~   & \underline{{\bf Line centre}} &~&~&~& \underline{{\bf Whole line}} &~&~&\\  
Source & Line & $v_{\rm min}$ & $v_{\rm
max}$ & $v_{\rm peak}$ & $\int T_{\rm
A}^{*}(v){\rm d}v \mid_{\rm centre}$ & $v_{\rm min}$
 & $v_{\rm max}$  & $\int T_{\rm A}^*(v){\rm d}v
\mid_{\rm whole \; line}$ \\
 &  & [km s$^{-1}$] & [km s$^{-1}$] & [km s$^{-1}$] & 
[K$\cdot$km s$^{-1}$]& [km s$^{-1}$] & [km s$^{-1}$] & 
[K$\cdot$km s$^{-1}$]\\

\hline
IRAS 12326-6245 & $J=2-1$ & -41.8 & -37.2 & -38.4 & 0.10(0.01)& -49.3 & -35.0 & 0.20(0.02)\\
                & $J=3-2$ &       &       & -40.0 & 0.23(0.02)& -48.4 & -32.0 & 0.40(0.03)\\
G326.64+0.61    & $J=2-1$ & -42.2 & -37.7 & -40.7 & 0.26(0.01)& -46.1 & -33.6 & 0.47(0.02)\\
                & $J=3-2$ &       &       & -41.0 & 0.17(0.02)& -45.5 & -33.6 & 0.31(0.03)\\
OH328.81+0.63   & $J=2-1$ & -44.6 & -38.3 & -41.9 & 0.51(0.01)& -55.3 & -24.9 & 0.99(0.03)\\
                & $J=3-2$ &       &       & -42.5 & 0.60(0.02)& -49.6 & -31.4 & 1.07(0.03)\\
IRAS 15566-5304 & $J=2-1$ & -46.2 & -41.0 & -45.8 & 0.41(0.01)& -53.1 & -33.4 & 0.85(0.03)\\
                & $J=3-2$ &       &       & -44.6 & 0.26(0.02)& -50.1 & -36.7 & 0.50(0.03)\\                
G330.95-0.19    & $J=2-1$ & -92.7 & -88.7 & -91.6 & 0.16(0.01)& -99.6 & -80.5 & 0.37(0.03)\\
                & $J=3-2$ &       &       & -88.4 & 0.13(0.02)& -97.5 & -85.0 & 0.20(0.03)\\
G345.01+1.8N    & $J=2-1$ & -16.8 & -10.6 & -16.4 & 0.31(0.02)& -23.5 & -6.5  & 0.50(0.03)\\
                & $J=3-2$ &       &       & -17.3 & 0.24(0.02)& -21.1 & -7.4  & 0.39(0.03)\\
IRAS 16562-3959 & $J=2-1$ & -14.9 &  -8.7 & -12.2 & 0.21(0.02)& -19.1 & -0.7  & 0.35(0.03)\\
                & $J=3-2$ &       &       & -11.3 & 0.27(0.02)& -21.8 & -2.4  & 0.62(0.03)\\
G345.00-0.23    & $J=2-1$ & -29.7 & -22.4 & -28.2 & 0.43(0.02)& -40.4 & -13.9 & 0.88(0.03)\\
                & $J=3-2$ &       &       & -26.9 & 0.48(0.02)& -40.7 & -11.2 & 1.09(0.04)\\
NGC 6334 FIR-V  & $J=2-1$ &  -9.3 &  -2.9 &  -6.7 & 0.23(0.02)& -20.7 & -3.7  & 0.51(0.03)\\
                & $J=3-2$ &       &       &  -7.0 & 0.29(0.02)& -17.7 & 1.4   & 0.56(0.03)\\
NGC 6334F       & $J=2-1$ &  -8.5 &  -2.7 &  -8.8 & 0.22(0.01)& -11.1 & -2.5  & 0.34(0.01)\\
                & $J=3-2$ &       &       & -10.1 & 0.20(0.01)& -13.8 & -0.1  & 0.37(0.02)\\
G351.77-0.54    & $J=2-1$ &  -5.1 &   0.8 &  -3.9 & 0.72(0.01)& -19.0 & 11.0  & 1.85(0.03)\\
                & $J=3-2$ &       &       &  -3.7 & 0.91(0.02)& -17.8 & 11.0  & 2.29(0.03)\\
G353.41-0.36    & $J=2-1$ & -19.7 & -13.6 & -16.7 & 0.39(0.01)& -24.7 & -6.2  & 0.66(0.02)\\
                & $J=3-2$ &       &       & -17.1 & 0.30(0.01)& -21.7 & -9.2  & 0.42(0.02)\\
W28 A2(2)       & $J=2-1$ &   6.5 &  12.8 &  10.2 & 0.23(0.02)& -2.5  & 26.6  & 0.69(0.04)\\
                & $J=3-2$ &       &       &  11.9 & 0.35(0.02)& 0.1   & 37.9  & 1.34(0.05)\\
W31 (2)         & $J=2-1$ &  -5.2 &   1.6 &  -4.5 & 0.39(0.02)& -12.8 & 7.2   & 0.68(0.03)\\
                & $J=3-2$ &       &       &  -2.1 & 0.41(0.02)& -12.5 & 5.7   & 0.67(0.03)\\
W33 CONT        & $J=2-1$ &  31.9 &  39.2 &  35.0 & 0.24(0.02)& 28.9  & 48.0  & 0.33(0.03)\\
                & $J=3-2$ &       &       &  35.4 & 0.28(0.02)& 27.1  & 47.7  & 0.48(0.03)\\
\hline 
\end{tabular} 
\end{table*}

\begin{table*}
\caption{CH$_3$CCH line parameters obtained from Gaussian fits.}
\centering
\label{table:CH3CCH_line_parameters}
\small
\begin{tabular}{lcccccccc}
\hline \hline
Source & $v_{\rm LSR}$ [km s$^{-1}$] & $\Delta v_{1/2}$ [km s$^{-1}$] & ~ & $\int T_{\rm A}^{*}(v){\rm d}v$ [K$\cdot$km s$^{-1}$]\\
~ & ~ & ~ & K=0 & K=1 & K=2 & K=3\\
\hline
IRAS 12326-6245 & -39.54(0.04) & 3.68(0.02) & 0.69(0.02)&0.58(0.02)&0.29(0.02)&0.13(0.02)\\
G326.64+0.61 & -39.76(0.02) & 3.35(0.02) & 1.41(0.02)&1.10(0.02)&0.47(0.02)&0.19(0.02)\\
OH328.81+0.63 & -41.72(0.01) & 3.07(0.02) & 1.89(0.02)&1.57(0.02)&0.63(0.01)&0.30(0.01)\\
IRAS 15566-5304 & -43.64(0.03) & 3.73(0.03) & 0.95(0.03)&0.77(0.03)&0.31(0.03)&0.11(0.02)\\
G330.95-0.19 & -90.95(0.05) & 5.42(0.06) & 1.31(0.02)&1.00(0.02)&0.54(0.02)&0.21(0.02)\\
G345.01+1.8N & -13.78(0.02) & 3.60(0.02) & 1.23(0.02)&0.96(0.02)&0.48(0.01)&0.24(0.01)\\
IRAS 16562-3959 & -12.17(0.01) & 3.65(0.01) & 4.92(0.02)&4.16(0.02)&1.92(0.02)&1.06(0.001)\\
G345.00-0.23 & -27.69(0.04) & 5.56(0.05) & 1.53(0.02)&1.36(0.02)&0.46(0.02)&0.26(0.02)\\
NGC 6334 FIR-V & -6.63(0.01) & 3.45(0.01) & 2.08(0.02)&1.61(0.02)&0.76(0.02)&0.39(0.02)\\ 
NGC 6334F & -7.00(0.03) & 5.08(0.03) & 2.17(0.02)&1.65(0.02)&0.83(0.02)&0.45(0.02)\\
G351.77-0.54 & -3.16(0.02) & 4.88(0.02) & 3.10(0.03)&2.56(0.02)&1.13(0.02)&0.62(0.02)\\
G353.41-0.36 & -16.78(0.02) & 4.02(0.02) & 1.89(0.02)&1.57(0.02)&0.58(0.02)&0.19(0.02)\\
W28 A2(2) & 8.88(0.01) & 3.65(0.01) & 3.73(0.02)&3.21(0.02)&1.51(0.02)&0.76(0.02)\\
W31 (2) & -2.78(0.03) & 5.00(0.03) & 2.08(0.02)&1.76(0.02)&0.81(0.02)&0.43(0.02)\\
W33 CONT & 34.80(0.01) & 3.79(0.01) & 4.20(0.02)&3.77(0.02)&1.77(0.02)&1.05(0.02)\\
\hline
\end{tabular}
\end{table*}

\begin{table*}
\caption{Fractional SiO and CH$_3$CCH abundances. The SiO column
densities derived from the $^{29}$SiO($2-1$) line centres and the
CH$_3$CCH column densities derived from the $J=5_K-4_K$ lines are used for
these estimates. The H$_2$ column densities are derived from the 1.2 mm
dust emission intensities assuming that $T_{\rm dust} = T_{\rm rot}$,
where $T_{\rm rot}$ is derived from the CH$_3$CCH($5_K - 4_K$) lines.}
\centering
\label{table:abundances}
\footnotesize
\begin{tabular}{l r r r r r r}
\hline\hline
Source & $T_{\rm rot}$ & $N({\rm H_2})$ & $N({\rm CH_3CCH})$ & 
         $\chi({\rm CH_3CCH})$ & $N({\rm SiO})$ & $\chi({\rm SiO})$   \\
       & [K] & [$10^{22}\, {\rm cm}^{-2}$] & [$10^{14}\, {\rm cm}^{-2}$] &
          [$10^{-9}$] & [$10^{13}\, {\rm cm}^{-2}$] & [$10^{-10}$]\\
\hline
IRAS 12326-6245$^a$ &  34.0&  6.4&    2.0&  3.1&  0.8&  1.3\\
G326.64+0.61    &  29.0&  5.7&    3.4&  6.0&  2.0&  3.5\\
OH328.81+0.63   &  30.5&  7.4&    4.9&  6.6&  3.8&  5.1\\
IRAS 15566-5304 &  27.2&  4.7&    2.2&  4.7&  3.1&  6.6\\
G330.95-0.19    &  31.9&  8.0&    3.5&  4.4&  1.2&  1.5\\
G345.01+1.8N    &  34.5&  7.3&    3.5&  4.8&  2.3&  3.2\\
IRAS 16562-3959 &  35.9&  8.3&   15.0& 18.0&  1.6&  1.9\\
G345.00-0.23    &  30.6&  6.0&    4.0&  6.7&  3.3&  5.5\\
NGC 6334 FIR-V       &  33.6&  9.2&    5.8&  6.3&  1.8&  2.0\\
NGC 6334F       &  35.8& 16.3&    6.3&  3.9&  1.6&  1.0\\
G351.77-0.54    &  34.4& 13.7&    8.9&  6.5&  5.5&  4.0\\
G353.41-0.36    &  25.2& 13.7&    4.2&  3.1&  3.0&  2.2\\
W28 A2(2)       &  34.6&  7.7&   11.2& 14.6&  1.7&  2.2\\
W31 (2)         &  35.0&  9.2&    6.2&  6.8&  2.9&  3.2\\
W33 CONT        &  38.6& 16.4&   14.3&  8.7&  1.8&  1.1\\ 
\hline
\end{tabular}

$^a$The rotational temperature and CH$_3$CCH column density derived from
the $J=6_K-5_K$ line are 37.2 K and $2.5 \cdot 10^{14} \, {\rm cm}^{-2}$,
respectively.

\end{table*}

\begin{table*}
\caption{Linear radii, core masses
estimated from 1.2 mm continuum ($M_{\rm cont}$) and assuming virial
equilibrium ($M_{\rm vir}$) and H$_2$ average number densities.}
\centering
\label{table:properties}
\small
\begin{tabular}{c c c c c}
\hline\hline
Source & $R$ & $M_{\rm cont}$  & $M_{\rm vir}$  & $n(\rm H_2)$\\
       & [pc] & [M$_{\sun}$] & [M$_{\sun}$] & [cm$^{-3}$]\\
\hline
IRAS 12326-6245 & 0.1 & 2200 &  410 & $ 1.2 \cdot 10^7$\\
G326.64+0.61    & 0.2 &  850 &  430 & $ 6.0 \cdot 10^5$\\
OH328.81+0.63   & 0.2 & 1200 &  320 & $ 8.4 \cdot 10^5$\\
IRAS 15566-5304 & 0.2 &  810 &  740 & $ 5.7 \cdot 10^5$\\
G330.95-0.19    & 0.2 & 8100 & 1200 & $ 5.7 \cdot 10^6$\\
G345.01+1.8N    & 0.1 &  410 &  550 & $ 2.3 \cdot 10^6$\\
IRAS 16562-3959 & 0.1 &  400 &  400 & $ 2.2 \cdot 10^6$\\
G345.00-0.23    & 0.1 &  960 &  870 & $ 5.4 \cdot 10^6$\\
NGC 6334 FIR-V  & 0.2 &  500 &  420 & $ 3.5 \cdot 10^5$\\
NGC 6334F       & 0.1 & 830 &   360 & $ 4.7 \cdot 10^6$\\
G351.77-0.54    & 0.03 & 120 &  150 & $ 2.5 \cdot 10^7$\\
G353.41-0.36    & 0.3 & 3500 &  960 & $ 7.3 \cdot 10^5$\\
W28 A2(2)       & 0.1 &  970 &    - & $ 5.5 \cdot 10^6$\\
W31 (2)         & 0.7 & 1600 & 3500 & $ 2.6 \cdot 10^4$\\
W33 CONT        & 0.3 & 4300 &  900 & $ 9.0 \cdot 10^5$\\
\hline
\end{tabular}
\end{table*}

\end{document}